\documentclass[11pt]{article}

\usepackage{a4}
\usepackage{epsfig}
\usepackage{amssymb}

\newif\iffigframe
\newif\ifbigfig
\bigfigtrue         % bigger version of figures at end of paper
\makeatletter
\newcommand{\mkfig}[5]{%
  \ifbigfig
    \expandafter\def\expandafter\mk@figs\expandafter{\mk@figs
      \begin{figure}[p]%
        \hbox to\hsize{\hss\figbox{#3\hsize}{#4}{#1}\hss}%
        \caption[#1]{\label{#1}#5}%
      \end{figure}}%
    \ignorespaces
  \else
    \begin{figure}[ht]%
      \hbox to\hsize{\hss\figbox{#2\hsize}{#4}{#1}\hss}%
      \caption[#1]{\label{#1}#5}%
    \end{figure}%
  \fi}
\newcommand{\clearfigs}{%
  \ifbigfig
    \expandafter\def\expandafter\mk@figs\expandafter{\mk@figs
      \clearpage}%
    \ignorespaces
  \fi}
\def\mk@figs{}
\def\mkfigs{\mk@figs\def\mk@figs{}}
\makeatother

\newcommand{\figbox}[3]{\hbox to#1\bgroup
  \dimen0=1bp \dimen1=#1\relax
  \def\a##1 ##2 ##3 ##4 ##5\\{\if!##1!\a##2 ##3 ##4 ##5 .\\\else
    \dimen3=##3\dimen0 \advance\dimen3 -##1\dimen0
    \dimen4=##4\dimen0 \advance\dimen4 -##2\dimen0
    \dimen5=\dimen4 \divide\dimen5 \dimen3
    \dimen2=\dimen1 \multiply\dimen2 \dimen5
    \multiply\dimen5 \dimen3 \advance\dimen4 -\dimen5
    \dimen5=\dimen1
    \loop \advance\dimen4 \dimen4 \divide\dimen5 2
    \ifnum\dimen5>0 \ifnum\dimen4<\dimen3 \else
      \advance\dimen4 -\dimen3 \advance\dimen2 \dimen5 \fi
    \repeat
    \dimen5=10\dimen1 \divide\dimen5 \dimen0
    \includegraphics{#3.eps}%
    \iffigframe \vrule\hss \else \hfil \fi
    \vbox to\dimen2\bgroup
      \iffigframe \hrule width\dimen1\vss \hrule \else \vfil \fi
      \egroup
    \iffigframe \vrule\hss \fi
    \egroup\fi}%
  \a#2 . . . .\\}

\newcounter{subequation}[equation]

\makeatletter
\expandafter\let\expandafter\reset@font\csname reset@font\endcsname
\newenvironment{subeqnarray}
  {\arraycolsep=1pt
    \def\@eqnnum\stepcounter##1{\stepcounter{subequation}{\reset@font\rm
      (\theequation\alph{subequation})}}\eqnarray}%
  {\endeqnarray\stepcounter{equation}}
\makeatother

\newcommand{\NoAlign}[1]{\noalign{\vskip\belowdisplayskip
  \noindent#1\par\vskip\abovedisplayskip}}

\newcounter{statement}
\newenvironment{statement}[4]
  {\par\penalty-100
    \vspace{1mm}\refstepcounter{statement}
    \noindent\vadjust{\nobreak}#1#2 \arabic{statement} #4\unskip: #3}
  {\par\vspace{1mm}}
\newenvironment{Thm}
  {\begin{statement}{\bf}{Theorem}{\sl}}{\end{statement}}
\newenvironment{Prop}
  {\begin{statement}{\bf}{Proposition}{\sl}}{\end{statement}}
\newenvironment{Lemma}
  {\begin{statement}{\bf}{Lemma}{\sl}}{\end{statement}}

\newenvironment{statement*}[4]
  {\par\penalty-100
    \noindent\vadjust{\nobreak}#1#2 #4\unskip: #3}
  {\par\vspace{2mm}}

\newenvironment{Rem}
  {\begin{statement*}{\sl}{Remark}{\rm}}{\end{statement*}}

\newcommand{\PrfMark}{\leavevmode\vrule width1.5exheight1.5exdepth0pt}
\newenvironment{Prf}
  {\begin{statement*}{\sl}{Proof}{\rm}}
  {\PrfMark\end{statement*}}
\newenvironment{PrfItem}
  {\begin{statement*}{\sl}{Proof}{\rm}}
  {\PrfMark\end{itemize}\end{statement*}}

\newcommand{\subs}[1]{_{\scriptscriptstyle\rm#1}}
\newcommand{\smsub}[1]{_{\scriptscriptstyle#1}}

\newcommand{\dd}{d}

\begin{document}

\thispagestyle{empty}

\hbox to\hsize{%
  \vbox{%
%       \hbox{DRAFT}%
%       \hbox{Submitted to}%
%       \hbox{\sl ???}%
        }\hfil
  \vbox{%
        \hbox{MPP-2004-169}%
        \hbox{December 16, 2004}%
        \hbox{gr-qc/0412067}%
        }}

\vspace{1.5cm}
\begin{center}
{\LARGE\bf
Classification of Static, Spherically Symmetric Solutions of the
Einstein-Yang-Mills Theory with Positive Cosmological Constant\\}

\vspace{1.5 cm}
\large
Peter Breitenlohner${}^a$,
Peter Forg\'acs${}^b$,
Dieter Maison${}^a$\\

\vspace{1 cm}

${}^a$
{\small\sl
Max-Planck-Institut f\"ur Physik\\
--- Werner Heisenberg Institut ---\\
F\"ohringer Ring 6\\
D--80805 Munich, Germany\\}

 \vspace{0.3 cm}

${}^b$
{\small\sl
Laboratoire de Math\'emathiques et Physique Th\'eorique\\
CNRS UMR6083\\
Universit\'e de Tours,
Parc de Grandmont\\
F--37200 Tours, France\\}

\end{center}

\vspace{20 mm}
\begingroup \addtolength{\leftskip}{1cm} \addtolength{\rightskip}{1cm}
\begin{abstract}\noindent
We give a complete classification of all static, spherically symmetric
solutions of the $SU(2)$ Einstein-Yang-Mills theory with a positive
cosmological constant. Our classification proceeds in two steps. We first
extend solutions of the radial field equations to their maximal interval of
existence. In a second step we determine the Carter-Penrose diagrams of all
4-dimensional space-times constructible from such radial pieces. Based on
numerical studies we sketch a complete phase space picture of all solutions
with a regular origin.

\end{abstract}

\endgroup
\newpage

\section{Introduction}

In contrast to the coupling of Abelian gauge fields to gravity, the
interaction of non-Abelian gauge fields with gravity yields a rich structure
of interesting solutions even in the spherically symmetric sector. While the
Reissner-Nordstr{\o}m black holes are the only static, spherically
symmetric, asymptotically flat solutions of the Einstein-Maxwell equations,
the non-Abelian SU(2) Einstein-Yang-Mills (EYM) theory has a much larger
class of such solutions. There is an infinite sequence (indexed by the
number $n$ of zeros of the Yang-Mills potential) of regular `soliton-type'
solutions, known as the Bartnik-McKinnon (BK) solutions
\cite{Bart,Smoller,BFM}. Besides there are corresponding families of black
holes \cite{Kuenzle,Bizonbh,Volkov,SWBH,BFM} and `limiting' solutions, when
$n$ tends to infinity \cite{BML}.

If a cosmological constant, $\Lambda$, is added to the theory, the
asymptotic behaviour of the regular solutions changes from Minkowskian to
deSitter for $\Lambda>0$ or Anti-deSitter for $\Lambda<0$. Furthermore for
$\Lambda>0$ the regular solutions have a cosmological horizon.

Numerical investigations of the static, spherically symmetric EYM equations
with a cosmological constant have led to a rather clear picture of the most
important properties of the solutions
\cite{Strau,Winstanley,Hosotani,neglam}. Let us briefly recall here some of
the most relevant numerical results of Refs.~\cite{Strau,Winstanley,Hosotani}.
In Ref.~\cite{Strau} families of solutions with a regular origin and a horizon
with $n=1,2,\ldots$ zeros were found. For $0<\Lambda<\Lambda_c(n)$ the
solutions are asymptotically deSitter. At $\Lambda=\Lambda_c(n)$ the global
structure of the spatial sections of the solutions changes. The area of the
2-spheres has a maximum (`equator') behind the horizon and then decreases
back to zero, where the solutions have a singularity, implying that the
spatial sections have the topology of 3-spheres. As $\Lambda$ increases the
horizon and the equator exchange their positions and as $\Lambda$ approaches
some maximal value $\Lambda_r(n)$ the radius $r_h$ of the horizon shrinks to
zero. The solutions for $\Lambda=\Lambda_r(n)$ are globally regular with
compact spatial sections of 3-sphere topology. The numerical results of
Ref.~\cite{Winstanley,Hosotani} for $\Lambda<0$ indicate the existence of
generic families of asymptotically Anti-deSitter soliton-type and black hole
solutions.

Besides the numerical results there are also a few analytical ones.
Unfortunately the well developed general theory of dynamical systems,
although very useful for local questions does not provide the tools required
to answer global questions in our case. Thus special properties of the
equations have to be exploited. For negative cosmological constant
Winstanley \cite{Winstanley} has given an existence proof for asymptotically
Anti-deSitter black holes with {\sl sufficiently large} radius $r_h$. On the
other hand for positive cosmological constant Linden \cite{Linden} has
proven the existence of smooth solutions with a regular origin for {\sl
sufficiently small\/} values of $\Lambda$. Clearly it would be desirable to
extend Linden's existence proof to the numerically found families up to the
3-sphere solutions and to black holes, as well as Winstanley's proof to all
asymptotically Anti-deSitter solutions.

Experience with the existence proofs for $\Lambda=0$ \cite{Smoller,BFM}
suggests that it is quite important to classify the possible {\sl global\/}
behaviour of solutions with a regular origin or a black hole horizon. Such a
classification is one of the fundamental problems in dynamical systems, and
no general method is known to solve it.

The main result of this paper is a complete classification of {\sl all\/}
static, spherically symmetric solutions of the EYM equations with a positive
cosmological constant. One should also mention here that a partial result in
this direction has been already obtained by Linden \cite{Lindencl}. Since
the field equations correspond to a rather complicated dynamical system, it
is highly remarkable that a complete global classification of the solutions
is possible at all.  It should, however, be remarked that the numerical
results show that the structure of the phase space as characterized by the
dependence of the solutions on the initial conditions is rather complicated.
This is in particular true inside black hole or outside cosmological
horizons%
\footnote{i.e.\ in the region where the Killing time is a space-like
variable}
as shown in \cite{Galtsov,BLM}.

Our classification proceeds in two steps. We first extend solutions of the
radial field equations to their maximal interval of existence. In a second
step we determine the Carter-Penrose diagrams of all 4-dimensional
space-times constructible from such radial pieces.

It may be appropriate to stress the importance of an interplay between
analytical and numerical methods. On the one hand it seems rather hopeless
to extract sufficient information on the complex structure of the phase
space just from studying the field equations analytically. On the other
hand precise and reliable numerics in particular concerning `critical'
solutions is impossible without some analytical understanding of the
behaviour of the solutions near singular points.

In Section~\ref{chappre} we set up the field equations. It turns out that
through the introduction of an auxiliary variable they take the form of a
system of coupled Riccati equations. In Section~\ref{chapsing} we study all
singular points and relate them to fixed points (f.p.s) of the dynamical
system. This requires the introduction of different independent and
dependent variables in each case, allowing to prove the existence and
uniqueness of local solutions. In Section~\ref{chapclass} we show that any
solution starting at an arbitrary regular point and integrated in both
directions either ends at one of the singular points described in
Section~\ref{chapsing} or runs towards $r=0$ performing a kind of
quasi-periodic motion around a repulsive focal point as described in
\cite{Galtsov,BLM}.  In Section~\ref{chap4d} we show how to glue together 2d
space-times corresponding to solutions of Eqs.~(\ref{taueq}) interpolating
between two singular points and to construct their Carter-Penrose diagrams
describing 4d inextensible (geodesically complete or singular) space-times.

Finally, Section~\ref{chapnum} contains numerical results illustrating
various types of solutions. In particular we illustrate how all cases of our
classification fit into a complete phase space picture.

The Appendix contains a proposition stating local existence and uniqueness
of solutions of a class of dynamical systems, allowing to parametrize the
solutions in terms of values at f.p. In addition we formulate several Lemmata
used in the proof of the classification theorems.

\section{Preliminaries}\label{chappre}

The line element of a spherically symmetric space-time can be written as
\begin{equation}\label{line}
\dd s^2=\dd s^2_2-r^2\dd\Omega^2\;,
\end{equation}
where $\dd s^2_2$ is the metric on the 2d orbit space factorizing out the
action of the rotation group and $\dd\Omega^2$ the invariant line element of
$S^2$. The 2d metric can always be brought to the diagonal form
\begin{equation}\label{2line}
\dd s_2^2=\sigma \Bigl(e^{2\nu}\dd t^2-e^{2\lambda}\dd R^2\Bigr)\;,
\end{equation}
where $\sigma=\pm1$ refers to regions, where $t$ is a time-like resp.\
space-like coordinate. Since we consider time-independent solutions it is
natural to choose $t$ to be the Killing time. This means that we can perform
a further dimensional reduction to a 1d theory. The metric functions $\nu$,
$\lambda$ and $r$ then only depend on the radial coordinate $R$. The latter
can be chosen arbitrarily exhibiting $\lambda$ as a gauge degree of freedom.

For the {\sl SU(2)\/} Yang-Mills field $W_\mu^a$ we use the standard minimal
spherically symmetric (purely `magnetic') ansatz
\begin{equation}\label{Ans}
W_\mu^a T_a \dd x^\mu=
  W(R) (T_1 \dd\theta+T_2\sin\theta \dd\varphi) + T_3 \cos\theta
\dd\varphi\;,
\end{equation}
where $T_a$ denote the generators of {\sl SU(2)\/}. The reduced EYM action
including a cosmological constant $\Lambda$ can be expressed (in suitable
units) as
\begin{eqnarray}\label{action}
S&=&-\int dR e^{(\nu+\lambda)}\Bigl[
  {1\over2}\Bigl(\sigma+e^{-2\lambda}((r')^2+\nu'(r^2)')\Bigr)
\nonumber\\&&\qquad\qquad\qquad\quad
  -e^{-2\lambda}W'^2-\sigma\Bigl({(W^2-1)^2\over2r^2}
   +{\Lambda r^2\over 2}\Bigr)\Bigr]\;.
\end{eqnarray}
In order to write the Euler-Lagrange equations in first order form we
introduce
\begin{equation}\label{first}
N\equiv e^{-\lambda}r',\quad \kappa\equiv re^{-\lambda}\nu'+N,\quad
U\equiv e^{-\lambda}W'\;,
\end{equation}
and obtain the equations of motion
\begin{subeqnarray}\label{feq}
   re^{-\lambda}N'&=&(\kappa-N)N-2U^2\;,\\
   re^{-\lambda}\kappa'&=&\sigma(1-2\Lambda r^2)+2U^2-\kappa^2\;,\\
   re^{-\lambda}U'&=&\sigma WT+(N-\kappa)U\;,
\end{subeqnarray}
where
\begin{equation}\label{defT}
T={W^2-1\over r}\;.
\end{equation}
In addition we have the only remaining diffeomorphism constraint
Eq.~(\ref{kappeq}) below.

We still have to fix the gauge, i.e.\ choose a radial coordinate. A simple
choice is to use the geometrical radius $r$ for $R$ (Schwarzschild
coordinates), yielding $e^{-\lambda}=N$. Putting $\mu\equiv\sigma N^2$,
$A\equiv e^{\nu}/N$, and $\Lambda=0$ Eqs.~(\ref{feq}) become equal to
Eqs.~(6) in~\cite{BFM}. This coordinate choice has the disadvantage that the
equations become singular at stationary points of $r$. This problem is
avoided using the gauge $e^{\lambda}=r$ introduced in \cite{BFM}, which will
also be used here. In order to stress the dynamical systems character of
Eqs.~(\ref{first},\ref{feq}) we denote the corresponding radial coordinate
by $\tau$ and $\tau$ derivatives by a dot. We thus get
\begin{subeqnarray}\label{taueq}
  \dot r&=&rN\;,\\
  \dot W&=&rU\;,\\
  \dot N&=&(\kappa-N)N-2U^2\;,\\
  \dot\kappa&=&\sigma\left(1-2\Lambda r^2\right)+2U^2-\kappa^2\;,\\
  \dot U&=&\sigma WT+(N-\kappa)U\;,
\end{subeqnarray}
and the constraint (invariant hypersurface)
\begin{equation}\label{kappeq}
2\kappa N-N^2=2U^2+\sigma(1-T^2-\Lambda r^2)\;.
\end{equation}
The metric function $\nu$ obeys the equation
\begin{equation}\label{nueq}
  \dot \nu=\kappa-N\;,
\end{equation}
that can be solved once the other functions are known. Using the constraint
Eq.~(\ref{kappeq}), the equation for $N$ can also be written as
\begin{equation}\label{Neq}
\dot N={\sigma\over2}(1-\Lambda r^2-T^2)-{1\over2}N^2-U^2\;.
\end{equation}
For $\Lambda=0$ and $\sigma=+1$ these equations coincide with the
Eqs.~(49,50) of~\cite{BFM}.

The quantity $T$, introduced in Eq.~(\ref{defT}) as an abbreviation, obeys
\begin{equation}\label{Teq}
\dot T=2UW-NT\;.
\end{equation}
Viewing $T$ as an additional dependent variable we obtain a system of
coupled Riccati equations~(\ref{taueq}) and~(\ref{Teq}) with two algebraic
constraints. This could explain, why solutions of this system stay bounded
except for simple poles at finite $\tau$-values (as is true for a single
Riccati equation).

An important quantity is the `mass function'
\begin{subeqnarray}\label{mass}
m&=&{r\over2}\Bigl(1-{\Lambda r^2\over3}-\sigma N^2\Bigr)\;,\\
\NoAlign{obeying the equation}
\dot m&=&
 \Bigl(\sigma U^2+{T^2\over2}\Bigr)rN\;.
\end{subeqnarray}

Furthermore there is a kind of $\tau$-dependent `energy' for the dynamical
system Eqs.~(\ref{taueq})
\begin{subeqnarray}\label{energy}
 E&=&{r^2\over4}\Bigl(2\kappa N-N^2-\sigma(1-\Lambda r^2)\Bigr)\\
 &=&{\dot W^2\over2}-\sigma{(W^2-1)^2\over4}\;,\\
\NoAlign{obeying}
\dot E&=&(2N-\kappa)\dot W^2\;,
\end{subeqnarray}
which will prove useful as a `Lyapunov Function'.

Due to the presence of the cosmological constant the vacuum solution
obtained for $W^2=1$ is no longer Minkowski space but deSitter space
\begin{equation}\label{dsmu}
\dd s^2=\mu\,\dd t^2-{\dd r^2\over\mu}-r^2\dd\Omega^2\;,
\end{equation}
with
\begin{equation}\label{deSitter}
\mu\subs{dS}(r)=1-{\Lambda r^2\over 3}\;.
\end{equation}
Because the radial direction changes from space-like to time-like at the
cosmological horizon $r_c=\sqrt{3/\Lambda}$ we have to subdivide the radial
domain into the intervals $0<r<r_c$ and $r_c<r<\infty$ if we use the
variables of Eqs.~(\ref{taueq}). The required sign change of $\sigma$ going
from one domain to the other can be achieved by the substitution
$(\dd\tau,N,\kappa,U)\to(-i\dd\tau,iN,i\kappa,iU)$.

The counterpart of the Schwarzschild solution for $\Lambda>0$ is the
Schwarz\-schild-deSitter (SdS) solution Eq.~(\ref{dsmu}) with
\begin{equation}\label{SdeSitter}
\mu\subs{SdS}(r)=1-{2M\over r}-{\Lambda r^2\over 3}\;,
\end{equation}
describing a black hole inside a cosmological horizon%
\footnote{We call a horizon where $r$ grows, while $\sigma$ in
Eq.~(\ref{2line}) becomes positive a black hole horizon and a horizon, where
$r$ grows, while $\sigma$ becomes negative, a cosmological horizon.}
as long as $\mu\subs{SdS}$ has two positive zeros, i.e.\
$M<1/3\sqrt{\Lambda}$. For $M=1/3\sqrt{\Lambda}$ the function
$\mu\subs{SdS}$ has a double zero at $r=1/\sqrt{\Lambda}$. Geometrically
this is the position of an extremal (degenerate) cosmological horizon. It
corresponds to the fixed point of Eqs.~(\ref{taueq})
$(W,U,r,N,\kappa)=(\pm1,0,1/\sqrt{\Lambda},0,\pm1)$. As for the extremal
Reissner-Nord\-str{\o}m solution the horizon is at infinite geodesic
distance in the space-like direction.

Likewise there is the Reissner-Nord\-str{\o}m-deSitter (RNdS) solution for
$W=0$ with
\begin{equation}\label{RNdeSitter}
\mu\subs{RN}(r)=1-{2M\over r}+{1\over r^2}-{\Lambda r^2\over 3}\;,
\end{equation}
carrying an abelian magnetic charge. For $\Lambda>1/4$ the function
$\mu\subs{RN}$ has only one positive zero, while for $\Lambda<1/4$ it has
three positive zeros as long as $M_-(\Lambda)<M<M_+(\Lambda)$ for certain
functions $M_\pm(\Lambda)$. For $M=M_-(\Lambda)$ the two smaller zeros merge
to a minimum of $\mu\subs{RN}$ at $r_-^2=(1-\sqrt{1-4\Lambda})/2\Lambda$,
while for $M=M_+(\Lambda)$ the two larger ones merge to a maximum of
$\mu\subs{RN}$ at $r_+^2=(1+\sqrt{1-4\Lambda})/2\Lambda$. For
$\Lambda=1/4$ and $M=2\sqrt{2}/3$ all three zeros merge to a triple zero at
$r=\sqrt{2}$. As long as $\mu\subs{RN}$ has three positive zeros the
solution describes a charged black hole located inside a cosmological
horizon. The horizon at the smallest value of $r$, a cosmological horizon
according to our terminology, corresponds to the Cauchy horizon of the
Reissner-Nord\-str{\o}m black hole. The two possible double zeros at
$r=r_\pm$ are the positions of extremal horizons of the black hole resp.\
cosmological type. Only the extremal black hole horizon has a limit for
$\Lambda\to0$.

Apart from these general solutions of deSitter type there are some
degenerate ones with $r$ and $W$ constant (and hence $N=U=0$) and a
space-time of the form $M_2\times S^2$. In fact these solutions correspond
to double zeros of $\mu\subs{SdS}$ or $\mu\subs{RN}$, resp.\ to the SdS and
RN fixed points (f.p.s) discussed in Section~\ref{sectfp}. From
Eqs.~(\ref{taueq},\ref{kappeq}) one finds that either $W^2=1$ (SdS) or $W=0$
(RN) and $\dot\kappa=\kappa_s^2-\kappa^2$ with
\begin{equation}
  \kappa_s^2=\sigma(1-2\Lambda r^2)\;,
  \qquad{\rm and}\qquad
  2\Lambda r^2=\cases{2&for $W^2=1\;,$\cr
  1\pm\sqrt{1-4\Lambda}&for $W=0\;.$}
\end{equation}
For $\kappa_s^2>0$ there are two solutions
\begin{subeqnarray}\label{Nari}
&\kappa(\tau)=\kappa_s\tanh(\kappa_s\tau)\;,\qquad\hfil
  &e^{2\nu}=\cosh^2(\kappa_s\tau)\;,
\\
&\kappa(\tau)=\kappa_s\coth(\kappa_s\tau)\;,\qquad\hfil
  &e^{2\nu}=\kappa_s^{-2}\sinh^2(\kappa_s\tau)\;,\\
\NoAlign{with the line element $\dd s_2^2=\sigma(e^{2\nu}\dd
t^2-r^2\dd\tau^2)$ known as the Nariai solutions \cite{Strau,Nariai}. For
$\kappa_s^2=-a^2<0$ one obtains}
&\kappa(\tau)=a\cot(a\tau)\;,\qquad\hfil
  &e^{2\nu}=a^{-2}\sin^2(a\tau)\;,\\
\NoAlign{and for $W=0$ in the limit $\Lambda\to1/4$, $\kappa_s\to0$}
&\kappa(\tau)={1\over\tau}\;,\qquad\hfil
  &e^{2\nu}=\tau^2\;.
\end{subeqnarray}
Geometrically the first three are $SO(2,1)/SO(1,1)$, while the last one is
flat space (in unusual coordinates).

\section{Classification of Singular Points}\label{chapsing}

Our aim is to give a complete classification of all solutions starting at an
arbitrary regular point. We may integrate Eqs.~(\ref{taueq}) in both
directions until either one of the dependent variables diverges for finite
$\tau$ or $\tau\to\pm\infty$ where the solution may converge (to a finite or
infinite value) or not. As the experience with a single Riccati equation
shows we have to expect singularities for finite values of $\tau$. In fact,
a typical example is given by non-degenerate horizons, where the function
$\kappa$ has a simple pole. Natural candidates for the convergent case for
$\tau\to\pm\infty$ are solutions running into one of the fixed points of
Eqs.~(\ref{taueq}). The latter are easily found due to the simple structure
of these equations. With exception of one case discussed in
Section~\ref{sectosz}, the singularities occurring at finite values of $\tau$
turn out to be simple poles. These singularities can all be transformed into
f.p.s by suitable changes of the independent and dependent variables.
Furthermore all the f.p.s except the RN~f.p.\ for $\Lambda=1/4$ are hyperbolic
ones. Using a technique known as `blow-up' \cite{Arnold,Dumortier} we are
able to reduce also this exceptional case to the hyperbolic one.

Reducing the study of singular points to that of hyperbolic f.p.s has the
advantage that the theory of dynamical systems provides powerful theorems
about the local behaviour of solutions in their neighbourhood
\cite{Codd,Hart,Arnold}. The theorem of Hartman and Grobman about the
existence of a stable (and unstable) manifold guarantees local existence and
uniqueness of solutions. In order to parametrize the solutions by values
determined at the f.p.\ itself, we rewrite the equations in the
(non-autonomous and non-hyperbolic) form required by Prop.~\ref{local} in
the appendix, whenever possible.

Most of the singularities are actually of geometrical origin. Since our
independent variable $\tau$ parametrizes the space of orbits of the isometry
group $G=R\times SO(3)$ of time-translations and rotations acting on the 4d
space-time $M_4$, we expect singularities at the fixed points of the group
action. For rotations this is $r=0$, the center of symmetry. It turns out
that the point $r=0$ can be reached for finite $\tau$, if $|N|\to\infty$
(curvature singularity) or for $\tau\to\pm\infty$ with $N\to\mp1$ (regular
origin). Regular (non-degenerate) horizons are represented in the quotient
space by their `bifurcation surface' \cite{hawking} consisting of f.p.s of
the time-translations. The required vanishing of $e^\nu$ is possible only if
$\kappa$ diverges at some finite $\tau$. In contrast, the degenerate
horizons we find require $\tau\to\pm\infty$ and finite $\kappa$. A deSitter
type singularity reached for finite $\tau$ occurs for $r\to\infty$ in the
region $\sigma=-1$ due to the presence of the cosmological constant.

In the subsequent paragraphs we shall give a characterization of the
solutions near the singular points discussed above, providing the basis for
our classification theorems.

\subsection{Regular Origin}

As a first case there are solutions for $\sigma=+1$ with a regular origin
$r\to0$, $W^2\to1$, and $N\to\pm1$ for $\tau\to\pm\infty$. These are
f.p.s of Eqs.~(\ref{taueq},\ref{kappeq}) with two dimensional stable
manifolds. In order to characterize the corresponding one parameter families
of solutions by a value $b$ determined at the f.p., one rewrites
Eqs.~(\ref{taueq},\ref{kappeq}) in the form required by Prop.~\ref{local} as
in \cite{BFM}. Choosing $W\to+1$ without restriction one thus obtains unique
solutions which are analytic in $b$ and $r$ with
\begin{subeqnarray}\label{bcor1}
  W(r)&=&1-br^2+O(r^4)\;,\\
  N^2(r)&=&1-\Bigl(4b^2+{\Lambda\over3}\Bigr)r^2+O(r^4)\;.
\end{subeqnarray}

\subsection{Reissner-Nordstr{\o}m Singularity at \boldmath$r=0$}

The generic behaviour at $r=0$ is singular in complete analogy to the case
$\Lambda=0$ \cite{BFM}, since the terms proportional to $\Lambda$ may be
neglected for $r\to 0$. The behaviour near $r=0$ depends on the sign of
$\sigma$. For $\sigma=+1$ one finds a Reissner-Nord\-str{\o}m type
singularity of the metric
\begin{subeqnarray}\label{RN}
  W(r)&=&W_0-{W_0\over 2(W_0^2-1)}r^2+W_3r^3+O(r^4)\;,\\
  N^2(r)&=&{(W_0^2-1)^2\over r^2}-{2M_0\over r}+O(1)\;,
\end{subeqnarray}
with three parameters $M_0$, $W_0$, and $W_3$. As was shown in \cite{BLM}
the singularity may be described as a f.p.\ introducing a suitable set of
dependent variables. Here we shall give an alternative formulation using
Schwarzschild coordinates and the dependent variables already introduced in
Prop.~13 of~\cite{BFM}.
\begin{subeqnarray}\label{RNsing}
r{\dd\over\dd r}\bar N&=&r\Bigl(\kappa\bar N-2rU^2\Bigr)/\bar N\;,\\
r{\dd\over\dd r}\kappa&=&r\Bigl(1-2\Lambda r^2+2U^2-\kappa^2\Bigr)/\bar N\;,\\
r{\dd\over\dd r}W&=&r^2U/\bar N\;,\\
r{\dd\over\dd r}U&=&r\Bigl(\lambda-\kappa U\Bigr)/\bar N\;,\\
r{\dd\over\dd r}\lambda&=&r\Bigl(3W^2-2U^2-1\Bigr)U/\bar N\;,
\end{subeqnarray}
with $\bar N=rN$ and $\lambda=WT+NU$. These equations are precisely of the
form required for the application of Prop.~\ref{local} implying the
regularity of local solutions as a function of $r$ obeying the constraints
\begin{equation}\label{conds}
\bar N(0)=\pm(W^2(0)-1)\;,\qquad \bar N(0)U(0)=W(0)(W^2(0)-1)\;,
\end{equation}
following from the finiteness of $\lambda(0)$ and the constraint
Eq.~(\ref{kappeq}). Geometrically this behaviour leads to a
Reissner-Nord\-str{\o}m type curvature singularity at which the space-time
is geodesically incomplete.

\subsection{Pseudo-Reissner-Nordstr{\o}m Singularity at \boldmath$r=0$}

There is an analogous singularity for $\sigma=-1$ called pseudo-RN
singularity in \cite{BLM}, with a rather different behaviour of the metric
and of the YM potential $W$ compared to the case $\sigma=+1$.
\begin{subeqnarray}\label{pRNsing}
W(r)&=&W_0\pm r+O(r^2)\;,\\
N^2(r)&=&{(W_0^2-1)^2\over r^2}\mp{4W_0(W_0^2-1)\over r}+O(1)\;,
\end{subeqnarray}
The description as a f.p.\ was already given in \cite{BLM}, but here we give
an alternative description, providing a unified treatment of the two
different signs of $W'$ at $r=0$. We introduce the dependent variables $\bar
U=U/N$, $\bar T=T/N$,
\begin{equation}\label{pRNvar}
  x={\bar U^2-1\over r}+{2W\bar U\bar T^2\over W^2-1}\;,
\qquad{\rm and}\qquad
  y={\bar T^2-1\over r}\;.
\end{equation}
Using Schwarzschild coordinates one finds
\begin{subeqnarray}\label{RNs}
r{\dd\over\dd r}W&=&rf\smsub{W}\;,\qquad\qquad\qquad\quad
  f\smsub{W}=\bar U\;,\\
r{\dd\over\dd r}\bar U&=&rf\smsub{U}\;,\qquad\qquad\qquad\quad
  f\smsub{U}=-{W\bar T^2\over W^2-1}+2z\bar U\;,\\
r{\dd\over\dd r}\bar T&=&rf_T\;,\qquad\qquad\qquad\quad
  f_T=\Bigl[x+z-{2W\bar U(\bar T^2-1)\over W^2-1}\Bigr]\bar T\;,\\
\noalign{\penalty-100}
r{\dd\over\dd r}x&=&-x-2y+rf_x\;,\qquad~
  f_x=-2xy+2\Bigl[2(x+y+2z)W\bar U+\bar U^2
\nonumber\\\noalign{\nobreak}&&\qquad
    +{(1-\Lambda r^2)\bar U^2+W^2(2\bar U^2-\bar T^2-4\bar U^2\bar T^2)
      \over W^2-1}
  \Bigr]{\bar T^2\over W^2-1}\;,\qquad\\
\noalign{\penalty-100}
r{\dd\over\dd r}y&=&2x-2y+rf_y\;,\qquad~~
  f_y=2(x+z)y
\nonumber\\\noalign{\nobreak}&&\qquad\qquad\qquad\qquad\qquad\quad
  +\Bigl[{1-\Lambda r^2\over W^2-1}
    -4W\bar Uy\Bigr]{\bar T^2\over W^2-1}\;,\\
\NoAlign{with $z=[r(1-\Lambda r^2)\bar T^2/(W^2-1)^2-y]/2$. In their
linearized form these equations have zero eigenvalues for $W$, $\bar U$, and
$\bar T$, while $x$ and $y$ contribute with the complex conjugate
eigenvalues $(-3\pm\sqrt{15}i)/2$ representing divergent modes. Introducing
the linear combinations (eigenmodes) $x_\pm=(1\pm\sqrt{15}i)x-4y$ we obtain
}
r{\dd\over\dd r}x_\pm&=&{-3\pm i\sqrt{15}\over2}x_\pm+rf_\pm\;,
\qquad
f_\pm=(1\pm\sqrt{15}i)f_x-4f_y\;,
\end{subeqnarray}
satisfying the requirements of Prop.~\ref{local}. We thus obtain local
solutions of Eqs.~(\ref{RNs}) with $x(0)=y(0)=0$ depending analytically on
$W(0)\ne\pm1$, $\bar U(0)$, $\bar T(0)$, and $r$. Imposing
Eqs.~(\ref{pRNvar}) as constraints we then obtain the conditions $\bar
U^2(0)=\bar T^2(0)=1$ and from the constraint Eq.~(\ref{kappeq}) we find
$\kappa/N=1+\bar U^2-rz\to2$ for $r\to0$.

\subsection{Schwarzschild Singularity at \boldmath$r=0$}

The form of the RN~type singularities requires that $W^2(0)\ne1$, while for
$W^2(0)=1$ we obtain a Schwarzschild type behaviour ($N^2\sim1/r$) with the
sign of the mass determined by $\sigma$. This case requires a different
linearization, which we shall give subsequently. Assuming without
restriction $W(0)=+1$, we introduce the dependent variables $\bar
W=(W-1)/r^2$, $\bar U=U/rN$, $k=\kappa/N$, $\bar N=\sigma rN^2$, and
$y=(1-2\kappa/N+2r^2\bar U^2)/r$, use the constraint Eq.~(\ref{kappeq}) to
express $y=\bar N^{-1}(1-r^2(2+r^2\bar W)^2\bar W^2-\Lambda r^2)$, and find
\begin{subeqnarray}\label{SSsing}
r{\dd\over\dd r}\bar N&=&rf\smsub{N}\;,\qquad
  f\smsub{N}=(y-2r\bar U^2)\bar N\;,\\
r{\dd\over\dd r}\bar W&=&\bar U-2\bar W\;,\\
r{\dd\over\dd r}\bar U&=&rf\smsub{U}\;\qquad~
  f\smsub{U}={(1+r^2\bar W)(2+r^2\bar W)\bar W\over \bar N}+y\bar U\;.\\
\NoAlign{In order to bring Eq.~(\ref{SSsing}b) into the form required by
Prop.~\ref{local}, we further introduce $\hat W=\bar W-\bar U/2$ with}
r{\dd\over\dd r}\hat W&=&-2\hat W-{rf\smsub{U}\over2}\;,
\end{subeqnarray}
and obtain solutions with $(\bar W,\bar U,\bar N)$ tending to finite values
$(-b,-2b,-2M)$ with $\sigma M<0$. The corresponding asymptotic behaviour at
$r=0$ is
\begin{subeqnarray}\label{SSasy}
W&=&1-br^2+O(r^4)\;,\\
\sigma N^2&=&-{2M\over r}+1+O(r)\;,\\
\kappa&=&{N\over 2}(1+O(r))\;.
\end{subeqnarray}

The RN and S~type singularities at $r=0$ are different singular points with
different asymptotic behaviour. The S~case (for $\sigma=+1$) can, however,
be seen as the limit $W_0\to\pm1$ of the more general RN case.
\mkfig{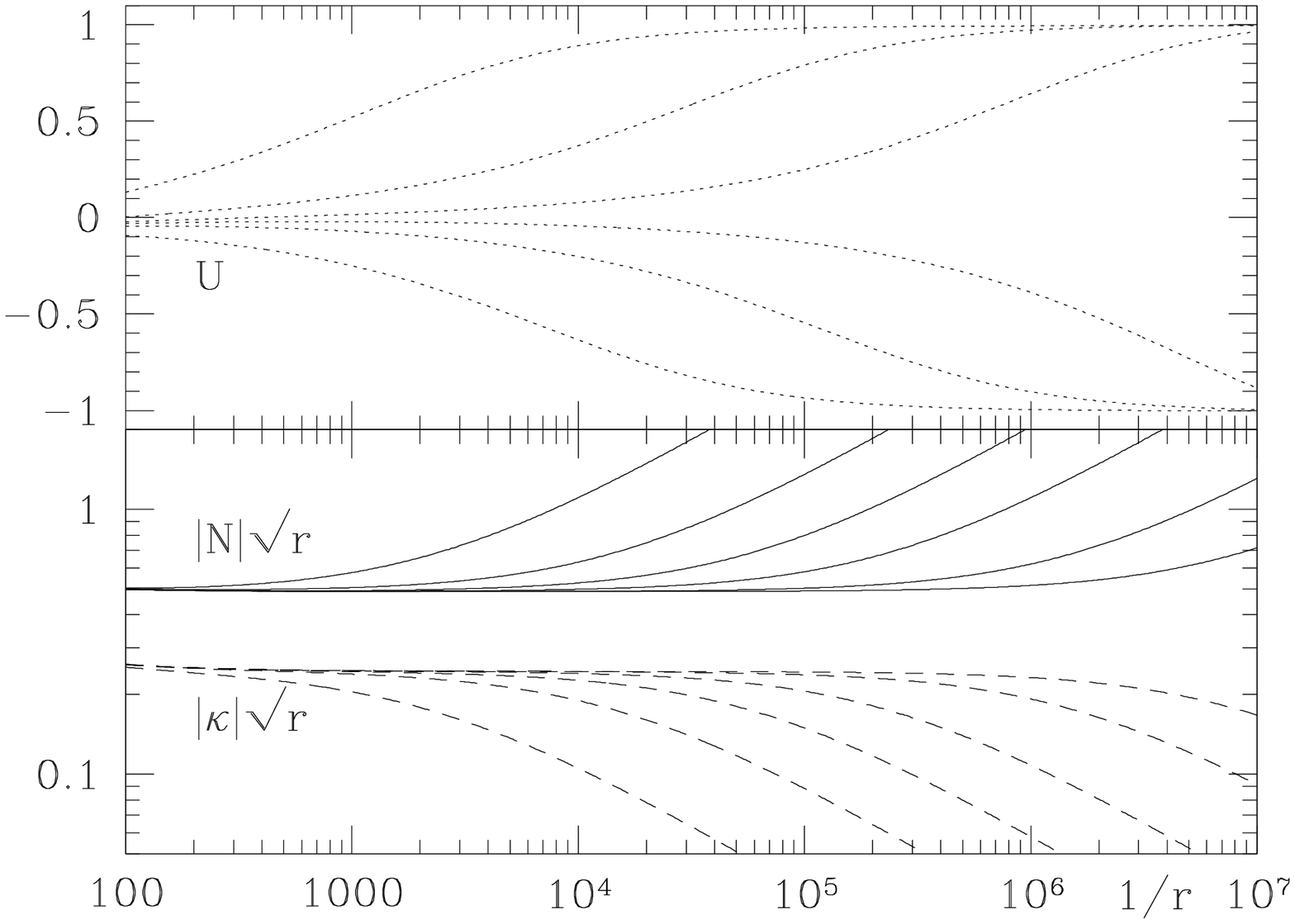}{0.6}{0.9}{63 179 562 565}
{Solutions near a singular origin with $|W_0|\approx1$}
In Fig.~\ref{figwto1} we present numerical solutions for $\Lambda=0.8$
starting at a regular origin with $b\sim0.237$ as they approach the RN~type
singularity with $W_0\approx-1$. $N$ and $\kappa$ grow like $1/\sqrt{r}$ and
$U\approx0$ as for the S~type singularity in an interval that increases as
$|W_0^2-1|$ decreases until eventually $N$ starts to grow like $1/r$ whereas
$\kappa$ has a finite limit and $U$ tends to $\pm W_0$. Thus in the limit
$|W_0|\to1$ the functions $W(r)$, $U(r)$, $N(r)$, and $\kappa(r)$ converge
pointwise to those for the S~type singularity.

\subsection{Unbounded Oscillations at \boldmath$r=0$}\label{sectosz}

The nature of the f.p.\ is rather different in the RN and pseudo-RN case.
While for $\sigma=+1$ the f.p.\ is a simple attractor it is a partially
repulsive (in the $\bar U$, $\bar T$ subsystem) focal point for $\sigma=-1$.

The RN singularity at the origin for $\sigma=+1$ is therefore generic (i.e.\
persistent for sufficiently small changes in the initial data and
$\Lambda$), as is the deSitter asymptotics for $\sigma=-1$ described below.
All other singular points discussed above and below are not generic because
they require one or more divergent modes to be suppressed.

The generic behaviour for $\sigma=-1$ and $r$ decreasing is somewhat more
complicated. As described in \cite{Galtsov,BLM} this leads to an oscillatory
behaviour of the solutions as they approach $r=0$ with $\kappa\to-\infty$
and unbounded oscillating $U$ and $N$. We introduce new dependent variables
$n=N/\kappa$, $u=U/\kappa$, and $t=T/\kappa$ which are bounded due to
Eq.~(\ref{kappeq}) since $\kappa$ diverges. In addition we introduce a new
independent variable $\rho$ with $\dd\rho=|\kappa|\dd\tau$ such that
$\rho\to\infty$ as $\kappa\to-\infty$ (see the proof of
Prop.~\ref{divergent-} for details) and denote the derivative with respect
to $\rho$ by a prime. Using $z=1/|\kappa|$ we find
\begin{subeqnarray}\label{rutnz}
r'&=&-rn\;,\\
u'&=&(2u^2-n+(2\Lambda r^2-1)z^2)\,u-Wzt\;,\\
t'&=&(2u^2+n-1+(2\Lambda r^2-1)z^2)\,t+2Wzu\;,\\
n'&=&(2u^2+n-2+(2\Lambda r^2-1)z^2)\,n+2u^2\;,\\
z'&=&(2u^2-1+(2\Lambda r^2-1)z^2)\,z\;,\\
\NoAlign{and the constraint}
2n&=&n^2+2u^2+t^2+z^2(\Lambda r^2-1)\;,
\end{subeqnarray}
with $|W|\sqrt{z}=\sqrt{z-rt}$ bounded. Neglecting terms that vanish as
$z\to0$ we obtain the asymptotic equations
\begin{subeqnarray}\label{utn}
u'&=&(2u^2-n)\,u\;,\\
t'&=&(2u^2+n-1)\,t\;,\\
n'&=&(2u^2+n-2)\,n+2u^2\;,\\
\NoAlign{with the constraint}
2n&=&n^2+2u^2+t^2\;,
\end{subeqnarray}
or, in terms of new dependent variables $\xi=n=N/\kappa$,
$\eta=2u^2/n=2U^2/N\kappa$, and $\zeta=t^2/n=T^2/N\kappa$
\begin{subeqnarray}\label{xez}
{\xi'\over\xi}=\xi-2+(\xi+1)\eta
    &=&\xi(2-\xi)-(\xi+1)\zeta\;,\\
{\eta'\over\eta}=2-\eta+(\eta-3)\xi
    &=&-(2-\eta)^2+(3-\eta)\zeta\;,\\
{\zeta'\over\zeta}=\zeta-2+(4-\xi-\zeta)\xi
    &=&2-\zeta-(\eta+\zeta)\eta\;,\\
\NoAlign{with the constraint}
2=\xi+\eta+\zeta\;.&&
\end{subeqnarray}
Fig.~\ref{figchaos} shows four trajectories as they emerge from the focal
point $(\xi,\eta,\zeta)=(1/2,1,1/2)$ corresponding to the pseudo-RN
singularity and rapidly approach the bounding triangle consisting of the
three separatrices $\xi=0$, $\eta=0$, and $\zeta=0$ connecting the three
hyperbolic fixed points $A=(2,0,0)$ corresponding to the S singularity,
$B=(0,0,2)$ corresponding to a regular horizon described below, and
$C=(0,2,0)$.
\mkfig{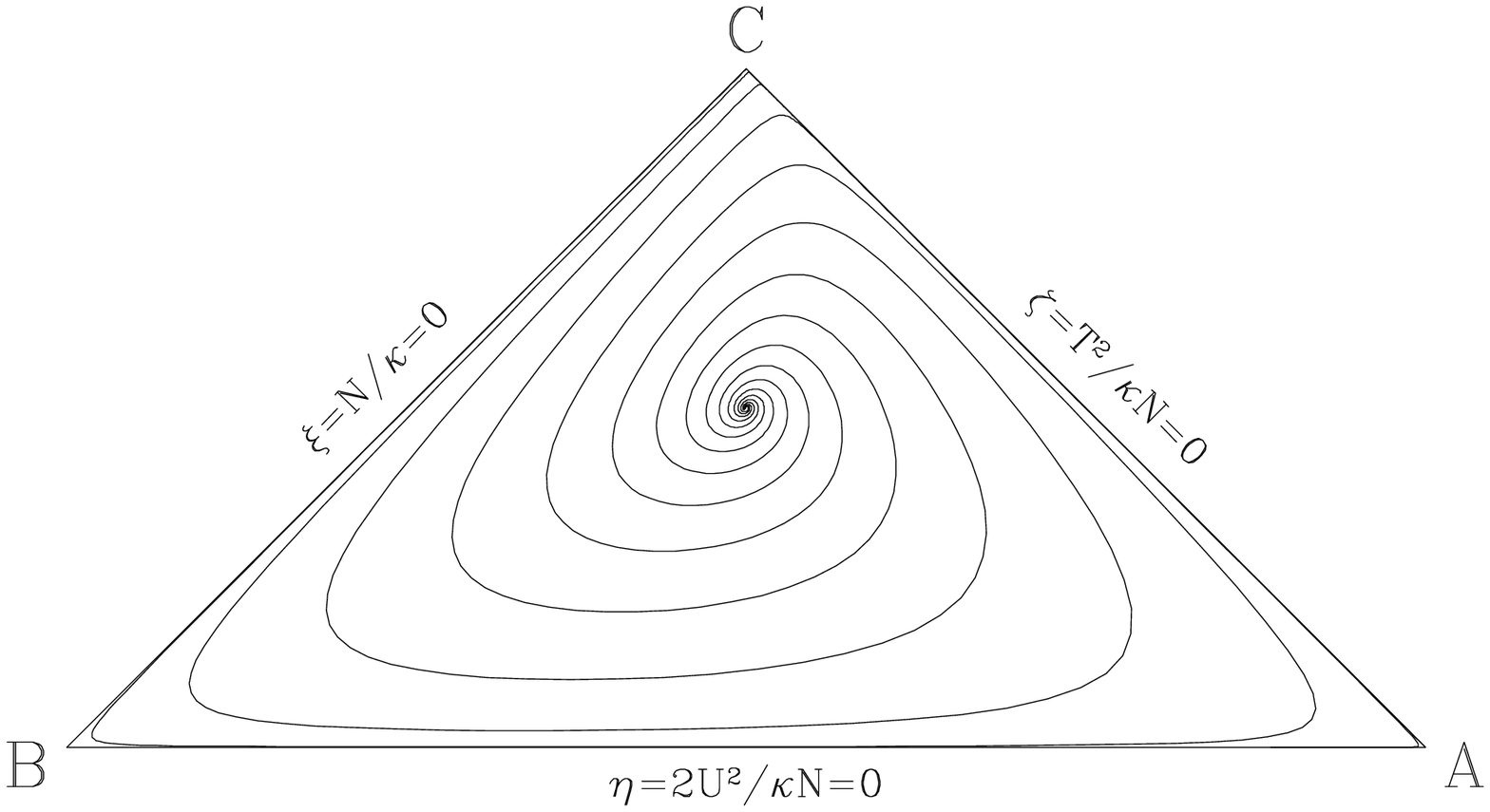}{0.5}{0.9}{39 166 571 454}
{Four trajectories of Eqs.~(\ref{xez}) emerging from the focal point
$(\xi,\eta,\zeta)=(1/2,1,1/2)$ and rapidly approaching the bounding
triangle}

The behaviour referred to as `chaotic' in \cite{BLM} is due to this rapid
approach to the boundary. The terms proportional to $W$ in the full
Eqs.~(\ref{rutnz}) may become relevant when $\eta\ll z$ or $\zeta\ll z$. A
solution may then either end with an S singularity or regular horizon or
nearly miss them with a resulting substantial change in $t$ or $u$. There is
no such instability for $\xi\ll z$.

\subsection{Regular Horizons}

In the case of a regular, i.e.\ non-degenerate horizon we may use
$s=1/\kappa$ as the independent variable tending to zero. We thus replace
Eqs.~(\ref{taueq}) by
\begin{subeqnarray}\label{hor}
s{\dd\over\dd s}\tau&=&-{\kappa\over{\dot\kappa}}=
  s\left(1+s^2F(r,U)\right)\;,\\
s{\dd\over\dd s}r&=&s\left(1+s^2F(r,U)\right)rN\;,\\
s{\dd\over\dd s}W&=&s\left(1+s^2F(r,U)\right)rU\;,\\
s{\dd\over\dd s}U&=&-U+s\left(1+s^2F(r,U)\right)
  (\sigma WT+NU)-s^2FU\;,\quad
\end{subeqnarray}
with
\begin{equation}\label{horF}
F(r,U)={1\over1-s^2f}\;,
\qquad
f=\sigma\left(1-2\Lambda r^2\right)+2U^2\;,
\end{equation}
while the function $N$ can be computed from the constraint
Eq.~(\ref{kappeq}). These equations have precisely the form required in
Prop.~\ref{local} with $U\to 0$ and $\tau$, $r$, $W$ tending to finite
values $\tau_h$, $r_h$, $W_h$.

In terms of the coordinate $\tau$ the behaviour near the horizon
is (performing a shift in $\tau$ we arrange for $\tau_h=0$)
\begin{subeqnarray}\label{bcbh}
  r(\tau)&=&r_h\left(1+N_1{\tau^2\over2}\right)+O(\tau^4)\;,\\
  N(\tau)&=&N_1(\tau+O(\tau^3))
    -W_1^2\tau^3+O(\tau^5)\;,\\
  W(\tau)&=&W_h+r_h W_1{\tau^2\over2}+O(\tau^4)\;,
\end{subeqnarray}
where
\begin{equation}\label{w1def}
  N_1={\sigma\over2}(1-\Lambda r^2-T^2)\Bigr|_h\;,\qquad
  W_1={\sigma\over2}WT\Bigr|_h\;.
\end{equation}

According to our definitions we have a black hole horizon if $\sigma N_1>0$
or a cosmological horizon if $\sigma N_1<0$. If $N_1$ vanishes the expansion
of $N$ starts with the cubic term with a corresponding change in the
expansion of $r$. Geometrically this describes the situation that the
horizon coincides with an equator, i.e.\ a maximum of $r$. The condition
$\sigma N_1>0$ determines the domains $C_+=\{1-\Lambda r^2-T^2>0\}$ in the
$(r,W)$ plane. In terms of $r^2$ and $W^2$ the boundary curves are (parts
of) ellipses. In Fig.~\ref{figellip} these domains are shown for various values
of $\Lambda$, for $\Lambda>1/4$ they become disconnected.
\mkfig{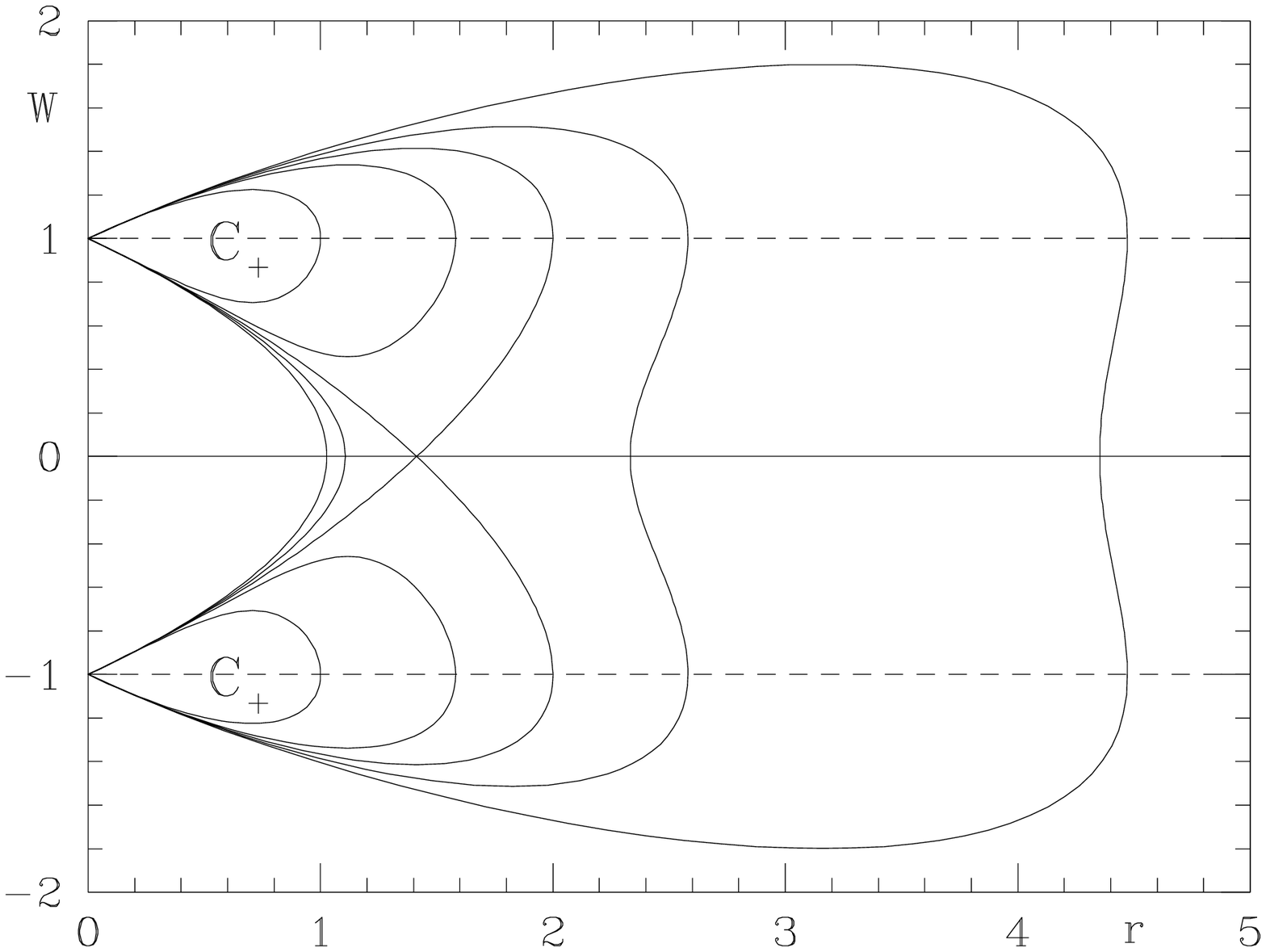}{0.5}{0.9}{50 173 574 570}
{Allowed domains $C_+$ for black hole boundary conditions for $\Lambda=0.05$,
$0.15$, $0.25$, $0.4$, and $1$}

Solutions with a regular origin can be seen as a limit $r_h\to0$ of those
with a regular horizon.
\mkfig{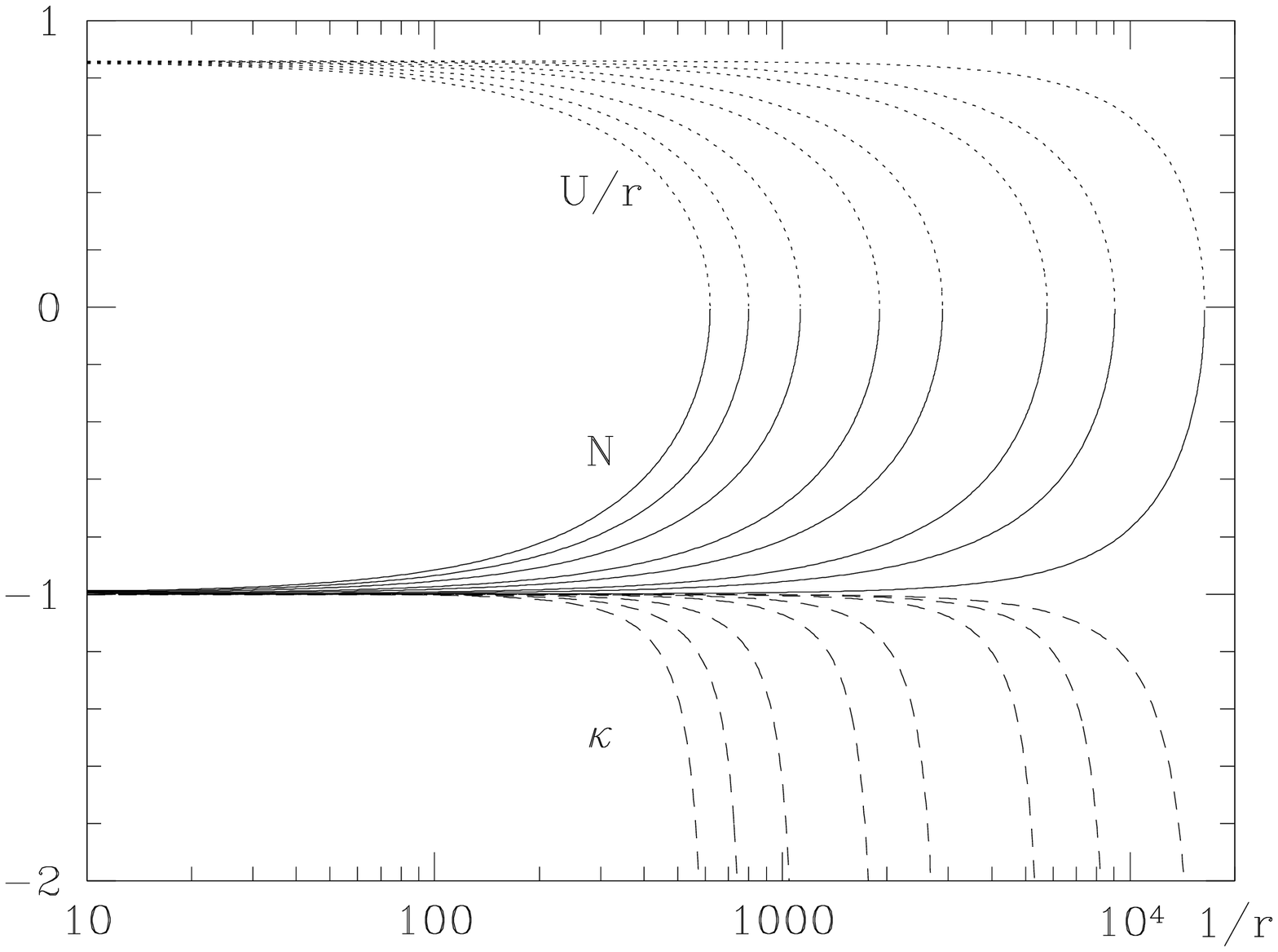}{0.6}{0.9}{63 179 562 565}
{Solutions near a regular horizon with $r_h\ll1$}
In Fig.~\ref{figrhto0} we present solutions for $\Lambda\sim0.3642$ starting
at a regular origin with $b\sim0.4296$ (i.e., near the point $R_2$ of
Fig.~\ref{figblam}) as they approach a regular horizon of decreasing radius.
$N\approx\kappa\approx-1$ and $U/r\approx2b$ as for the regular 3-sphere
solution with 2 zeros of $W$ in an interval that increases as $r_h\to0$
until eventually $\kappa$ starts to diverge whereas $N$ and $U$ tend to
zero. Thus the black hole solutions converge to the one with a regular
origin.

\subsection{deSitter Asymptotics}

For $\sigma=-1$ the radial variable $r$ can tend to $\infty$ leading to
deSitter asymptotics
\begin{subeqnarray}\label{asydS}
W&=&W_0+{W_1\over r}+O(1/r^2)\;,\\
N^2&=&{\Lambda\over 3}r^2-1+{2M_0\over r}+O(1/r^2)\;.
\end{subeqnarray}
We use $s=1/r$ as independent variable and introduce the dependent variables
$\bar U=UN$, $\bar N=N/r$, and the `mass function' Eq.~(\ref{mass}).
\begin{subeqnarray}\label{lindS}
s{\dd\over\dd s}W&=&-s{\bar U\over {\bar N}^2}\;,\\
s{\dd\over\dd s}\bar U&=&s\Bigl(W(W^2-1)+2s^3{\bar U^3\over\bar N^4}\Bigr)\;,\\
s{\dd\over\dd s}m&=&s\Bigl({\bar U^2\over\bar N^2}-(W^2-1)^2\Bigr)\;,
\end{subeqnarray}
with $\bar N^2=\Lambda/3-s^2+2s^3m$. Thus all the requirements of
Prop.~\ref{local} are fulfilled.

\subsection{The Schwarzschild-deSitter and
  Reissner-Nordstr{\o}m\\Fixed Points}\label{sectfp}

Besides the singular points reached at finite $\tau$ there are the f.p.s of
Eqs.~(\ref{taueq}) with $(r,W,U,N,\kappa)\to(r_s,W_s,U_s,N_s,\kappa_s)$ as
$\tau\to\infty$. One such f.p.\ is the regular origin $r_s=0$ already
discussed. For finite $r_s\neq0$ this requires $U_s=N_s=0$ and either
$W_s^2=1$ or $W_s=0$, which we denote as Schwarzschild-deSitter (SdS) resp.\
Reissner-Nordstr{\o}m (RN) f.p. The finite values of $\kappa_s$ and $r_s$
are determined by Eqs.~(\ref{taueq}d) and~(\ref{kappeq}). In order to prove
the local existence of solutions running into these f.p.s we will apply the
usual linearization technique, using variables $\bar r=r/r_s-1$, $\bar
W=(W-W_s)/r$, and $\bar\kappa=\kappa-\kappa_s$.

The SdS f.p.\ requires $\sigma=-1$ in order to get $\kappa_s^2>0$ and we find
$r_s=1/\sqrt{\Lambda}$ and $\kappa_s=\pm1$. From Eqs.~(\ref{taueq}) we obtain
\begin{subeqnarray}\label{fp1}
  \dot N&=&\kappa_sN+f_1\;,\\
  \dot{\bar \kappa}&=&-2\kappa_s(\bar\kappa+N)+f_2\;,\\
  \dot{\bar W}&=&U+f_3\;,\\
  \dot U&=&-2\bar W-\kappa_sU+f_4\;,
\end{subeqnarray}
and the constraint Eq.~(\ref{kappeq}) yields $\bar r=N/\kappa_s+f_5$. The
functions $f_i$, $i=1,\ldots,5$ are analytic and at least quadratic in the
dependent variables $\bar W$, $U$, $N$, and $\bar\kappa$.
The eigenvalues of the linear system are $\kappa_s=\pm1$, $-2\kappa_s=\mp2$
for the $N$, $\kappa$ part and $(-\kappa_s\pm\sqrt{7}i)/2$ in the $\bar W$,
$U$ sector. For $\kappa_s=1$ there are three convergent modes for
$\tau\to\infty$, while for $\kappa_s=-1$ there is only one convergent mode.
For $\kappa_s=1$ the function $W$ performs damped oscillations of period
$4\pi/\sqrt{7}$ around the limiting value $W_s=\pm1$. For $\kappa_s=-1$
both divergent modes of the $W$, $U$ subsystem have to be suppressed.
However, the r.h.s.\ of Eqs.~(\ref{fp1}c,d) vanish for $\bar W=U=0$ and thus
the stable manifold of the f.p.\ contains only the SdS solution
Eq.~(\ref{SdeSitter}) with $W\equiv\pm1$ and a double zero of
$\mu\subs{SdS}$. Similarly for $\kappa_s=+1$ the divergent mode of $N$ has
to be suppressed and thus $N(\tau)\approx\int_\tau^\infty2U^2e^{\tau-\tau'}\dd\tau'\ge0$.
Except for the Nariai solutions~(\ref{Nari}a,b) with $\bar W\equiv\bar r\equiv0$,
the f.p.\ can only be reached from $r<r_s$ with $N>0$.

The RN f.p.\ with $W_s=0$  leads to
$r_s^2=r_\pm^2=(1\pm\sqrt{1-4\Lambda})/2\Lambda$ and
$\kappa_s^2=\sigma(1-2\Lambda r_\pm^2)$. It requires $\Lambda\leq1/4$ and
the combinations $\sigma=+1$, $r_-$ resp.\ $\sigma=-1$, $r_+$ to obtain
$\kappa_s^2\geq0$. For the linearization we assume $\Lambda<1/4$, i.e.\
$\kappa_s\ne0$ and obtain
\begin{subeqnarray}\label{fp2}
  \dot N&=&\kappa_sN+f_1\;,\\
  \dot{\bar \kappa}&=&-4\sigma\Lambda r_s^2{N\over\kappa_s}
               -2\kappa_s\bar\kappa+f_2\;,\\
  \dot{\bar W}&=&U+f_3\;,\\
  \dot U&=&-\sigma\bar W-\kappa_sU+f_4\;,
\end{subeqnarray}
where $\bar r=N/\kappa_s+f_5$ is again obtained from the constraint.
The eigenvalues of the linear system are $\kappa_s$, $-2\kappa_s$ for
$N$, $\kappa$ and $(-\kappa_s\pm\sqrt{\kappa_s^2-4\sigma})/2$ for the
$\bar W$, $U$ system. Since $\kappa_s^2<1$ the latter eigenvalues are
complex conjugate for $\sigma=+1$ and real for $\sigma=-1$. For $\sigma=+1$
there are three convergent modes for $\kappa_s>0$ and only one convergent
mode for $\kappa_s<0$, whereas for $\sigma=-1$ there are always two
convergent modes. The need to suppress the divergent modes for $\sigma=+1$
and $\kappa_s<0$ again allows only the local (and thus global) solution
$W\equiv 0$, i.e.\ the stable manifold contains only the RNdS solution. The
period of the damped oscillations of $W$ around zero in the case
$\sigma=+1$, $\kappa_s>0$ is $4\pi/\sqrt{4-\kappa_s^2}$ and, as in the SdS
case, the f.p.\ can only be reached from $r<r_s$ with $N>0$, except for the
Nariai solutions with $W\equiv\bar r\equiv0$. For $\sigma=-1$, $\kappa_s>0$
there are again the Nariai solutions and solutions with $W\not\equiv0$,
$N>0$, and $r<r_s$. For $\sigma=-1$, $\kappa_s<0$ there is the RNdS solution
and there are solutions with $W\not\equiv0$.

\subsection{The Reissner-Nordstr{\o}m Fixed Point for \boldmath$\Lambda=1/4$}
\label{specialRN}

For $\Lambda=1/4$ some of the eigenvalues of the linearized Eqs.~(\ref{fp2})
vanish and this degenerate case requires special consideration. First we
introduce new dependent variables
\begin{equation}\begin{array}{c}%\displaystyle
y\smsub{W}\equiv\bar W=W/r\;,\quad
y\smsub{U}\equiv\bar U=U+\kappa\bar W/2\;,
\\%\displaystyle
y_\kappa\equiv\kappa\;,\quad
y_r\equiv\bar r=1-r^2/2\;,\quad
y\smsub{N}\equiv\bar N=N+\bar W\bar U\;,
\end{array}\end{equation}
and in addition
\begin{equation}
y\smsub{A}\equiv A=\sigma\bar W^2+\bar U^2\;.
\end{equation}
Next we rewrite Eqs.~(\ref{taueq}) and the constraint Eq.~(\ref{kappeq}) as
\begin{subeqnarray}\label{wukrnAy}
\dot y_i&=&F_i(y)+f_i(y)\;,\quad i=W,U,\kappa,r,N,A\;,\\
0&=&F_c(y)+f_c(y)\;,
\end{subeqnarray}
with the `leading' contributions
\begin{subeqnarray}\label{FwukrnA}
\left(\begin{array}{c}F\smsub{W}\\F\smsub{U}\end{array}\right)&=&
  \left(\begin{array}{cc}-\kappa/2&1\\
    -\sigma&-\kappa/2\end{array}\right)
  \left(\begin{array}{c}\bar W\\\bar U\end{array}\right)\;,\\
F_\kappa&=&\sigma\bar r-\kappa^2\;,\\
F_r&=&-2\bar N\;,\\
F\smsub{N}&=&-{\sigma\bar r^2\over4}\;,\\
F\smsub{A}&=&-\kappa A\;,\\
F_c&=&A-\kappa\bar N-{\sigma \bar r^2\over4}\;,
\end{subeqnarray}
dominating the behaviour near the f.p., and with the `non-leading' corrections
\begin{subeqnarray}\label{fwukrnA}
\left(\begin{array}{c}f\smsub{W}\\f\smsub{U}\end{array}\right)&=&
  \left(\begin{array}{cc}-N&0\\\bar f\smsub{U}&N\end{array}\right)
  \left(\begin{array}{c}\bar W\\\bar U\end{array}\right)\;,\quad
  \bar f\smsub{U}={\sigma\bar r\over2}-{\kappa^2\over4}
    +\sigma r^2\bar W^2+U^2-\kappa N\;,\qquad\\
f_\kappa&=&2U^2\;,\\
f_r&=&2(\bar W\bar U+\bar rN)\;,\\
f\smsub{N}&=&-{\sigma\bar r^3\over 2r^2}-{N^2\over2}
  +\Bigl(\bar f\smsub{U}-{\kappa^2\over4}
    -{\sigma r^2\bar W^2\over2}\Bigr)\bar W^2\;,\\
f\smsub{A}&=&2N(\bar U^2-\sigma\bar W^2)+2\bar f\smsub{U}\bar W\bar U\;,\\
f_c&=&-{\sigma\bar r^3\over2r^2}+{N^2\over2}
  +\Bigl({\kappa^2\over4}
    -{\sigma r^2\bar W^2\over2}\Bigr)\bar W^2\;.
\end{subeqnarray}
From the linearized form of these equations we get vanishing eigenvalues
from $\kappa$, $\bar r$, $\bar N$, and $A$, whereas $\bar W$ and $\bar U$
contribute the eigenvalues $\pm i$ for $\sigma=+1$ or $\pm1$ for
$\sigma=-1$. We therefore expect that (at least some of) the functions
decrease with powers of $\tau$ instead of exponentially as for
non-vanishing eigenvalues.

In order to analyze the behaviour of solutions tending to the f.p.\ as
$\tau\to\infty$ we perform a `quasi-homogeneous blow-up', following
\cite{Dumortier} with some minor variations. In terms of rescaled variables
\begin{equation}\label{blowupvar}
\tilde y_i=\lambda^{-\alpha_i}y_i\;,\quad
\alpha_\kappa=1\;,~
\alpha\smsub{W}=\alpha\smsub{U}=\alpha_r=2\;,~
\alpha\smsub{N}=3\;,~
\alpha\smsub{A}=4\;,
\end{equation}
we find
\begin{subeqnarray}
F_i(y)&=&\lambda^{\alpha_i+1}F_i(\tilde y)\;,
  \qquad
  f_i(y)=O(\lambda^{\alpha_i+2})\;,
  \qquad{\rm for}\quad i=\kappa,r,N,A\;,\qquad\\
F_c(y)&=&\lambda^4F_c(\tilde y)\;,
  \qquad\quad~
  f_c(y)=O(\lambda^5)\;.
\end{subeqnarray}
Furthermore the leading contributions $F_i$, $i=\kappa,r,N,A$ and $F_c$
depend only on $\kappa$, $\bar r$, $\bar N$, and $A$. Hence they define a
quasi-homogeneous dynamical system for $(\kappa,\bar r,\bar N,A)$ of type
$\alpha=(1,2,3,4)$ and degree~2 in the sense of \cite{Dumortier}, whereas the
non-leading corrections yield at least one additional power of $\lambda$
upon rescaling. The non-leading correction terms, $f_i$ depend on $\bar W$
and $\bar U$ individually, not only on the combination $A$. One can easily
find bounds for $|\tilde W|$ and $|\tilde U|$: for $\sigma=+1$ they are
bounded by $\sqrt{\tilde A}$, whereas for $\sigma=-1$ with eigenvalues
$\pm1$ they are bounded by $\exp(-\delta\tau)$ for any $\delta<1$ due to
the special form of Eqs.~(\ref{FwukrnA}a,\ref{fwukrnA}a).

Next we introduce $\mu_i=M/\alpha_i$ with some $M$ such that $\mu_i\ge2$,
new dependent variables $\lambda$ and $\tilde y$
\begin{equation}
\lambda^M=\sum_i\gamma_i\left|y_i\right|^{\mu_i}\;,
\qquad
\tilde y_i=\lambda^{-\alpha_i}y_i\;,\quad i=\kappa,r,N,A\;,
\end{equation}
with some constants $\gamma_i\ge0$ such that $\sum_i\gamma_i>0$, and a new
independent variable $t$ with $\dd t=\lambda\dd\tau$. For $\gamma_i>0$ this
is a `polar' blow-up whereas e.g., $\gamma=(1,0,0,0)$ yields a `directional'
blow-up in the $\kappa$-direction. Neglecting for a moment the non-leading
correction terms we obtain
\begin{subeqnarray}\label{blowupeq}
{1\over\lambda}{\dd\lambda\over\dd t}&=&
  \sum_i\gamma_i\left|\tilde y_i\right|^{\mu_i}
  {F_i(\tilde y)\over\alpha_i\tilde y_i}\;,\\
{\dd\tilde y_i\over\dd t}&=&
  F_i(\tilde y)-\alpha_i\tilde y_i{1\over\lambda}{\dd\lambda\over\dd t}\;,
\end{subeqnarray}
i.e., an $\tilde y$-dependent `radial' flow with the f.p.\ $\lambda=0$ and a
$\lambda$-independent `tangential' flow with f.p.s determined by the
conditions
\begin{equation}\label{blowupcond}
\alpha_i\tilde y_iF_j(\tilde y)=\alpha_j\tilde y_jF_i(\tilde y)\;,\quad
i,j=\kappa,r,N,A\;.
\end{equation}
There are three pairs of tangential f.p.s $P_{i\pm}$, $i=1,2,3$ with
$\tilde\kappa\gtrless0$ and
\begin{subeqnarray}\label{blowupfp}
P_{1\pm}:&\qquad&
  \tilde r=\tilde N=\tilde A=0\;,\\
P_{2\pm}:&\qquad&
  \tilde r={3\sigma\over4}\tilde\kappa^2\;,\quad
  \tilde N={3\sigma\over16}\tilde\kappa^3\;,\quad
  \tilde A={21\sigma\over64}\tilde\kappa^4\;,\\
P_{3\pm}:&\qquad&
  \tilde r={4\sigma\over3}\tilde\kappa^2\;,\quad
  \tilde N=-{4\sigma\over9}\tilde\kappa^3\;,\quad\tilde A=0\;.
\end{subeqnarray}
Inserting these values into Eq.~(\ref{blowupeq}a) yields
$\dd\lambda/\dd t=-\tilde\kappa\lambda$ at $P_{1\pm}$, $-\tilde\kappa\lambda/4$
at $P_{2\pm}$, and $+\tilde\kappa\lambda/3$ at $P_{3\pm}$. Thus $P_{1+}$,
$P_{2+}$, and $P_{3-}$ are attractors for the radial flow, whereas $P_{1-}$,
$P_{2-}$, and $P_{3+}$ are repulsive. On the 2-dimensional surface $\tilde
y$ subject to the constraint $F_c(\tilde y)=0$ the f.p.s $P_{1-}$ and $P_{3+}$
are attractors for the tangential flow, $P_{1+}$ and $P_{3-}$ are repulsive,
and $P_{2\pm}$ are saddle points.
\mkfig{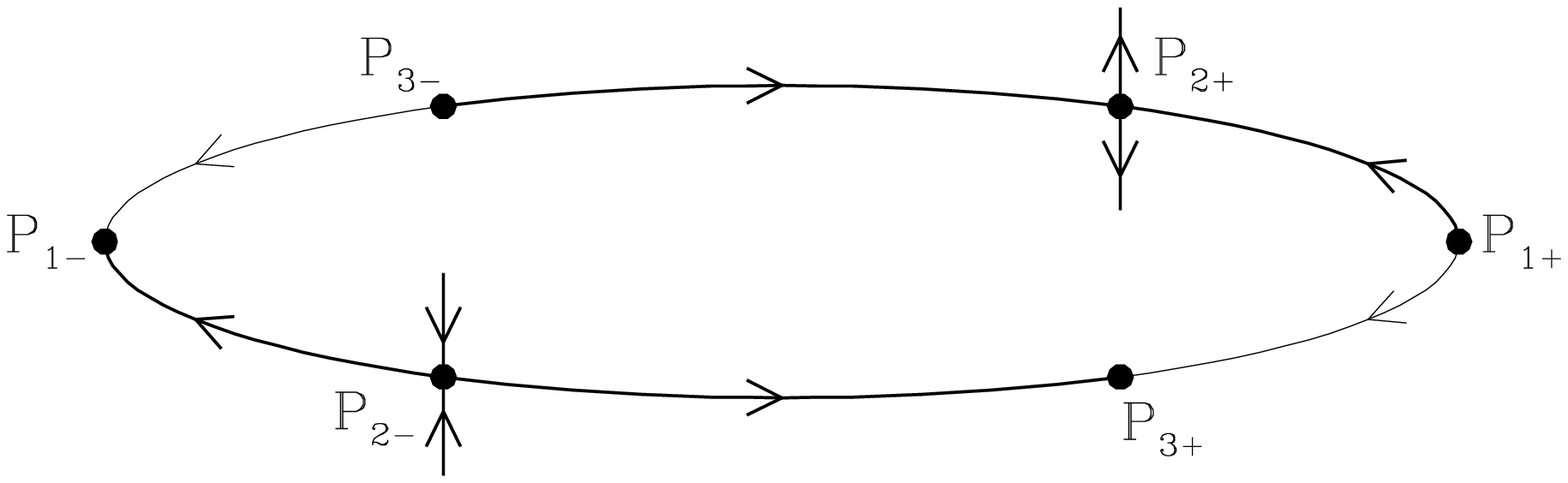}{0.6}{0.9}{43 197 568 354}
{RN f.p.\ with $\Lambda=1/4$: Tangential flow after the blow-up}
Furthermore there are trajectories along the curve $\tilde N^2=\sigma\tilde
r^3/12$ connecting all six f.p.s as shown in Fig.~\ref{figflow}: separatrices
from $P_{1+}$ and $P_{3-}$ to $P_{2+}$, separatrices from $P_{2-}$ to
$P_{1-}$ and $P_{3+}$, as well as trajectories from $P_{1+}$ to $P_{3+}$ and
from $P_{3-}$ to $P_{1-}$. This leaves no room for limit or separatrix
cycles, therefore according to the Poincar\'e-Bendixson theory the f.p.s are
the only limit points of the tangential flow.

With this blow-up we have transformed the degenerate f.p.\ $y=0$ into a set
of hyperbolic f.p.s $(\lambda,\tilde y)=(0,P_{i\pm})$ with one dimensional
stable manifolds for $(0,P_{1+})$, $(0,P_{2-})$, $(0,P_{3-})$ and two
dimensional stable manifolds for $(0,P_{1-})$, $(0,P_{2+})$, $(0,P_{3+})$. 
The non-leading correction terms, $f_i$ in Eqs.~(\ref{wukrnAy}) yield
$O(\lambda)$ corrections to the r.h.s.\ of Eqs.~(\ref{blowupeq}) and thus
preserve the position and character of the six f.p.s. The structural
stability of the system~(\ref{wukrnAy}) guarantees that the flow pattern is
not changed by the non-leading corrections. Moreover the stable manifolds
for $(0,P_{1-})$, $(0,P_{2-})$, and $(0,P_{3+})$ are purely tangential at
$\lambda=0$ and therefore excluded for trajectories tending to the f.p.\
$y=0$ as $\tau\to\infty$.

We have chosen $\bar W$, $\bar U$, and $\bar N$ such that
$1/\lambda-c_i\tilde\kappa\tau$ has a finite limit on each of the stable
manifolds with $c_1=1$ for $P_{1+}$, $c_2=1/4$ for $P_{2+}$, and $c_3=-1/3$
for $P_{3-}$. Consequently we choose yet another independent variable
$s=\lambda=1/\tau$ as well as dependent variables $\tilde y$ as in
Eq.~(\ref{blowupvar}) to obtain
\begin{subeqnarray}\label{rescaled}
s{\dd\over\dd s}\left(\begin{array}{c}\tilde W\\\tilde U\end{array}\right)
  &=&\left(\begin{array}{cc}
      \tilde\kappa/2-2&-1/s\\\sigma/s&\tilde\kappa/2-2
    \end{array}\right)
  \left(\begin{array}{c}\tilde W\\\tilde U\end{array}\right)
  -s\left(\begin{array}{cc}g\smsub{W}&0\\\bar g\smsub{U}&g\smsub{U}\end{array}\right)
  \left(\begin{array}{c}\tilde W\\\tilde U\end{array}\right)\;,
\nonumber\\&&\qquad\qquad
  g\smsub{W}=-{N\over s^2}\;,\quad
  g\smsub{U}={N\over s^2}\;,\quad
  \bar g\smsub{U}={\bar f\smsub{U}\over s^2}\;,\\
s{\dd\over\dd s}\tilde\kappa&=&\tilde\kappa^2-\tilde\kappa-\sigma\tilde r
  -sg_\kappa\;,\qquad\qquad\qquad\quad~
  g_\kappa={f_\kappa\over s^3}\;,\\
s{\dd\over\dd s}\tilde r&=&2(\tilde N-\tilde r)
  -sg_r\;,\qquad\qquad\qquad\qquad~~\,
  g_r={f_r\over s^4}\;,\\
s{\dd\over\dd s}\tilde N&=&{\sigma\tilde r^2\over4}-3\tilde N
  -sg\smsub{N}\;,\qquad\qquad\qquad\qquad
  g\smsub{N}={f\smsub{N}\over s^5}\;,\\
s{\dd\over\dd s}\tilde A&=&(\tilde\kappa-4)\tilde A
  -sg\smsub{A}\;,\qquad\qquad\qquad\qquad~~
  g\smsub{A}={f\smsub{A}\over s^6}\;,\\
\NoAlign{as well as the constraint}
0&=&\tilde A-\tilde\kappa\tilde N-{\sigma\tilde r^2\over4}
  +sg_c\;,\qquad\qquad\qquad~
  g_c={f_c\over s^5}\;,
\end{subeqnarray}
with $g$'s that are analytic in $s$ and the rescaled functions. The
requirement that the r.h.s.\ of Eqs.~(\ref{rescaled}b--e) vanish at $s=0$
yields the conditions Eqs.~(\ref{blowupfp}) for the f.p.s with
$\tilde\kappa(0)=1$ for $P_{1+}$, $\tilde\kappa(0)=4$ for $P_{2+}$, and
$\tilde\kappa(0)=-3$ for $P_{3-}$. These three different solutions are
to be expected as the limit $\Lambda\to1/4$ of the Nariai
solution~(\ref{Nari}b), and the RN f.p.s\ with the combinations $r_+$,
$\kappa>0$ and $r_-$, $\kappa<0$. In terms of shifted variables $\hat
r=\tilde r-\tilde r(0)$ etc., that vanish at $s=0$ we thus obtain the
linearized equations
\begin{equation}
\arraycolsep=5pt
s{\dd\over\dd s}\left(\begin{array}{c}
    \hat r\\ \hat N\\ \hat\kappa\\ \hat A
  \end{array}\right)=\left(\begin{array}{cccc}
    -2& \phantom{-}2& 0& 0\\
    \displaystyle\sigma{\tilde r(0)\over2}& -3& 0& 0\\
    -\sigma& \phantom{-}0& 2\tilde\kappa(0)-1& 0\\
    \phantom{-}0& \phantom{-}0& \tilde A(0)& \tilde\kappa(0)-4
  \end{array}\right)\left(\begin{array}{c}
    \hat r\\ \hat N\\ \hat\kappa\\ \hat A
  \end{array}\right)\;,
\end{equation}
with eigenvalues $+1$ (corresponding to a shift in $\tau$),
$\tilde\kappa(0)-4$ (due to the constraint), as well as $-2$, $-3$ for
$P_{1+}$, $+7$, $-6$ for $P_{2+}$, and $-6$, $-7$ for $P_{3-}$.

We still need an additional equation for the YM functions $\tilde W$ and\slash
or $\tilde U$. For $\sigma=+1$ the solutions with $\tilde\kappa(0)<4$ yield
$W\equiv0$ whereas $\tilde\kappa(0)=4$ requires $W\not\equiv0$. Motivated by
the eigenvalues $\pm i$, we introduce in the latter case $\tilde
W=\sqrt{\tilde A}\sin(\tau+\theta)$ and $\tilde U=\sqrt{\tilde
A}\cos(\tau+\theta)$. The equation
\begin{equation}
s{\dd\over\dd s}\theta=s((g\smsub{U}-g\smsub{W})\cos(\tau+\theta)
  +\bar g\smsub{U}\sin(\tau+\theta))\sin(\tau+\theta)\;,
\end{equation}
implies that the phase $\theta$ has a finite limit as $s\to0$.

For $\sigma=-1$ with the eigenvalues $\pm1$ we introduce $\tilde
W_\pm=\tilde W\pm\tilde U$ and replace Eq.~(\ref{rescaled}f) by
\begin{equation}
s{\dd\over\dd s}\tilde W_\pm=
  \Bigl(\mp{1\over s}+{\tilde\kappa-4\over2}\Bigr)\tilde W_\pm
  -s(g_\pm\tilde W_\pm+\bar g_\pm\tilde W_\mp)\;,
\end{equation}
where $g_\pm$ and $\bar g_\pm$ are linear combinations of $g\smsub{W}$, $g\smsub{U}$, and
$\bar g\smsub{U}$. The divergent mode $\tilde W_+$ has to be suppressed and
consequently $\tilde W_+$ and $\tilde W_-$ decrease exponentially,
implying that $\tilde A=\tilde W_+\tilde W_-$ vanishes at $s=0$ and thus
excluding the solution with $\tilde\kappa(0)=4$. As for $\Lambda<1/4$, there
are solutions with $\kappa>0$ and $\kappa<0$ with both $W\equiv0$ and
$W\not\equiv0$

\section{Classification of Solutions}\label{chapclass}

Starting from any regular point $\tau=\tau_0$ of the Eqs.~(\ref{taueq}) we
may integrate these equations to both sides until we meet a singular point.
In case $\int r\,\dd\tau$ is finite the corresponding space-time is
geodesically incomplete, otherwise it is complete in the radial direction.
This does not necessarily mean completeness in all directions as the example
of the extremal RN solution demonstrates. A general recipe how to glue
together incomplete pieces of space-time to a geodesically complete manifold
is described in \cite{Walker}.

In the preceding section we have already discussed the possible singular
points we may encounter integrating Eqs.~(\ref{taueq}) and described the
local behaviour in their neighbourhood. The missing step is a classification
of the singular points reachable from some regular point integrating the
equations to both sides. Instead of integrating towards decreasing $\tau$ we
may as well reverse the sign of $N$, $\kappa$, and $U$ and always integrate
towards increasing $\tau$.

The global classification is different for the two possibilities
$\sigma=\pm1$, i.e.\ for regions with space-like or time-like radial
direction. Here we start with some general properties equally valid for both
cases.

\begin{Prop}{}\label{prop+-}
Consider a solution of Eqs.~(\ref{taueq},\ref{kappeq}).
\begin{itemize}
\item[i)]
If $N(\tau_0)<0$ for some $\tau_0$ then $N(\tau)<0$ for all $\tau>\tau_0$.

\item[ii)]
$r$ has no minima and at most one maximum.

\item[iii)]
$r$ has a limit $r_1$ as $\tau\to\tau_1$ with $\ln r_1$ finite if and only if
$\int^{\tau_1}N\,\dd\tau$ is finite.

\item[iv)]
If $N(\tau)$ and $\kappa(\tau)$ are finite at some $\tau$ then $W(\tau)$ and
$U(\tau)$ are finite.

\item[v)]
If $N<0$, $\kappa<0$, and $\kappa+N$ is unbounded, then $\kappa+N\to-\infty$
at some finite $\tau_1$.

Let us define $\zeta=(\kappa-N)/(\kappa+N)$ and
$\gamma=(\tau-\tau_1)(\kappa+N)$. If $\zeta$ is constant then
$\gamma=4/(1+3\zeta^2)+O((\tau-\tau_1)^2)$. If $\zeta$ is restricted to an
interval then $\gamma$ is some ($\tau$ dependent) average of this expression
over that interval, in the most general case $1\le\gamma\le4$ up to terms
$O((\tau-\tau_1)^2)$.

\end{itemize}
\end{Prop}
\begin{PrfItem}{}
\begin{itemize}
\item[i)]
This is an immediate consequence of Eq.~(\ref{taueq}c).

\item[ii)]
This is an immediate consequence of the previous point and
Eq.~(\ref{taueq}a).

\item[iii)]
Property ii) implies that $r$ is monotonic for sufficiently large $\tau$ and
therefore has a limit; the finiteness of $\ln r$ is a consequence of
Eq.~(\ref{taueq}a).

\item[iv)]
Due to the previous point $\ln r$ has a finite limit and $U^2$ is integrable
due to Eq.~(\ref{taueq}c). The Schwarz inequality then implies that $|U|$
and thus $r|U|$ are integrable, i.e.\ $W$ is bounded. The boundedness of $U$
finally follows from Lemma~\ref{linear} in the appendix applied to
Eq.~(\ref{taueq}e).

\item[v)]
From Eqs.~(\ref{taueq}c,d) we obtain
\begin{eqnarray}\label{kNinequ}
(\kappa+N)\dot{}&=&\sigma(1-2\Lambda r^2)
  -{1\over4}(\kappa+N)^2-{3\over4}(\kappa-N)^2
\nonumber\\
  &\le&\sigma(1-2\Lambda r^2)-{1\over4}(\kappa+N)^2\;,
\end{eqnarray}
with $1-2\Lambda r^2$ bounded since $N<0$. Therefore Lemma~\ref{Riccati}
implies that $\kappa+N$ diverges for some finite $\tau_1$ and we assume
$\tau_1=0$ without restriction. Introducing $\eta=1/(\tau(\kappa+N))$ we
find
\begin{equation}
\eta(\tau)=-{1\over4\tau}\int_\tau^0(1+3\zeta^2)\dd\tau'
  +\sigma\tau\int_\tau^0\Biggl({\tau'\over\tau}\Biggr)^2
    (1-2\Lambda r^2)\eta^2\dd\tau'\;.
\end{equation}
Since $\eta\le1+O(\tau^2)$ due to Lemma~\ref{Riccati}, the second integral
on the r.h.s.\ is $O(\tau^2)$, whereas the first integral yields $1/\gamma$
as the average value of $(1+3\zeta^2)/4$.
\end{PrfItem}

According to Prop.~\ref{prop+-} we can choose a suitable
$\bar\tau_0\ge\tau_0$ such that $N$ has a definite sign for
$\tau\ge\bar\tau_0$. We have to consider the cases that either $N$ and
$\kappa$ are bounded for all $\tau\ge\bar\tau_0$ or that $N$ and\slash or
$\kappa$ are unbounded.

For $N$ and $\kappa$ bounded we have the result:
\begin{Prop}{}\label{bounded+-}
Suppose a solution of Eqs.~(\ref{taueq},\ref{kappeq}) with $N$, $\kappa$,
$W$, $U$, and $\ln r$ bounded for $\tau\ge\tau_0$, then the solution tends
to one of the f.p.s of Eqs.~(\ref{taueq}) as $\tau\to\infty$:
\begin{itemize}
\item[i)]
for $\sigma=-1$, the SdS f.p.\ described by Eqs.~(\ref{fp1});

\item[ii)]
for $\sigma=\pm1$, the RN f.p.\ described by Eqs.~(\ref{fp2}).

\end{itemize}
\end{Prop}
\begin{Prf}{}
Since $\ln r$ is bounded, $|N|$ is integrable and therefore $U^2$ is
integrable due to Eq.~(\ref{taueq}c). Integrating
\begin{equation}\label{U3}
(U^3)\dot{}=3\Bigl(\sigma WT+(N-\kappa)U\Bigr)U^2\;,
\end{equation}
derived from Eq.~(\ref{taueq}e), we see that $U$ has a limit, which must be
zero since otherwise $W$ would diverge. Lemma~\ref{Riccati} applied to
Eq.~(\ref{taueq}d) implies that $\kappa$ has a limit. From
Eqs.~(\ref{energy}) then follows that the `energy' $E$ is bounded and
$|\dot E|$ is integrable, thus $E$ has a limit. This implies that $N$ has a
limit which must be zero since otherwise $\ln r$ would diverge, and that $W$
has a finite limit which must be one of the f.p.s of Eq.~(\ref{taueq}e),
$W\to\pm1$ for the SdS f.p.\ or $W\to0$ for the RN f.p.
\end{Prf}

For $N$ bounded and $\kappa$ unbounded we have the result:
\begin{Prop}{}\label{horizon+-}
Suppose a solution of Eqs.~(\ref{taueq},\ref{kappeq}) with $N$ bounded and
$\kappa$ unbounded, then the solution has a regular horizon at some finite
$\tau_1$ as described in Eqs.~(\ref{bcbh},\ref{w1def}).
\end{Prop}
\begin{Prf}{}
First we show that $\tau_1$ must be finite. For $\sigma=+1$ we use
Eq.~(\ref{kNinequ}), for $\sigma=-1$ we use
\begin{equation}
(\kappa+N)\dot{}=1-4U^2-2T^2-\kappa^2+5\kappa N-3N^2\le
  1+7N(\kappa+N)-(\kappa+N)^2\;.
\end{equation}
In both cases Lemma~\ref{Riccatilin} implies that $\kappa+N$ and thus
$\kappa$ diverges to $-\infty$ for some finite $\tau_1$, thus $\ln r$ has a
finite limit. We assume again $\tau_1=0$ without restriction.

Next, for $\sigma=+1$ Eq.~(\ref{Neq}) implies that $U^2$ is integrable,
since otherwise $N$ would be unbounded from below. Therefore $|U|$ is
integrable due to the Schwarz inequality and thus $W$ has a finite limit.
The constraint Eq.~(\ref{kappeq}) then implies that $U^2/\kappa$ is bounded.
For $\sigma=-1$ the constraint implies that $U^2/\kappa$ is bounded.

For $\sigma=\pm1$ we use $s=1/\kappa$ as in Eqs.~(\ref{hor},\ref{horF}) with
$\dd\tau=\left(1+s^2F\right)\dd s$ and find that $sf$ and hence $sF$ is bounded.
Therefore $\int^0|U|\dd\tau<\infty$ and thus $W$ has a finite limit. We now
turn to Eq.~(\ref{hor}d) for $U$ and obtain
\begin{equation}
{\dd\over\dd s}\left(se^{G}{U\over r}\right)=
   \sigma se^{G}\left(1+s^2F\right){W(W^2-1)\over r^2}\;,
\end{equation}
with $G(s)=-\int_s^0s'^2F\,ds'$ bounded, yielding
\begin{equation}
e^{G}{U\over r}={c\over s}-\sigma\int_s^0{s'\over s}
    e^{G}\left(1+s'^2F\right){W(W^2-1)\over r^2}ds'\;.
\end{equation}
The constant $c$ must vanish, otherwise $sU^2$ would not be bounded. This
implies $U\to0$ for $s\to0$. From the constraint we finally obtain $N(0)=0$,
thus the solution has a horizon.
\end{Prf}

\subsection{Solutions with \boldmath$\sigma=+1$}\label{positive}

To begin with we state some properties of the solutions required for their
global classification. Most of these are trivial generalizations of results
already proved for $\Lambda=0$ in \cite{BFM}, but we repeat their proofs for
completeness.

\begin{Prop}{}\label{prop+}
Consider a solution of Eqs.~(\ref{taueq},\ref{kappeq}) with $\sigma=+1$.
\begin{itemize}
\item[i)]
The function $W$ can have neither maxima if\/ $W>1$ or $0>W>-1$ nor minima
if\/ $W<-1$ or $0<W<1$.

\item[ii)]
$N$ is bounded from above and if $N(\tau_0)<1$ for some $\tau_0$
then $N(\tau)<1$ for all $\tau>\tau_0$.

\item[iii)]
$r$ is bounded from above.

\item[iv)]
$\kappa+N$ is bounded from above and if $(\kappa+N)(\tau_0)<2$ for some
$\tau_0$ then $(\kappa+N)(\tau)<2$ for all $\tau>\tau_0$.

\item[v)]
If $W^2(\tau_0)>1$ and $WU(\tau_0)\ge0$ for some $\tau_0$ and
$\kappa$ is bounded, then $N(\tau_1)<-1$ for some finite $\tau_1>\tau_0$.

\item[vi)]
If $N(\tau_0)<-1$ for some $\tau_0$ and $\kappa$ is bounded, then
$N(\tau)$ diverges to $-\infty$ for some finite $\tau_1>\tau_0$.

\end{itemize}
\end{Prop}
\begin{PrfItem}{}
\begin{itemize}
\item[i)]
This is an immediate consequence of Eqs.~(\ref{taueq}b,e).

\item[ii)]
This follows from Eq.~(\ref{Neq}).

\item[iii)]
In the following argument we can always assume $\Lambda r^2\ge3$ and
$N\ge0$ since otherwise there is nothing to prove. Thus $\dot N\le-1$ due
to Eq.~(\ref{Neq}) and therefore $N=0$ for some finite $\tau_1$ where
$r$ has a finite maximum.

\item[iv)]
Eq.~(\ref{kNinequ}) implies $(\kappa+N)\dot{}\le0$ for $\kappa+N\ge2$.

\item[v)]
In the following argument we can always assume $N\ge-1$ since otherwise
there is nothing to prove. Let $A>1$ be a constant such that $\kappa<2A$.
First we note that $W^2$ monotonically increases and $WU>0$ for
$\tau>\tau_0$. From $\ddot W=W(W^2-1)+(2N-\kappa)\dot W$ we obtain a
positive lower bound for $|\dot W|$ and thus there exists a
$\tilde\tau_0\ge\tau_0$ such that $|W|(\tilde\tau_0)>A$. Using the Schwarz
inequality we find
\begin{equation}
|TU|\,\dot{}=2|W|(2T^2+U^2)-\kappa|TU|\ge(\sqrt{8}-2)|TU|\;.
\end{equation}
Therefore $|TU|$ and, using again the Schwarz inequality, $2T^2+U^2$ grow
exponentially for $\tau>\tilde\tau_0$. Eq.~(\ref{Neq}) then implies that
$N<-1$ for some finite $\tau_1$.

\item[vi)]
This is a consequence of Eq.~(\ref{Neq}) and Lemma~\ref{Riccati}.
\end{PrfItem}

In addition there are new features compared to the case $\Lambda=0$, which
are mainly due to the possibility that the variable $\kappa$ may become negative.

\begin{Prop}{}\label{bounded+}
Suppose a solution of Eqs.~(\ref{taueq},\ref{kappeq}) for $\sigma=+1$ with
$N$ and $\kappa$ bounded for all $\tau\ge\tau_0$, then the solution runs
into
\begin{itemize}
\item[i)]
the RN f.p.\ described by Eqs.~(\ref{fp2}), if $r\to r_1>0$ (with
$\kappa_s<0$ only possible if $W\equiv0$);

\item[ii)]
a regular origin as described by Eqs.~(\ref{bcor1}), if $r\to0$.

\end{itemize}
\end{Prop}
\begin{PrfItem}{}
First we note that the limit $r_1$ of $r$ cannot exceed $1/\sqrt{\Lambda}$
since otherwise $N$ would diverge due to Eq.~(\ref{Neq}). Moreover the `mass
function' Eqs.~(\ref{mass}) is monotonic and bounded and therefore has a
limit. Furthermore $W$ is bounded due to Prop.~\ref{prop+}.
\begin{itemize}
\item[i)]
If $r_1>0$ then $U$ is bounded due to the constraint Eq.~(\ref{kappeq}) and
Prop.~\ref{bounded+-} implies that the solution runs into the RN f.p., as
discussed in Section~\ref{chapsing}. For reasons given there the case
$\kappa_s<0$ is only possible if $W\equiv0$.

\item[ii)]
For $r_1=0$ we first of all choose a $\bar\tau_0\ge\tau_0$ such that $N<0$
and $\Lambda r^2<1/2$ for all $\tau\ge\bar\tau_0$ and restrict the following
arguments to this subinterval.

Eq.~(\ref{kNinequ}) implies that $\kappa+N\ge-2$ since otherwise
$\kappa+N$ would diverge. From Eq.~(\ref{taueq}d) we see that if
$\kappa>-\sqrt{1-2\Lambda r^2}$ for some $\tilde\tau_0$, then $\kappa$ will
increase towards zero and stay positive afterwards. This is, however,
impossible since then $(rN)\dot{}=\kappa rN-2rU^2\le0$ and $rN$ would not
converge to zero as required. We thus obtain the bounds
$-2\le\kappa\le-\sqrt{1-2\Lambda r^2}$.

We then show that $W$ cannot have zeros for $r<1/2$. Assume there were such
a zero at $\tilde\tau_0$. This would imply $T^2>9/4$ for
$\tau\ge\tilde\tau_0$ as long as $|W|<1/2$. Thus $|W|=1/2$ for some
$\tau=\tilde\tau_0+\Delta$ since otherwise $N$ would diverge due to
Eq.~(\ref{Neq}). We then estimate, repeatedly using the Schwarz inequality
\begin{eqnarray}
N(\tilde\tau_0+\Delta)&\le&N(\tilde\tau_0)
  +\int_{\tilde\tau_0}^{\tilde\tau_0+\Delta}
  {1\over2}\Bigl(1-{2\dot W^2+(W^2-1)^2\over r^2}\Bigr)\dd\tau
\nonumber\\
  &\le&-{5\Delta\over8}-{1\over\Delta}<-1\;,
\end{eqnarray}
and thus $N$ would diverge due to Prop.~\ref{prop+}.

Since the extrema of $W$ are separated by zeros, $W$ is monotonic for
sufficiently large $\tau$ and therefore has a limit. This limit can only be
$\pm1$ since otherwise $N$ would diverge due to Eq.~(\ref{Neq}).

Next we estimate (using $TUW\le0$)
\begin{equation}\label{TUest}
|TU|\,\dot{}=-2|W|(T^2+2U^2)-\kappa|TU|\le-(\sqrt{8}|W|-2)|TU|\;,
\end{equation}
i.e., $|TU|$ decreases exponentially in $\tau$. Eq.~(\ref{TUest}) then
implies that $T$ and $U$ both must decreases exponentially since $TU$ cannot
change sign.

Applying Lemma~\ref{Riccati} to Eqs.~(\ref{taueq}d) and~(\ref{Neq}) we
finally obtain $\kappa\to-1$ and $N\to\-1$ for $\tau\to\infty$. Thus the
solution tends to a regular origin.
\end{PrfItem}

\begin{Prop}{}\label{divergent+}
Suppose a solution of Eqs.~(\ref{taueq},\ref{kappeq}) for $\sigma=+1$ with
$N$ unbounded, then the solution has an origin $r=0$ at some finite $\tau_1$
with
\begin{itemize}
\item[i)]
the S~type singularity as described in Eqs.~(\ref{SSsing},\ref{SSasy}), if
$\kappa$ is unbounded;

\item[ii)]
the RN~type singularity as described in Eqs.~(\ref{RN},\ref{RNsing}), if
$\kappa$ is bounded.

\end{itemize}
\end{Prop}
\begin{PrfItem}{}
In view of Prop.~\ref{prop+} $N$ must be unbounded from below. Integrating
Eq.~(\ref{Neq}) shows that $N\to-\infty$ for some finite $\tau_1$, and we
assume $\tau_1=0$ without restriction. In the following we will consider
sufficiently small $|\tau|$ such that $N(\bar\tau_0)<-1$. In order to
proceed we distinguish the cases that $\kappa$ is unbounded or bounded from
below.
\begin{itemize}
\item[i)]
For $\kappa$ unbounded from below we first note that $\kappa+N$ is unbounded
from below and thus $\kappa+N\to-\infty$ for $\tau\to0$ due to
Eq.~(\ref{kNinequ}).

From the equation
\begin{eqnarray}
\hbox to-5pt{}
(2N-\kappa)\dot{}&=&(N-\kappa)(2N-\kappa)+{1\over2}(
  3-3N^2-2U^2-5T^2-\Lambda r^2)
\nonumber\\
&<&(N-\kappa)(2N-\kappa)\;,
\end{eqnarray}
we then see that if $(2N-\kappa)(\bar\tau_0)\le0$ for some $\bar\tau_0$,
then $2N-\kappa\le0$ for all $\tau\ge\bar\tau_0$.

First assume $2N-\kappa>0$ for all $\tau$ and therefore $\kappa\to-\infty$.
Now consider the equation
\begin{eqnarray}\label{kNeq}
\Bigl({\kappa\over N}\Bigr)\dot{\vphantom{\Bigr)}}&=&
  (\kappa-2N)\Bigl(2+{T^2+\Lambda r^2\over N^2}\Bigr)
  +{3T^2+\Lambda r^2\over N}+3N-{1\over N}{\kappa\over N}
\nonumber\\
  &<&3N+{1\over|N|}{\kappa\over N}\;.
\end{eqnarray}
With $A(\tau)=\int_\tau^0\dd\tau'/|N|=O(\tau)$ we then find
$(e^A\kappa/N)\dot{}<3Ne^A<0$ and therefore $\kappa/N$, which is
non-negative, has a finite limit. Lemma~\ref{Riccati} then implies
$\int_\tau^0\kappa\,\dd\tau'=-\infty$ and therefore
$\int_\tau^0N\dd\tau'=-\infty$. This is, however, impossible since $\kappa/N$
would then become negative due to Eq.~(\ref{kNeq}).

Therefore $2N-\kappa\le0$ for $|\tau|$ sufficiently small. Next we want to
show that $\kappa<0$ for $|\tau|$ sufficiently small. For $r\to0$ this
follows from Eq.~(\ref{taueq}d). For $r\to r_1>0$ we first assume that $W$
is bounded. Lemma~\ref{linear} applied to the equation
\begin{equation}\label{rUeq}
(rU)\dot{}=W(W^2-1)+(2N-\kappa)rU\;,
\end{equation}
then implies that $U$ is bounded and the constraint Eq.~(\ref{kappeq})
implies $\kappa<0$. Then we assume that $W$ is unbounded and therefore, due
to Prop.~\ref{prop+}, $WU>0$ and $|W|\to\infty$. This requires that $U$ is
unbounded and the equation
\begin{equation}
\Bigl({U\over r^2}\Bigr)\dot{\vphantom{\Bigr)}}={W(W^2-1)\over r^3}-(\kappa+N){U\over r^2}\;,
\end{equation}
then implies that $U/r^2$ is monotonic and thus $U$ diverges.
Eq.~(\ref{taueq}d) then again implies $\kappa<0$.

Now consider
\begin{equation}
(r\kappa+rN)\dot{}=r\kappa(2N-\kappa)+r(1-2\Lambda r^2)\;.
\end{equation}
The r.h.s.\ is bounded from below  and therefore $r\kappa+rN$, which cannot be
positive, has a finite limit. This implies that $r\to0$ and that $r\kappa$
and $rN$ are bounded. From the equation
\begin{equation}
(r^2N)\dot{}=(r\kappa+rN)rN-2r^2U^2\;,
\end{equation}
then follows that $r^2U^2$ is integrable and, using the Schwarz inequality,
that $W$ has a finite limit $W_1$. Eq.~(\ref{rUeq}) finally implies that
$\dot W\equiv rU$ has a finite limit $\dot W_1$.

From Eqs.~(\ref{mass}b) we obtain
\begin{equation}
m(\tau)=m(\tau_0)
  -\int_{r(\tau)}^{r(\tau_0)}{2\dot W^2+(W^2-1)^2\over2r'^2}\dd r'\;,
\end{equation}
and therefore $\lim\limits_{\tau\to0}rm=-\dot W_1^2-(W_1^2-1)^2/2$. Thus
$r\kappa$ and $rN$ have finite limits $k_1$ resp.\ $n_1$ satisfying
$n_1^2=2\dot W_1+(W_1^2-1)^2$ and, due to the constraint Eq.~(\ref{kappeq}),
$2k_1n_1=n_1^2+2\dot W_1-(W_1^2-1)^2=4\dot W_1^2$.

We want to show that $n_1=0$. Assuming $n_1<0$, we find $2n_1-k_1<0$ and
$r(\tau)=n_1\tau+O(\tau^2)$. This implies
$\int_\tau^0(2N-\kappa)\dd\tau'=-\infty$ and Lemma~\ref{linear} applied to
Eq.~(\ref{rUeq}) yields $k_1=\dot W_1=0$. As a consequence
$\int_\tau^0\kappa\,\dd\tau$ is finite and Eq.~(\ref{taueq}d) implies that
$\kappa$ is bounded from below. This is a contradiction, therefore
$n_1=k_1=\dot W_1=0$ and $W_1^2=1$, and we may assume $W_1=+1$ without
restriction.

From the equation
\begin{equation}
(2\kappa-N)\dot{}=-(\kappa+N)(2\kappa-N)+6U^2+2(1-\Lambda r^2)
  >-(\kappa+N)(2\kappa-N)\;,
\end{equation}
we see that if $(2\kappa-N)(\bar\tau_0)>0$ for some $\bar\tau_0$, then
$2\kappa-N>0$ for all $\bar\tau_0\le\tau<0$. This is, however, impossible
because then $(2r\kappa-rN)\dot{}\ge0$ and $2r\kappa-rN$ could not converge
to zero as required. Therefore $2\kappa-N\le0$ for all sufficiently small
$|\tau|$.

We can now rewrite Eq.~(\ref{kNeq}) as
\begin{equation}
\Bigl({\kappa\over N}\Bigr)\dot{\vphantom{\Bigr)}}=(2\kappa-N)
  +{T^2\over N}\Bigl({\kappa\over N}+1\Bigr)
  -{\Lambda r^2\over N}+{\Lambda r^2-1\over N}{\kappa\over N}\;.
\end{equation}
The first two terms on the r.h.s.\ are both negative and the last two terms
are bounded. Therefore $2\kappa-N$ and $T^2/N$ are integrable and
$\kappa/N$ has a limit which must be $1/2$ since $\kappa+N\le3/\tau+O(\tau)$
due to Prop.~\ref{prop+-}. This then implies $\tau N\to2$ and
$\ln r/\ln|\tau|\to2$ for $\tau\to0$.

With
\begin{equation}\label{CDdef}
C(\tau)=\int_\tau^0\Bigl(\kappa-{N\over2}\Bigr)\,\dd\tau'=O(1)\;,
\qquad
D=e^Cr^{3\over2}\;,
\end{equation}
such that $\dot D=(2N-\kappa)D\le0$, we then derive
\begin{subeqnarray}\label{WrU}
W(\tau)&=&1-\int_\tau^0 rU\,\dd\tau'\;,\\
(rU)(\tau)&=&{D(\tau)\over D(\tau_0)}(rU)(\tau_0)
  +\sigma\int_{\tau_0}^\tau{D(\tau)\over D(\tau')}W(W^2-1)\,\dd\tau'\;.
\end{subeqnarray}
Starting with $W-1=O(1)$, these equations first imply $rU=O(\tau)$ and thus
$W-1=O(\tau^2)$, and finally $rU=O(\tau^3)$ and thus $W-1=O(\tau^4)$. As a
consequence $U$ and $T$ are bounded, $\kappa=1/\tau+O(\tau)$,
$N=2/\tau+O(\tau)$, and $r=-\sigma M\tau^2/2+O(\tau^3)$ with some
$\sigma M<0$ as required by Eqs.~(\ref{SSsing},\ref{SSasy}).

\item[ii)]
For $\kappa$ bounded from below we proceed like in \cite{BFM}, with some
modifications since we cannot assume $\kappa\ge1$. Let $A$ be a positive
constant such that $\kappa+A\ge1$. We introduce $B=e^{A\tau}$ and the new
dependent variable $\lambda=W(W^2-1)+NU$ and rewrite Eqs.~(\ref{taueq}c,e)
as
\begin{subeqnarray}\label{lameq}
(BrN)\dot{}&=&(\kappa+A)BrN-2BrU^2\;,\\
  \dot U&=&\lambda-\kappa U\;,\\
  \dot\lambda&=&(3W^2-2U^2-1)U\;.
\end{subeqnarray}
We want to show, that the r.h.s.\ of these equations stay bounded as
$\tau\to0$. In the following $\epsilon$ is an arbitrary, suitably small
positive number. Since $(BrN)\dot{}<0$ we see that $r^\epsilon$ and
$|BrN|^{-\epsilon}$ are bounded and monotonic and therefore their derivatives
are absolutely integrable at $\tau=0$. In particular we find
\begin{equation}
\int^0 r^\epsilon |N| \dd\tau<\infty,\quad
     \int^0{1\over r^{1-\epsilon}}\dd\tau<\infty,\quad
     \int^0{U^2\over(Br)^\epsilon |N|^{1+\epsilon}}\dd\tau<\infty\,.
\end{equation}

Next we want to show that $W|BrN|^{-\epsilon}$ is bounded. In view of
Prop.~\ref{prop+} this is obvious except when $W^2\ge1$ and $W\dot W\ge0$.
In that case we get, using the Schwarz inequality,
$\int^0|U||BN|^{-\epsilon}\dd\tau<\infty$ and hence
$\int^0|\dot W||BrN|^{-\epsilon}\dd\tau$ is finite for $0<\epsilon<1$.
Integration by parts yields
\begin{eqnarray}
&&\hbox{\hskip-3pt}
{W\over|BrN|^\epsilon}(\tau)+
  \epsilon\int_{\tau_0}^\tau{W\over|BrN|^\epsilon}
    \Bigl(\kappa+A-{2U^2\over N}\Bigr)\dd\tau'
\qquad\qquad\qquad\nonumber\\&&\qquad\qquad\qquad\qquad\qquad
={W\over|BrN|^\epsilon}(\tau_0)+
  \int_{\tau_0}^\tau{\dot W\over|BrN|^\epsilon}\dd\tau'\;,
\end{eqnarray}
with some suitably small negative $\tau_0$. Since the r.h.s.\ is bounded and
both terms on the l.h.s.\ have the same sign $W|BrN|^{-\epsilon}$ is
bounded.

The constraint Eq.~(\ref{kappeq}) implies that $(r(N+A))^2-(W^2-1)^2$ is
bounded from above and hence $W$ and with it $rN$ are bounded. As a
consequence we get $r(0)=0$.

Next we use the equation
\begin{equation}\label{Ueps}
(r^\epsilon U)\dot{}=\sigma{W(W^2-1)\over r^{1-\epsilon}}
  +\Bigl((1+\epsilon)N-\kappa\Bigr)r^\epsilon U\;.
\end{equation}
From Lemma~\ref{linear} we conclude that $r^\epsilon U$ is bounded and
consequently $|U|^n$ is integrable for any $n>0$ on the interval
$(\tau_0,0)$. Hence $\lambda$ and thus $U$ have a finite limit at $\tau=0$,
implying that also $\kappa$ is bounded from above. We thus get the RN
singularity of Eqs.~(\ref{RN},\ref{RNsing}).
\end{PrfItem}

Since Props.~\ref{horizon+-}, \ref{bounded+}, and~\ref{divergent+} exhaust
all possible cases, we obtain the following:
\begin{Thm}{{\boldmath(Classification for $\sigma=+1$)}}\label{class+}
A solution of Eqs.~(\ref{taueq},\ref{kappeq}) for $\sigma=+1$ starting at
some regular point $\tau_0$ (which is not a f.p.) can be extended in both
directions to one of the following singular points:
\begin{itemize}
\item[i)]
a regular origin $r=0$ at infinite $\tau$ as in Eqs.~(\ref{bcor1});

\item[ii)]
a regular horizon of the black hole or cosmological type at some finite
$\tau_1$ as in Eqs.~(\ref{bcbh},\ref{w1def});

\item[iii)]
the RN f.p.\ at infinite $\tau$ as in Eqs.~(\ref{fp2}), where
$\kappa_s\lessgtr0$ for $\tau\to\pm\infty$ requires $W\equiv0$;

\item[iv)]
$r=0$ at some finite $\tau_1$ with the S~type singularity as in
Eqs.~(\ref{SSsing},\ref{SSasy});

\item[v)]
$r=0$ at some finite $\tau_1$ with the RN~type singularity as in
Eqs.~(\ref{RN},\ref{RNsing}).

\end{itemize}
\end{Thm}

\subsection{Solutions with \boldmath$\sigma=-1$}\label{negative}

Due to the presence of the cosmological constant the asymptotics for large
$r$ is deSitter instead of Minkowski and this requires $\sigma=-1$. We shall
again give a complete classification of all possible solutions starting at
some regular point.

\begin{Prop}{}\label{dS}
Suppose a solution of Eqs.~(\ref{taueq},\ref{kappeq}) for $\sigma=-1$ with
$N(\tau)\ge0$ and unbounded, then $r\to\infty$ for some finite $\tau_1$ with
the deSitter asymptotics of Eqs.~(\ref{asydS},\ref{lindS}).
\end{Prop}
\begin{Prf}{}
The constraint Eq.~(\ref{kappeq}) implies $2\kappa N\ge N^2-1$ and therefore
\begin{equation}\label{kNbound}
\quad|N|>1\quad{\rm implies}\qquad
\kappa N>0\quad{\rm and}\quad|\kappa|\ge{N^2-1\over2|N|}>{|N|-1\over2}\;,
\end{equation}
i.e.\ $\kappa$ is unbounded from above. Eq.~(\ref{kNinequ}) implies that
$(\kappa+N)/r$ is bounded from above and therefore $r\to\infty$. We then
introduce $\bar N=N/r$ and $\bar\kappa=\kappa/r$ and see from the constraint
Eq.~(\ref{kappeq}) that $\bar N$ and $\bar\kappa$ are bounded from below by
positive constants. This implies that $\tau\to\tau_1$ stays finite as
$r\to\infty$. In order to be able to use $E/r^4$ as a Lyapunov function, we
introduce a new independent variable $\rho$ with $\dd\rho=r\dd\tau$ such that
$\rho\to\infty$ for $r\to\infty$, denote the derivative with respect to
$\rho$ by a prime, and obtain
\begin{equation}
\left({E\over r^4}\right)'=-{\bar\kappa U^2+\bar NT^2\over r^2}\;.
\end{equation}
According to Lemma 11.1 of~\cite{Hart} $(E/r^4)'$ vanishes on the limit
points of the solution, showing that $U/r$ and $W/r$ tend to zero for
$\rho\to\infty$. For $\bar N$ we find
\begin{equation}
\bar N'={1\over2}\left(\Lambda-3\bar N^2-{1\over r^2}
  +{T^2-2U^2\over r^2}\right)\;.
\end{equation}
Applying Lemma~\ref{Riccati} we obtain $\bar N\to\sqrt{\Lambda/3}$
and further from Eq.~(\ref{kappeq}) $\bar\kappa\to 2\sqrt{\Lambda/3}$.

Using $s=1/r$ as independent variable and introducing the new dependent
variables $y_\epsilon=(W_\epsilon,\bar U_\epsilon,m_\epsilon)$ where
$W_\epsilon=s^\epsilon W$, $\bar U_\epsilon=s^{2\epsilon}UN$,
$m_\epsilon=s^{3\epsilon}m$, and $0\le\epsilon\le1$, we can rewrite
Eqs.~(\ref{lindS}) as
\begin{subeqnarray}\label{lindSeps}
s{\dd\over\dd s}W_\epsilon&=&\epsilon W_\epsilon
  -s^{1-\epsilon}{\bar U_\epsilon\over\bar N^2}\;,\\
s{\dd\over\dd s}\bar U_\epsilon&=&2\epsilon\bar U_\epsilon
  +s^{1-\epsilon}\Bigl(W_\epsilon(W_\epsilon^2-s^{2\epsilon})
    +s^{3(1-\epsilon)}{\bar U_\epsilon^3\over \bar N^4}\Bigr)\;,\\
s{\dd\over\dd s}m_\epsilon&=&3\epsilon m_\epsilon
  +s^{1-\epsilon}\Bigl({\bar U_\epsilon^2\over\bar N^2}
    -{(W_\epsilon^2-s^{2\epsilon})^2\over2}\Bigr)\;,
\end{subeqnarray}
with $\bar N^2=\Lambda/3-s^2+2s^{3(1-\epsilon)}m_\epsilon$. The dynamical
system Eqs.~(\ref{lindSeps}) has three stable modes with real eigenvalues
$\ge\epsilon$. Thus $s^{-\eta}y_\epsilon$ is bounded near $s=0$ for any
solution vanishing at $s=0$ and $\eta<\epsilon$ (see e.g.\ \cite{Codd}).
Assuming that $y_\epsilon$ is bounded and vanishes at $s=0$ for some
$\epsilon=\epsilon_0>0$, which we know to be true for $\epsilon_0=1$, we
therefore conclude that the same is true for all $\epsilon>2\epsilon_0/3$,
and thus by induction for all $\epsilon>0$. For $\epsilon<1/4$ we can
finally convert Eqs.~(\ref{lindS}) into integral equations with convergent
integrals, i.e.\ $W$, $UN$, and $m$ are bounded as required by
Eqs.~(\ref{asydS}).
\end{Prf}

\begin{Prop}{}\label{bounded-}
Suppose a solution of Eqs.~(\ref{taueq},\ref{kappeq}) for $\sigma=-1$ with
$N$ and $\kappa$ bounded for all $\tau\ge\tau_0$, then the solution runs
into
\begin{itemize}
\item[i)]
the SdS f.p.\ described by Eqs.~(\ref{fp1}) (with $\kappa_s=-1$ only
possible if $W^2\equiv1$);

\item[ii)]
the RN f.p.\ described by Eqs.~(\ref{fp2}).

\end{itemize}
\end{Prop}
\begin{PrfItem}{}
The constraint Eq.~(\ref{kappeq}) implies that $r$, $U$, and $T$ and
thus $W$ are bounded. Moreover $r$ has a limit $r_1\ge1/\sqrt{2\Lambda}$
since otherwise $\kappa+N$ would diverge due to Eq.~(\ref{kNinequ}).
Prop.~\ref{bounded+-} then implies that the solutions runs into
\begin{itemize}
\item[i)]
the SdS f.p., as discussed in Section~\ref{chapsing}. For reasons given
there the case $\kappa_s=-1$ is only possible if $W^2\equiv1$.

\item[ii)]
the RN f.p., as discussed in Section~\ref{chapsing}.
\end{PrfItem}

\begin{Prop}{}\label{divergent-}
Suppose a solution of Eqs.~(\ref{taueq},\ref{kappeq}) for $\sigma=-1$ with
$N$ unbounded and $N(\bar\tau_0)<0$ for some $\bar\tau_0$, then the solution
has an origin ($r=0$) at some finite $\tau_1$. Depending on the behaviour of
$N/\kappa$ as $\tau\to\tau_1$ the solution has
\begin{itemize}
\item[i)]
an S singularity as described in Eqs.~(\ref{SSasy}), if $N/\kappa\to2$;

\item[ii)]
a pseudo-RN singularity as described in Eqs.~(\ref{pRNsing}), if
$N/\kappa\to1/2$;

\item[iii)]
unbounded oscillating $N$, $U$, and $T$ as described in~\cite{BLM}, if
$N/\kappa$ has no limit.

\end{itemize}
\end{Prop}
\begin{PrfItem}{}
From Eq.~(\ref{kNbound}) we find that $\kappa+N$ is unbounded from below
and therefore diverges to $-\infty$ at some finite $\tau_1$ due to
Eq.~(\ref{kNinequ}). Without restriction we may assume $\tau_1=0$. Moreover
Eq.~(\ref{kNbound}) implies
\begin{equation}\label{kkNbound}
\kappa<{\kappa+N+1\over3}\qquad{\rm for}\quad
  \kappa+N<-1\;,
\end{equation}
and consequently $\kappa$ diverges as well. Lemma~\ref{Riccati} applied to
Eq.~(\ref{taueq}d) then yields $\kappa\le1/\tau+O(\tau)$. In the following
we can assume $\kappa<-1$, restricting $|\tau|$ to sufficiently small
values.

Next we introduce new dependent variables $n=N/\kappa$, $u=U/\kappa$, and
$t=T/\kappa$ and note that $n$, $u$, and $t$ are bounded due to
Eqs.~(\ref{kkNbound}) and~(\ref{kappeq}). If $n$ has a limit $n_1>0$ then
$\int_{\tau_0}^0 N\,\dd\tau=-\infty$ and $r\to0$.
\begin{itemize}
\item[i)]
First consider the case that $n\to2$ as $\tau\to0$, and consequently $u\to0$
and $t\to0$ due to the constraint Eq.~(\ref{kappeq}). Prop.~\ref{prop+-}
then implies $\tau\kappa\to1$, $\tau N\to2$, and $\ln r/\ln|\tau|\to2$ as
$\tau\to0$. This implies $r\kappa\to0$, $rU\to0$, and $W^2\to1$ and we may
assume $W\to+1$ without restriction.

Defining
\begin{equation}
C(\tau)=\int_{\tau_0}^\tau\Bigl({N\over2}-\kappa\Bigr)\,\dd\tau'\;,
\qquad
D=e^Cr^{3\over2}\;,
\end{equation}
such that $r^\epsilon e^C$ is bounded for $\epsilon>0$ (compare
Eq.~(\ref{CDdef})), we proceed as in the proof of Prop.~\ref{divergent+}
and find that the solution has all properties required by
Eqs.~(\ref{SSsing},\ref{SSasy}).

\item[ii)]
Next consider the case that $n$ has a limit $n_1<2$. We introduce a new
independent variable $\rho$ with $\dd\rho=|\kappa|\dd\tau$ such that
$\rho\to\infty$ for $\tau\to0$ and denote the derivative with respect to
$\rho$ by a prime. We then obtain again Eqs.~(\ref{rutnz}) with
$z=1/|\kappa|$.

First assume $n_1=0$ and therefore $u\to0$ and $t\to0$.
Eqs.~(\ref{rutnz}c,e) then first imply that $e^\rho z$ has a finite
non-vanishing limit and that $e^\rho t$ and thus $T$ are bounded.
Eqs.~(\ref{rutnz}b,d) then imply that $e^{2\rho}u$ and $e^{2\rho}n$ are
bounded, i.e.\ $N\to0$ contrary to the assumption that $N$ is unbounded.

For $0<n_1<2$ we see from the constraint Eq.~(\ref{kappeq}) that $t^2+2u^2$
has the limit $n_1(2-n_1)>0$. From Eq.~(\ref{rutnz}b) we see that for any
$\epsilon>0$ and $\rho$ sufficiently large $u$ is monotonic as long as
$|2u^2-n_1|>\epsilon$ and $|u|>\epsilon$. Since $u$ is bounded it therefore
has a limit $u_1$ with $(2u_1^2-n_1)u_1=0$ and thus $t$ has a limit $t_1$.
From Eq.~(\ref{rutnz}d) we then obtain the condition
$2u_1^2(n_1+1)=n_1(2-n_1)$. This only leaves the possibility
$n_1=|u_1|=|t_1|=1/2$. Prop.~\ref{prop+-} then implies $\tau\kappa\to2$,
$\tau N\to1$, and $\ln r/\ln|\tau|\to1$ as $\tau\to0$. Thus $r|U|$ is
integrable and $W$ has a limit $W_1$, and consequently $-rN$, $-r\kappa/2$,
and $|rU|$ all have the limit $|W_1^2-1|$.

Integrating Eq.~(\ref{taueq}b) we then find $W=W_1+O(\tau)$ and therefore
$W_1^2\ne1$ since otherwise $t\to0$ for $\tau\to0$. Thus the solution has a
pseudo-RN singularity as described in Eqs.~(\ref{pRNsing}).

\item[iii)]
Finally assume that $n$ has no limit. Since $n$ is bounded this implies that
$n$ has an infinite sequence of maxima and minima where the r.h.s.\ of
Eq.~(\ref{rutnz}d) changes sign. Therefore $n$, $u$, and $t$ vary between
their respective maxima and minima.

If $\int_\rho^\infty n\,\dd\rho'$ is finite then the r.h.s.\ of
Eq.~(\ref{rutnz}d) is integrable since $u^2<n(2-n)+z^2$ due to the constraint
Eq.~(\ref{kappeq}), and therefore $n$ has a limit contrary to the
assumption. Thus $\int_\tau^0N\,\dd\tau'=-\int_\rho^\infty n\,\dd\rho'$ diverges
and $r\to0$.
\end{PrfItem}
\begin{Rem}{}
The proof of Prop.~\ref{divergent-} also applies to the case $\Lambda=0$
corresponding to Bartnik-McKinnon type black holes inside their horizon
discussed in \cite{BLM}. In fact, this is to be expected, because the
cosmological constant becomes less relevant as $r$ decreases.
\end{Rem}

Since Props.~\ref{horizon+-} and~\ref{dS}--\ref{divergent-} exhaust all
possible cases, we obtain the following:
\begin{Thm}{{\boldmath(Classification for $\sigma=-1$)}}\label{class-}
A solution of Eqs.~(\ref{taueq},\ref{kappeq}) for $\sigma=-1$ starting at
some regular point $\tau_0$ (which is not a f.p.) can be extended in both
directions to one of the following singular points:
\begin{itemize}
\item[i)]
$r=\infty$ at some finite $\tau_1$ with the deSitter asymptotics of
Eqs.~(\ref{asydS},\ref{lindS});

\item[ii)]
a regular horizon of the black hole or cosmological type at some finite
$\tau_1$ as in Eqs.~(\ref{bcbh},\ref{w1def});

\item[iii)]
the SdS f.p.\ at infinite $\tau$ as in Eqs.~(\ref{fp1}), where
$\kappa_s=\mp1$ for $\tau\to\pm\infty$ requires $W^2\equiv1$;

\item[iv)]
the RN f.p.\ at infinite $\tau$ as in Eqs.~(\ref{fp2});

\item[v)]
$r=0$ at some finite $\tau_1$ with the pseudo-RN~type singular behaviour of
Eqs.~(\ref{pRNsing},\ref{RNs});

\item[vi)]
$r=0$ at some finite $\tau_1$ with the S~type singular behaviour of
Eqs.~(\ref{SSsing},\ref{SSasy});

\item[vii)]
$r=0$ at some finite $\tau_1$ with unbounded oscillating $N$, $U$, and $T$
as described in~\cite{BLM}.

\end{itemize}
\end{Thm}
\noindent
The cases ii), v), and vi) correspond to the NARN, NAPRN, and NAS solutions
of~\cite{BLM}.

\section{4d Space-Times}\label{chap4d}

In the previous section we have determined all possible 1d radial solutions
extending between singular points. Taking into account the time independence
of the solutions this gives us pieces of 2d space-times with a metric as
given in Eq.~(\ref{2line}). Our aim in this section is to determine all
inextensible (in the sense of geodesic completeness) 4d space-times, which
can be constructed from these 2d pieces. A standard way for that is to
construct the Carter-Penrose diagrams of these extensions. Recipes for their
construction are given in \cite{Walker,Katanaev}. As a first step, we
have to bring the 2-metric to the required standard form
\begin{equation}\label{standard}
\dd s_2^2=F\dd t^2-{1\over F}\dd\hat{r}^2\;,
\end{equation}
by the coordinate change $\dd\hat{r}=re^\nu \dd\tau$, where $F=\sigma
e^{2\nu}$ in terms of our variables. Introducing Eddington-Finkelstein type
coordinates via $\dd u_{\pm}=\dd t\pm F^{-1}\dd\hat{r}$ this takes the form
\begin{equation}\label{Edding1}
\dd s_2^2=F\dd u_+\dd u_-\;,
\end{equation}
useful for a Kruskal type extension.

The next step is to express $F$ as a function of $\hat{r}$ and study its
behavior near the singular points. This will be done one by one:
\begin{itemize}
\item
At $r=0$ the solutions are geodesically incomplete. The singular solutions
have a genuine curvature singularity
and do not allow for an extension. Although solutions with a regular
origin are easily extended taking into account the angular dependence, we
treat them as inextensible in the 2d setting.

\item
At a regular horizon the space-time is geodesically incomplete in the
space-like as well as in the light-like direction and $F$ has a simple zero
at $\hat{r}=\hat{r}_h$.

\item
At the RN and SdS fixed points the space-time is always inextensible (i.e.\
geodesically complete) in the space-like direction. While for
$\kappa_s\gtrless0$ the coordinate $\hat{r}$ tends to infinity for
$\tau\to\pm\infty$ it stays finite for $\kappa_s\lessgtr0$. Since $\hat{r}$
plays the role of an affine parameter for light-like geodesics, the
space-time is geodesically complete in the first case and incomplete in the
second one. For $\kappa\tau\to-\infty$ one easily finds
$\hat{r}\approx\hat{r}_s+e^{\kappa_s\tau}r_s/\kappa_s$ and since
$e^{2\nu}\approx e^{2\kappa_s\tau}$ we get
\begin{equation}\label{normal}
F=\sigma e^{2\nu}\approx\sigma{\kappa_s^2\over r_s^2}(\hat{r}_s-\hat{r})^2\;
\end{equation}
near the f.p.  This shows that for $\kappa_s\lessgtr0$ the space-time has a
degenerate horizon.

There is an exceptional situation of the RNdS f.p.\ for $\kappa_s=0$, which
is only possible for $\Lambda=1/4$, where $F$ has a cubic zero at
$r=\sqrt{2}$. Again one has to distinguish the cases $\kappa\gtrless0$ and
$\kappa\lessgtr0$ for $\tau\to\pm\infty$, leading to geodesic completeness
resp.\ incompleteness at this singular point. For $\sigma=+1$ only the RNdS
solution with $W\equiv0$ is possible, but for $\sigma=-1$ there are
solutions with non-trivial $W$. They are $C^\infty$, but have an essential
singularity at the degenerate horizon.

\item
For solutions with dS asymptotics the space-time is geodesically complete at
$r=\infty$.

\end{itemize}
\mkfig{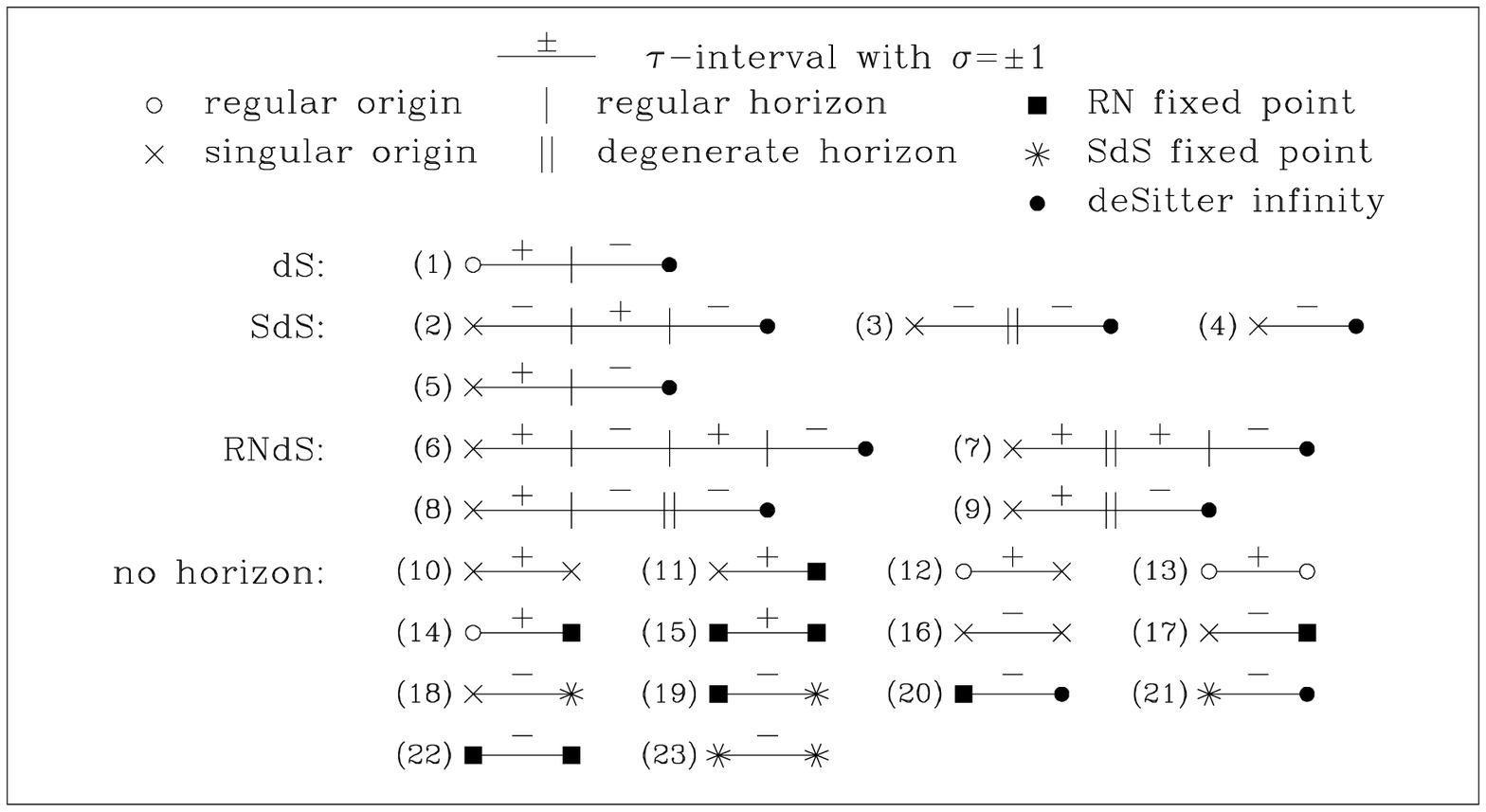}{0.6}{0.9}{43 197 568 482}
{1d graphs}
\mkfig{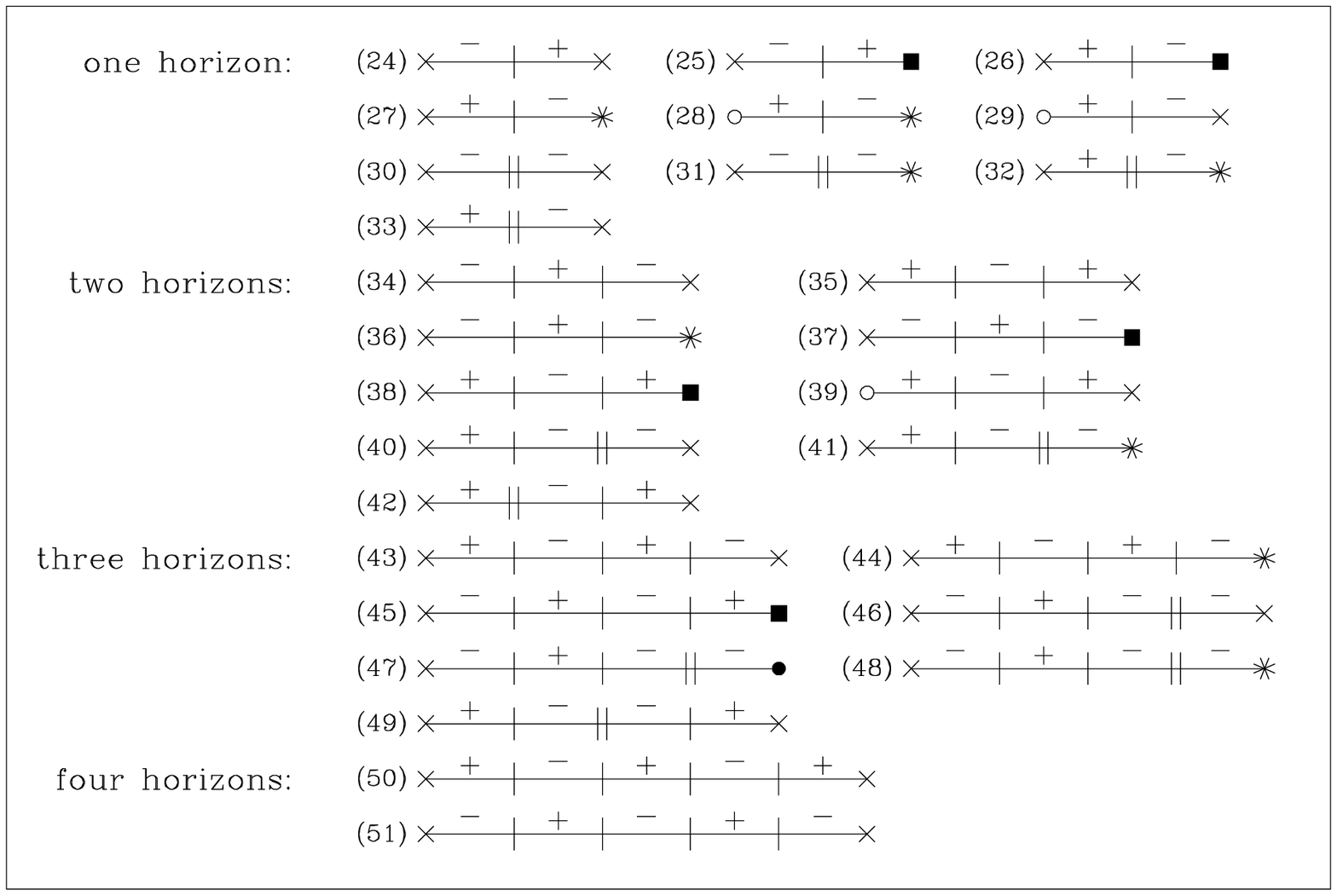}{0.6}{0.9}{43 197 568 547}
{1d graphs with horizons}

Before constructing the Carter-Penrose diagrams we present in
Figs.~\ref{figdiag1} and~\ref{figdiag2} a graphical representation of all
possible compositions of the pieces of 2d space-times discussed in the
previous section. In principle there is an infinite number of such possible
compositions, because one can always insert additional pairs of regular
horizons. We have, however, restricted ourselves to compositions which are
compatible with the number of parameters of a generic solution. Thus we take
into account the necessity to suppress the divergent modes at every singular
point. For example at every horizon we loose one parameter. In this counting
the singular solutions at $r=0$ are considered as f.p.s without divergent
mode (compare Sect.~\ref{chapsing}) as are the solutions with dS
asymptotics. For a given $\Lambda$ the five autonomous Eqs.~(\ref{taueq})
are reduced to four taking into account the constraint Eq.~(\ref{kappeq}).
This results in a 3-parameter family of orbits. Hence we can suppress
maximally four divergent modes varying also $\Lambda$, implying e.g.\ that
we can have a maximum of four horizons. This does, however, not necessarily
imply that all cases are realized, i.e.\ can be actually obtained
numerically. Nor does it mean that solutions with more than four horizons
are excluded. It just means that we see no way to construct them varying the
available parameters at hand. Actually we have not even found any solution
with four horizons.

The graphs in Figs.~\ref{figdiag1} and~\ref{figdiag2} are more or less self
explanatory, but some remarks are in order. We ignore the presence resp.\
position of an equator, which would just lead to a much larger number of
diagrams without a particular gain in information. Likewise we ignore the
type of the singularity at a singular origin.
\clearfigs
\mkfig{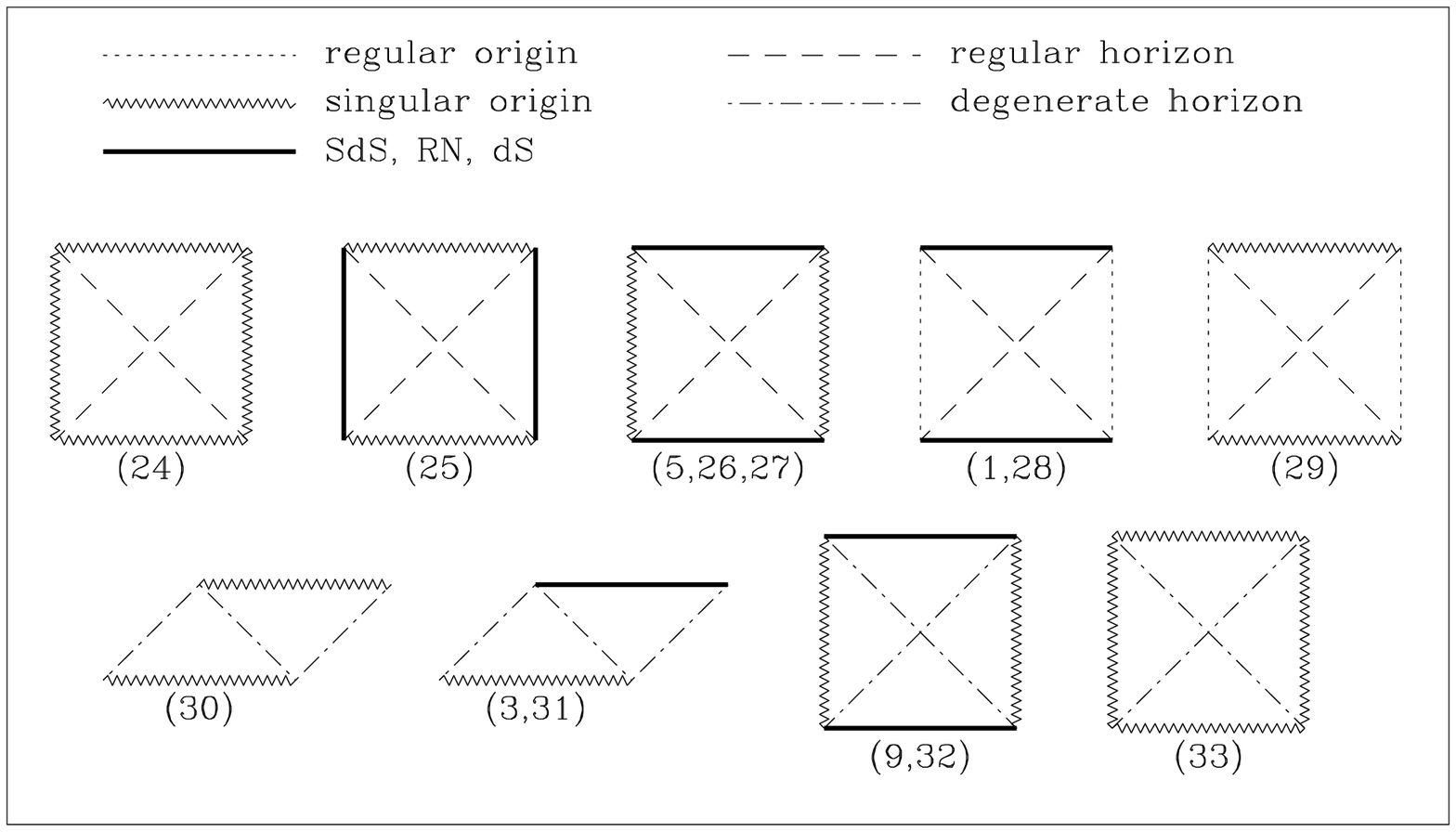}{0.6}{0.9}{43 197 568 494}
{Carter-Penrose diagrams: one regular or degenerate horizon}

Diagram~2 represents the generic SdS solution, while 3 corresponds to the
degenerate case. Diagram~5 is the case of SdS with negative mass. Diagram~6
is the generic RNdS solution, 9~corresponds to the most degenerate RNdS
solution for $\Lambda=1/4$ with a cubically degenerate horizon. Diagrams~4,
10, and~16 represent generic solutions. The space-times 10--23 without
horizons either have spatial sections with 3-sphere topology or have
infinite throats with dS resp.\ AdS~type asymptotic behaviour. Diagrams~3
and~8 represent the degenerate SdS resp.\ RNdS solutions as well as ones
with non-trivial YM field. Diagram~9 describes besides the most degenerate
RNdS solution also solutions joining the $\sigma=+1$ piece of the latter to
a non-trivial $\sigma=-1$ part at the cubically degenerate horizon. The same
property is shared by the diagrams~32, 33, and~42.
\mkfig{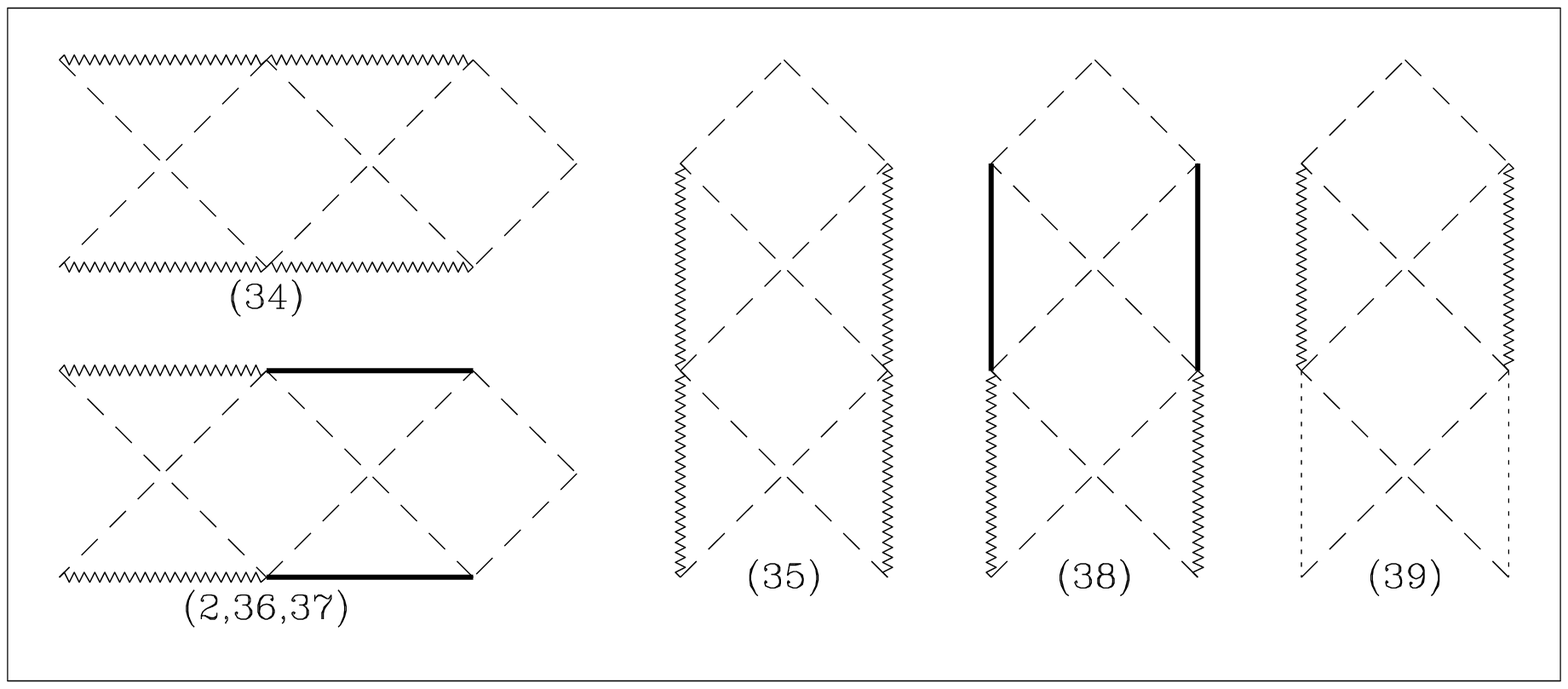}{0.6}{0.9}{43 197 568 424}
{Carter-Penrose diagrams: two regular horizons}

Diagrams~15, 22, and~23 describe the degenerate space-time of Eqs.~(\ref{Nari}a).
Diagram~41 describes a solution with a degenerate horizon running into the
SdS f.p. Such a solution could occur as a limit of a 1-parameter family of
solutions of the type~8 with YM hair.
\mkfig{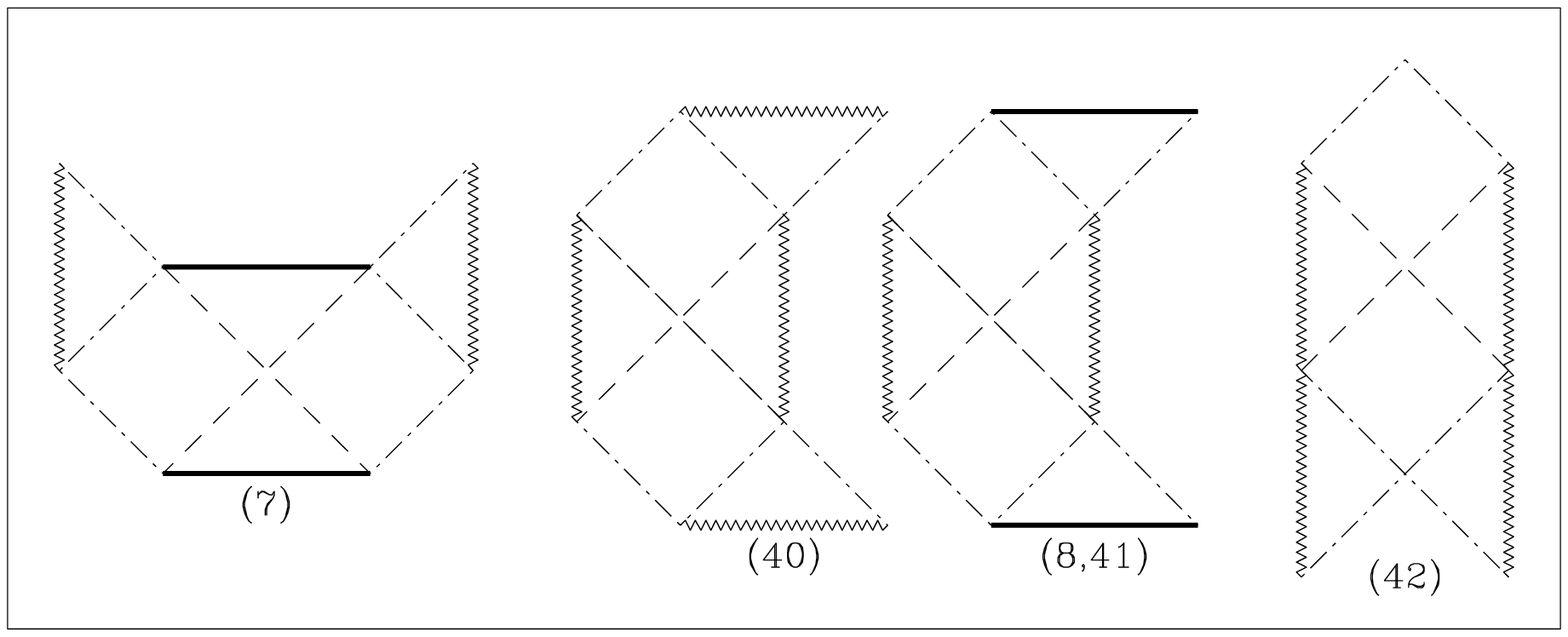}{0.6}{0.9}{43 197 568 407}
{Carter-Penrose diagrams: one degenerate and one regular horizon}

The solutions with a regular origin of \cite{Strau} are given by diagrams~1,
28, 29, and~13 in the order as they occur with increasing $\Lambda$.
Replacing the regular origin by a regular black hole horizon one obtains the
diagrams~2, 36, 34, and~29. Replacing the origin by two horizons as in the RN
black hole one obtains the diagrams~6, 44, 43, and~39.  The diagrams~19, 20,
and~21 are new types of regular solutions starting at the RN resp.\ SdS
f.p.\ and running into the SdS f.p.\ or to infinity.
\mkfig{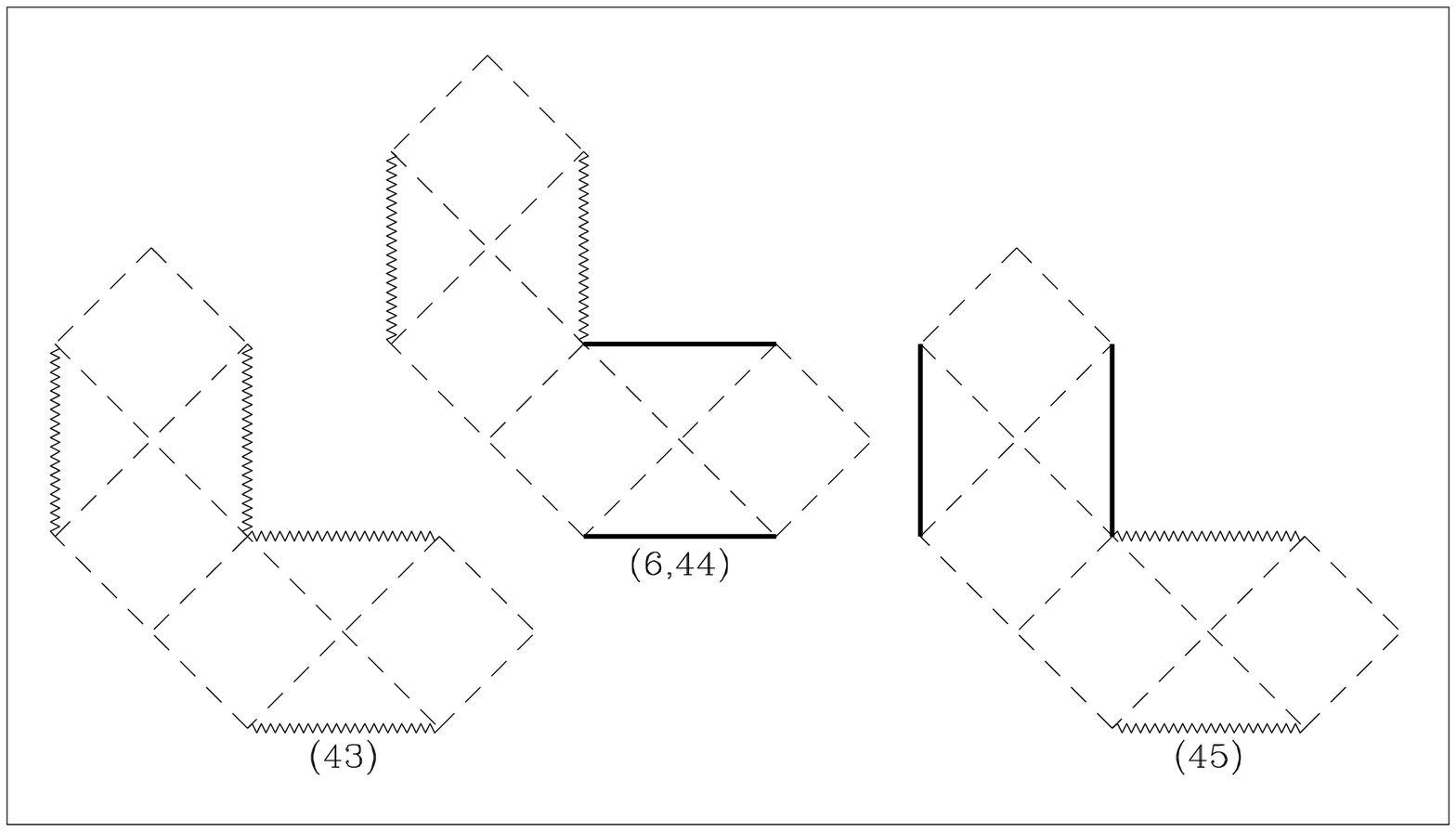}{0.6}{0.9}{43 197 568 494}
{Carter-Penrose diagrams: three regular horizons}
\mkfig{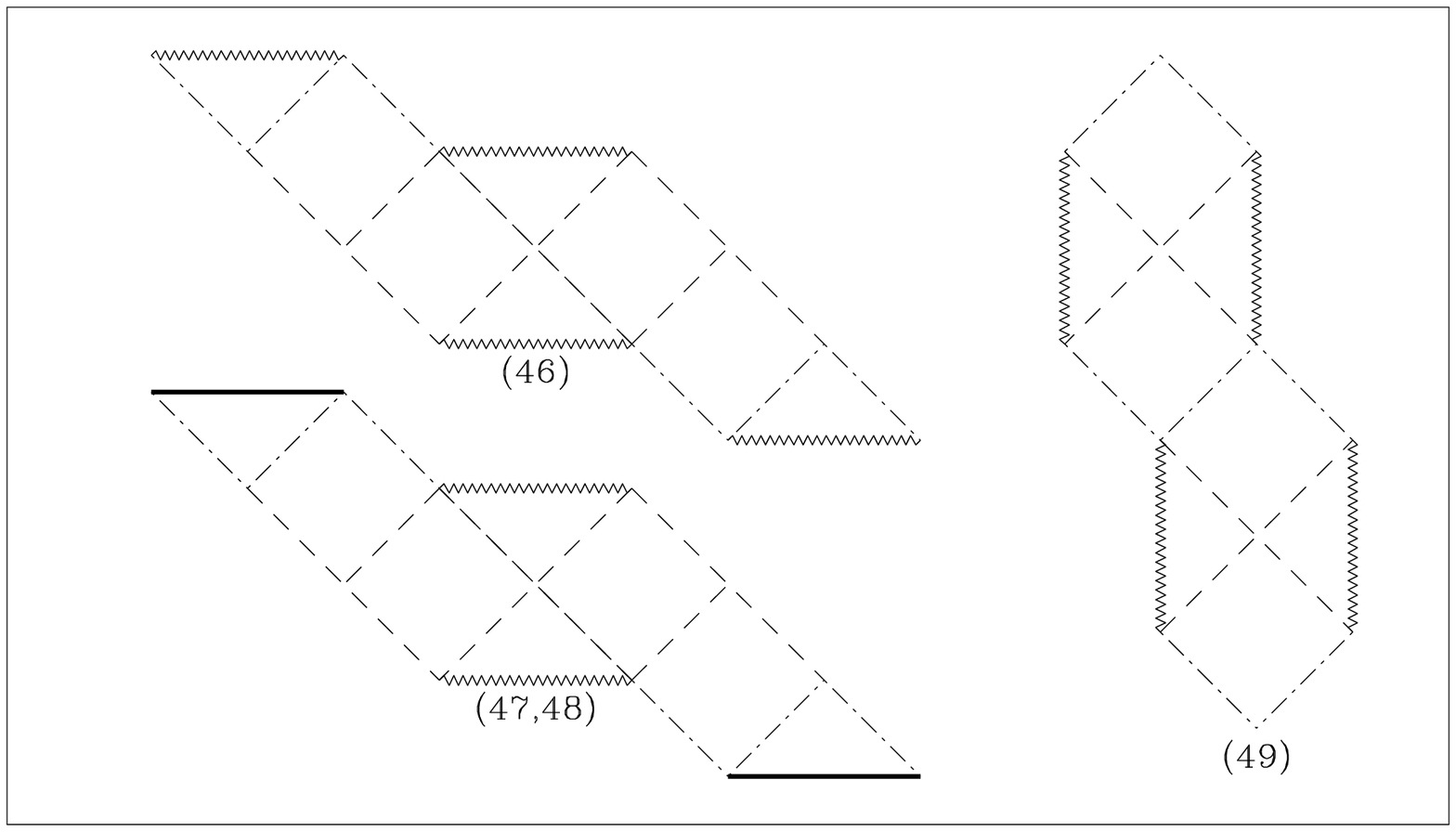}{0.6}{0.9}{43 197 568 494}
{Carter-Penrose diagrams: one degenerate and two regular horizons}
\mkfig{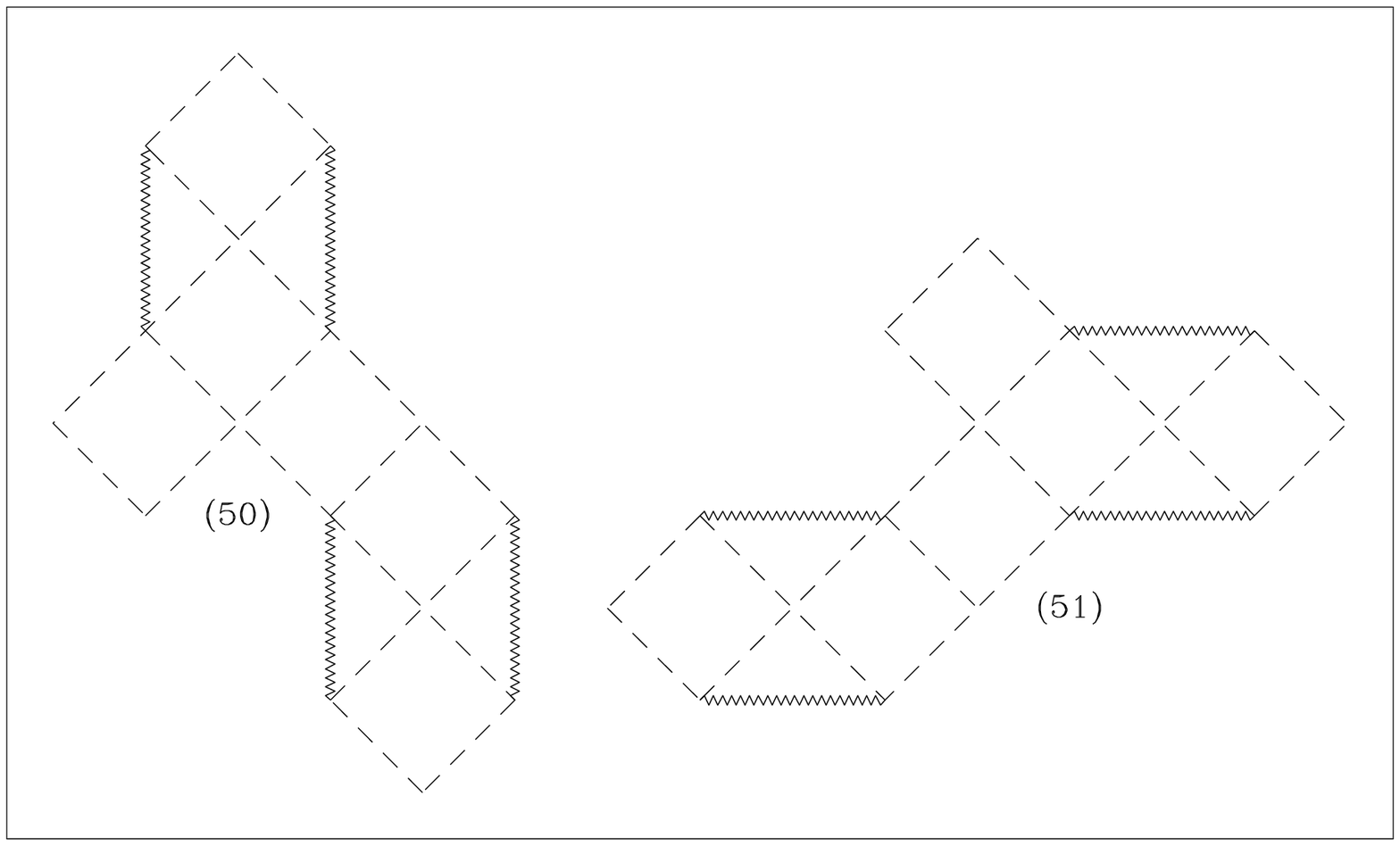}{0.6}{0.9}{43 197 568 513}
{Carter-Penrose diagrams: four regular horizons}

Figs.~\ref{figcart1}--\ref{figcart6} show the elementary building blocks of
the Carter-Penrose diagrams corresponding to the diagrams of
Figs.~\ref{figdiag1} and~\ref{figdiag2}, which have at least one horizon.
The complete diagrams are obtained from those by periodic repetition, gluing
them along suitable common horizons \cite{Walker,Katanaev}. As usual
space-like boundaries are represented by horizontal lines, time-like ones
are vertical.

\section{Numerical Results}\label{chapnum}

In this section we present a selection of numerical results, some of which
we would like to emphasize. We perform a quantitative study of the limit
$n\to\infty$ of the 3-sphere solutions. We present a whole new class of
solutions with two and three regular horizons. We give a complete
description of the phase space of solutions for $\sigma=+1$ with a regular
origin. An analogous description for $\sigma=-1$ exhibits some important
aspects. One cannot expect a complete description due to the very
complicated structure of solutions with a singular origin caused by
unbounded oscillations \cite{Galtsov,BLM}.

\subsection{Solutions with a regular origin and a regular horizon}

\mkfig{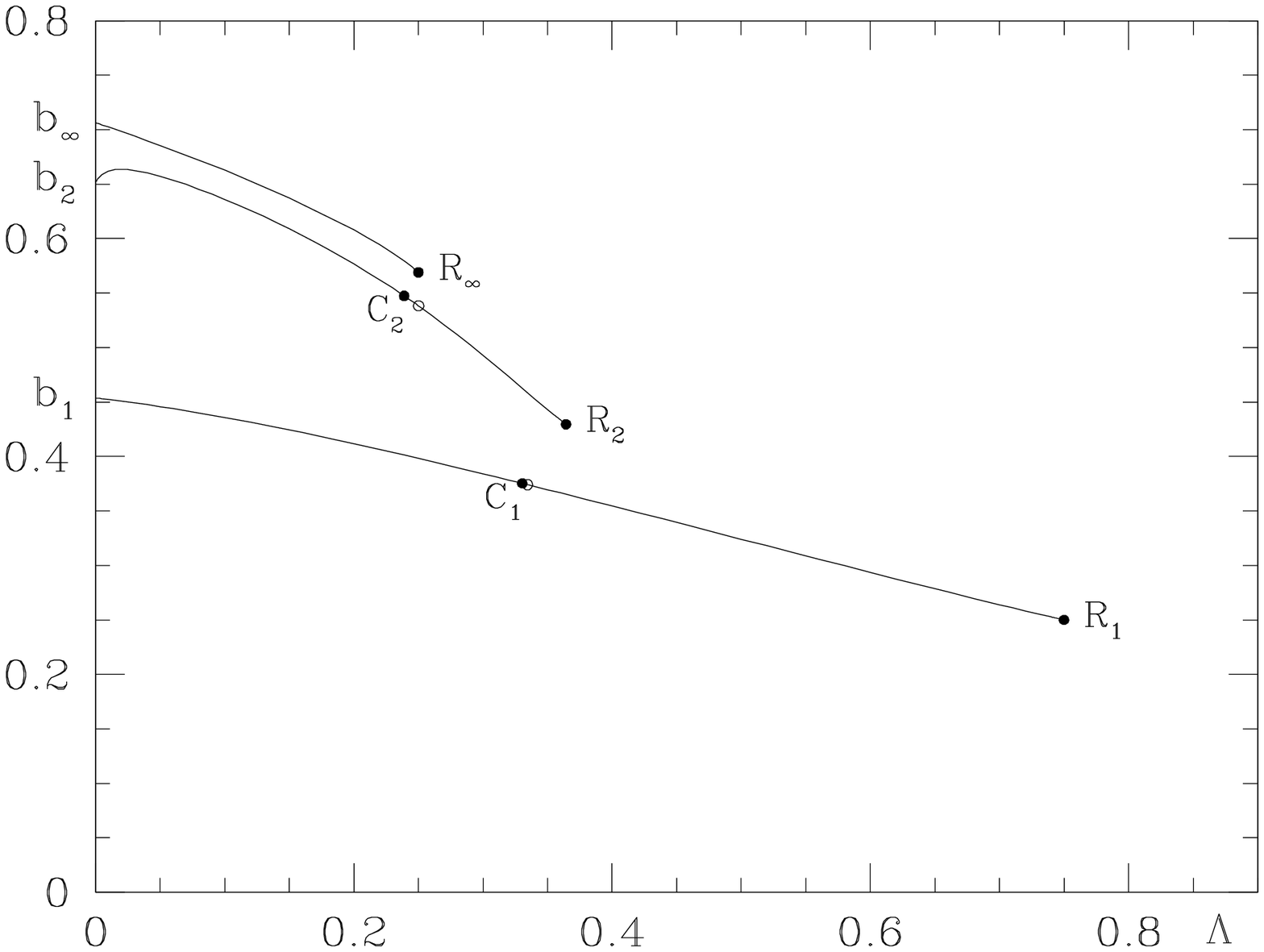}{0.5}{0.9}{50 173 574 570}
{Regular solutions with $n=1$, $2$, and~$\infty$ zeros of $W$}
In Fig.~\ref{figblam} we show parameters of solutions with a regular origin
and a regular horizon with $n=1$ and $2$ zeros of $W$ as well as for the
limiting solutions ($n=\infty$) connecting the origin and the RN~f.p.
\begin{table}
\caption[tabpllam]{
Parameters of regular solutions
\label{tabpllam}}
\begin{center}\begin{tabular}{c|cc|cc|c}
 $n$&$\Lambda_c$&$r_c$& $\Lambda_*$&$r_*$& $\Lambda_r$\\
  &$b_c$&$W_c$& $b_*$&$W_*$& $b_r$\\\hline\hline
%
% data from ../mc.dat, ../rb.dat, and ../xb.dat
%
 1& $0.33049695$& $1.729083$& $0.33448289$& $1.706678$& $0.75000000$\\
  & $0.37534807$& $-0.85135$& $0.37417940$& $-0.85218$& $0.25000000$\\\hline
 2& $0.23896851$& $1.934703$& $0.25004507$& $1.812673$& $0.36424423$\\
  & $0.54743193$& $+0.45963$& $0.53831284$& $+0.48411$& $0.42959769$\\\hline
 3& $0.23712384$& $1.832376$& $0.24706870$& $1.686850$& $0.29321764$\\
  & $0.57517232$& $-0.23626$& $0.56568105$& $-0.28415$& $0.50882906$\\\hline
 4& $0.24027732$& $1.746936$& $0.24790151$& $1.606049$& $0.27032753$\\
  & $0.57699432$& $+0.13396$& $0.56950069$& $+0.18872$& $0.54048970$\\\hline
 5& $0.24275653$& $1.688873$& $0.24861452$& $1.556666$& $0.26089512$\\
  & $0.57578215$& $-0.08158$& $0.56992909$& $-0.13598$& $0.55401931$\\\hline
 6& $0.24445557$& $1.648272$& $0.24907014$& $1.524611$& $0.25638498$\\
  & $0.57448821$& $+0.05216$& $0.56981254$& $+0.10345$& $0.56042575$\\\hline
\end{tabular}\end{center}

\end{table}
In Table~\ref{tabpllam} we present parameters of these solutions for $n\le6$
and some special values of $\Lambda$. When continued beyond the horizon into
the region with $\sigma=-1$ they are asymptotically deSitter for $\Lambda$
up to an $n$-dependent critical value $\Lambda_c$. For $\Lambda=\Lambda_c$,
indicated by the points $C_n$ in Fig.~\ref{figblam} they run into the SdS
f.p.\, while for $\Lambda>\Lambda_c$ they have an equator and end at $r=0$.
The equator coincides with the horizon for $\Lambda=\Lambda_*$ (the open
circles in Fig.~\ref{figblam}) and occurs for $\sigma=\mp1$ if
$\Lambda\lessgtr\Lambda_*$. The radius of the horizon decreases as $\Lambda$
increases and vanishes for $\Lambda=\Lambda_r$, the points $R_n$,
corresponding to the regular 3-sphere solutions.

\subsection{Regular 3-sphere solutions}

\begin{table}
\caption[tabplreg]{
Parameters of 3-sphere solutions
\label{tabplreg}}
\begin{center}\begin{tabular}{r|l|l|l}
 $n$&$\Lambda$&$b$&$r_{\rm max}$\\\hline\hline
%
% data from ../rb.dat
%
  1& 0.7500000000& 0.2500000000& 1.4142136\\\hline
  2& 0.3642442330& 0.4295976874& 1.4355566\\\hline
  3& 0.2932176375& 0.5088290567& 1.5213897\\\hline
  4& 0.2703275292& 0.5404896987& 1.4826529\\\hline
  5& 0.2608951239& 0.5540193079& 1.4829530\\\hline
 10& 0.2512790802& 0.5674090979& 1.4396422\\\hline
 20& 0.2501165095& 0.5688910784& 1.4224344\\\hline
 30& 0.2500264064& 0.5690006373& 1.4181848\\\hline
 40& 0.2500089730& 0.5690215671& 1.4165412\\\hline
 50& 0.2500038394& 0.5690276981& 1.4157401\\\hline
%
% data from ../cb.dat
%
\hline
 $\infty$& 0.25  & 0.5690322659& $\sqrt{2}$\\
\end{tabular}\end{center}

\end{table}
Table~\ref{tabplreg} contains the parameters of some regular 3-sphere
solutions with up to 50 zeros of $W$. As $n$ increases the values
$\Lambda_n$, $b_n$ rapidly converge to $\Lambda_\infty=1/4$ and
$b_\infty\approx0.569032$, i.e.\ the point $R_\infty$ of Fig.~\ref{figblam}.
The corresponding limiting solution connects a regular origin and the RN f.p.\ with $\Lambda=1/4$ and
asymptotic behaviour (compare Sect.~\ref{sectfp}).
\begin{equation}
\kappa\approx{4\over\tau-\tau_0}\;,\qquad
W\approx\sqrt{84}r{\sin(\tau+\theta)\over(\tau-\tau_0)^2}\;.
\end{equation}
Numerically we find $\theta\approx1.7884$ and $\tau_0\approx-3.8717$
(normalizing $\tau$ such that $\tau-\ln r\to0$ as $r\to0$). Solutions with
$(\Lambda_\infty+\delta\Lambda,b_\infty+\delta b)$ for small
$\delta\Lambda$, $\delta b$ will miss the RN f.p.\ but yield a regular
3-sphere provided $\kappa$, $N$, and $W$ or $U$ simultaneously vanish for
some $\tau$. Assuming that the phase of $W$ remains essentially unchanged
this requires that $\kappa$ and $N$ vanish at $\tau+\theta\approx(n+1)\pi/2$
in order to have $n$ zeros of $W$. For solutions starting from the origin we
first have to choose $\delta b\approx-1.2\delta\Lambda$ to suppress the
unstable mode (with eigenvalue $+6$) near the f.p., where the solution
approximately obeys
\begin{subeqnarray}\label{krnL}
\dot\kappa&=&\bar r-\kappa^2\;,\\
\dot{\bar r}&=&-2\bar N\;,\\
\dot{\bar N}&=&-{\bar r^2\over4}-\delta\Lambda\;,
\end{subeqnarray}
with the consequence that $\bar r^3/12+\delta\Lambda\bar r-\bar N^2$ is
constant. Moreover Eqs.~(\ref{krnL}) are invariant under the rescaling
$(\dd\tau,\kappa,\bar r,\bar
N,\delta\Lambda)\to(\dd\tau/\lambda,\lambda\kappa,\lambda^2\bar
r,\lambda^3\bar N,\lambda^4\delta\Lambda)$. Starting from some $\tau_1$ with
$\kappa=\bar N=0$ and $\bar r=-1$ we can thus integrate
\begin{subeqnarray}\label{krL}
\dot\kappa&=&\bar r-\kappa^2\;,\\
\dot{\bar r}&=&-\sqrt{{\bar r^3+1\over3}+4\delta\Lambda(\bar r+1)}\;,
\end{subeqnarray}
towards decreasing $\tau$, adjusting $\tau_1$ and $\delta\Lambda$ such that
$\kappa$ and $\bar r$ diverge at $\tau=\tau_0$ with $\bar r/\kappa^2\to3/4$.
Numerically we find $\delta\Lambda\approx0.81575$ and $\tau_1-\tau_0\approx
3.8288$ and thus obtain for $\Lambda_n$ and $\bar r_n$, the value of $\bar
r$ at the equator, the asymptotic expressions
\begin{subeqnarray}
\left((n+1){\pi\over2}-\theta-\tau_0\right)\root4\of{\Lambda_n-0.25}
  &\approx&3.8288\root4\of{0.81575}\approx3.6387\;,\\
{\bar r_n\over\sqrt{\Lambda_n-0.25}}
  &\approx&-{1\over\sqrt{0.81575}}\approx-1.1072\;,
\end{subeqnarray}
in good agreement with the numerical results shown in Fig.~\ref{figrbfit}
for $n\le50$.
\mkfig{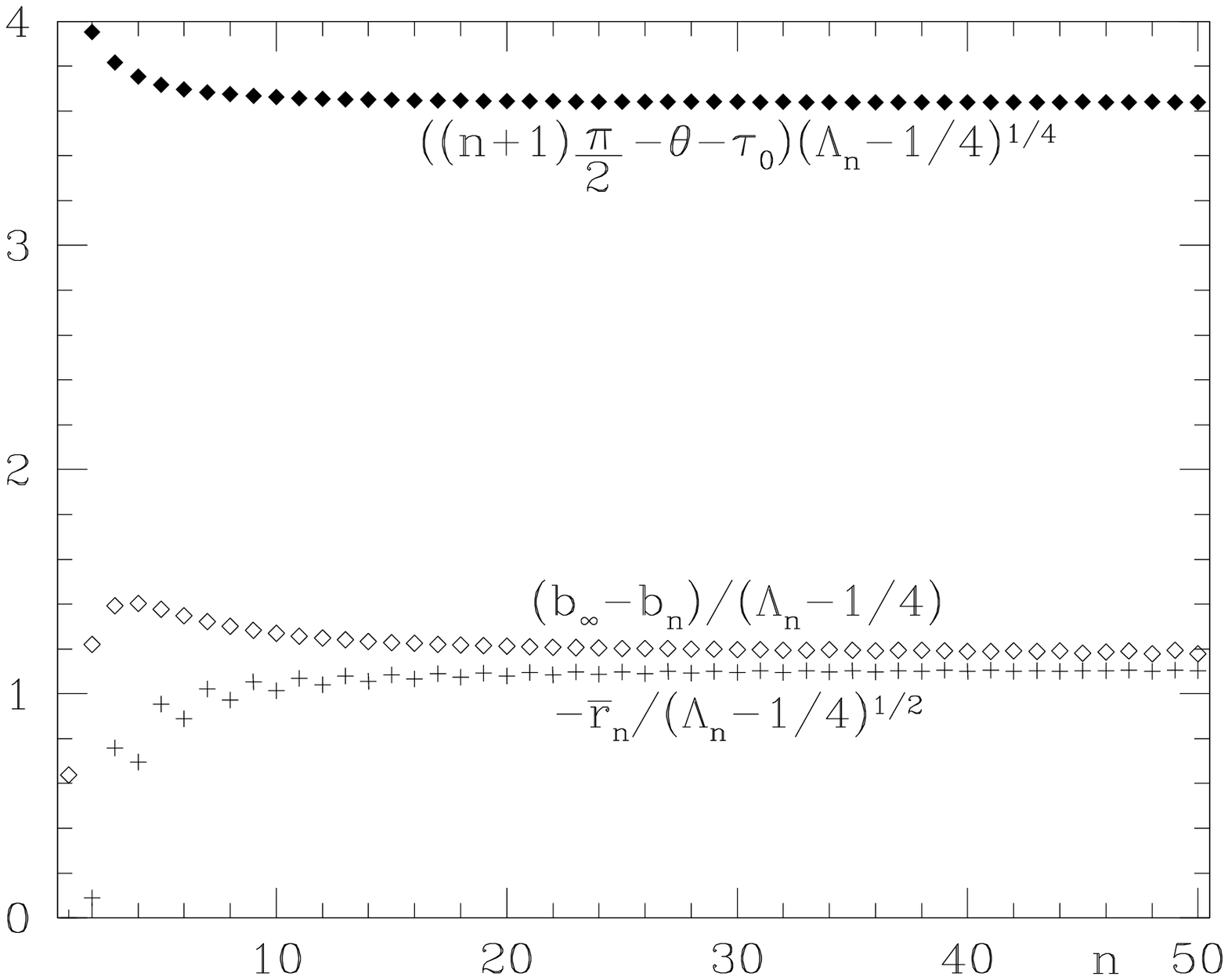}{0.6}{0.9}{63 179 562 565}
{Asymptotic behaviour of the parameters of regular 3-sphere solutions}

\subsection{Solutions with two or more regular horizons}

Fig.~\ref{figrhrh} shows parameters for solutions with two horizons for
$\sigma=+1$ and various values of $\Lambda$.
\mkfig{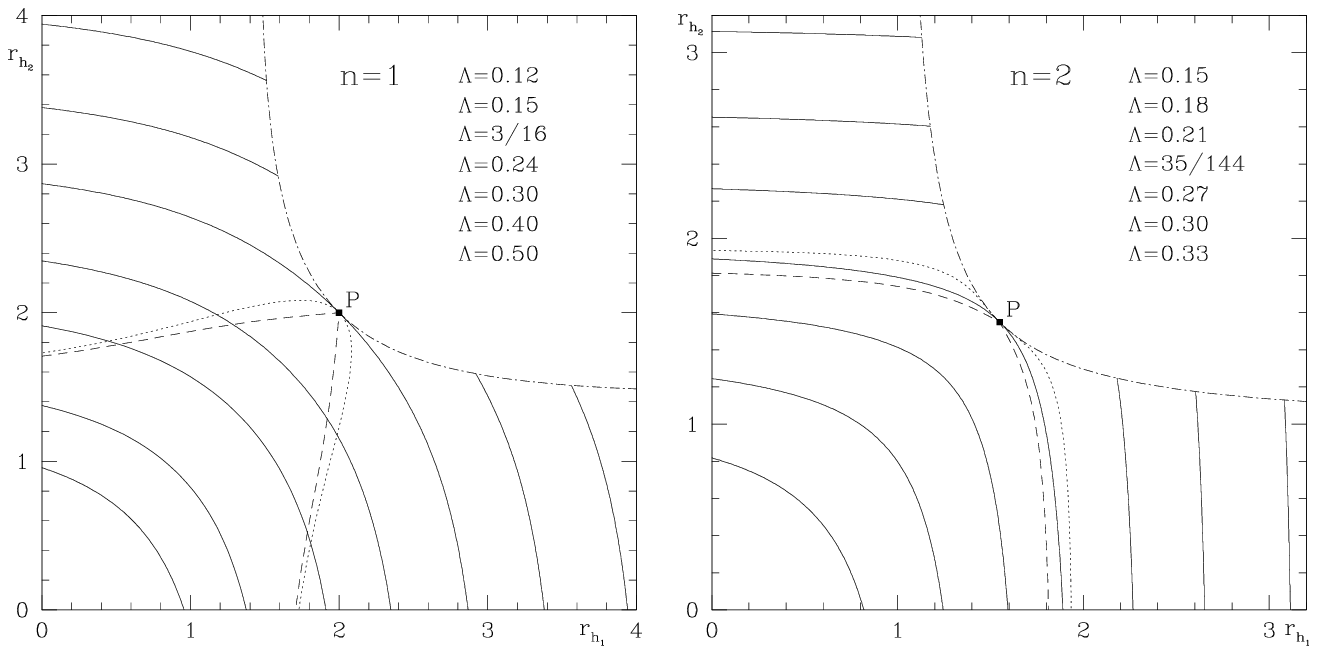}{0.8}{0.9}{105 551 479 734}
{$r_{h_1}$ vs.\ $r_{h_2}$ for solutions with $n=1$ and~$2$ zeros of $W$}
In the limit $r_h\to0$ for one or both horizons they reproduce the solutions
with an origin and a horizon and the regular 3-spheres discussed above. The
dashed-dotted curve delimits the region of existence; on it the value of
$W_h$ vanishes and the solutions bifurcate with
Reissner-Nordstr{\o}m-deSitter solutions. The dotted curves correspond to
the critical $\Lambda$ values, giving solutions running into the
Schwarzschild f.p.\ in the region $\sigma=-1$. The solutions in the region
outside the dotted curves are asymptotically deSitter.  The dashed curves
correspond to solutions with coinciding equator and horizon. Solutions in
the region inside the dashed curves have an equator between their horizons
in contrast to those outside of it.  The RN-deSitter and the critical
solutions have no equator and thus the dashed-dotted and the dotted curves
must lie outside the dashed ones.  Since on the diagonal the two horizons
are equal, these solutions have an equator and thus the diagonal lies inside
the dashed curves. Thus the dashed curves must go through the point $P$ on
the intersection of the diagonal with the boundary curve, where the two
horizons merge. Since the solutions on the boundary curve are asymptotically
deSitter, the dotted curve must also go through $P$. At this point the
solution can be obtained exactly through a suitable rescaling. Putting
\begin{equation}\label{scaling}
W=\epsilon \bar W, U=\epsilon\bar U,N=\epsilon^2\bar N \quad{\rm and}\quad
r=R(1+\epsilon^2\bar r)\;,
\end{equation}
where $\epsilon$ is a small parameter we obtain in leading order from the
constraint Eq.~(\ref{kappeq}) $\Lambda=(R^2-1)/r^4$ and from
Eqs.~(\ref{taueq})
\begin{subeqnarray}\label{scaledeq}
\dot{\bar W}&=&R\bar U\\
\dot{\bar U}&=&-{\bar W\over R}-\kappa\bar U\\
\dot\kappa&=&1-2\Lambda R^2-\kappa^2\;.
\end{subeqnarray}
Solving Eq.~(\ref{scaledeq}c) we get $\kappa=a\cot(a\tau)$ as in
Eq.~(\ref{Nari}c) with $a^2=2\Lambda R^2-1$ and plugging the result into
Eqs.~(\ref{scaledeq}a,b) we arrive at the second order equation
\begin{equation}\label{second}
\ddot{\bar W}+a\cot(a\tau)\dot{\bar W}+\bar W=0\;.
\end{equation}
Introducing $x=\cos(a\tau)$ we find the Legendre equation
\begin{equation}\label{Legendre}
{\dd\over\dd x}\left((1-x^2){\dd\bar W\over\dd x}\right)
  +{1\over a^2}\bar W=0\;.
\end{equation}
The acceptable solutions are those regular at $x=\pm1$ given by the
Legendre polynomials, obtained for $a^{-2}=n(n+1)=2$, $6$, $12,~\ldots$
yielding
\begin{subeqnarray}\label{rational}
R^2&=&{2\over 1-a^2}=4,~{12\over 5},~{24\over 11},~\ldots\;,\\
\Lambda&=&{1-a^4\over 4}={3\over 16},~{35\over 144},~{143\over 576},~\ldots\;.
\end{subeqnarray}
The cases $n=1$ and~$2$ are shown in Fig.~\ref{figrhrh}.

Combining numerical results for $\sigma=+1$ and $\sigma=-1$ we have found a
one parameter family of solutions with three regular horizons and two zeros
of $W$.
\begin{table}
\caption[tabpl3h]{
Some solutions with three regular horizons and two zeros of $W$
\label{tabpl3h}}
\hbox to\hsize{\hss\begin{tabular}{c|c|cc|cc|rc|c}
 $\Lambda$&$W_0$& $r_1$&$W_1$& $r_2$&$W_2$& $r_3~~~$&$W_3$&$W_\infty$\\\hline\hline
%
% data from ../emun2.dat, ../../iymc/mun2.dat, , and ../../iymc/gn2mu.dat
%
 $0.0045$& $1.3655$& $0.0948$& $1.3582$& $1.0256$& $0.1000$
  & $24.6488$& $0.0862$& $0.1637$\\\hline
 $0.0643$& $1.4072$& $0.1207$& $1.3964$& $1.0459$& $0.0800$
  & $5.6355$& $0.0263$& $0.0638$\\\hline
 $0.1187$& $1.4388$& $0.1417$& $1.4249$& $1.0918$& $0.0600$
  & $3.7151$& $0.0197$& $0.0610$\\\hline
 $0.1655$& $1.4737$& $0.1664$& $1.4556$& $1.1502$& $0.0400$
  & $2.8234$& $0.0152$& $0.0647$\\\hline
 $0.2061$& $1.5141$& $0.1968$& $1.4904$& $1.2351$& $0.0200$
  & $2.2350$& $0.0098$& $0.0716$\\\hline
 $0.2247$& $1.5378$& $0.2155$& $1.5103$& $1.3051$& $0.0100$
  & $1.9718$& $0.0059$& $0.0762$\\\hline
 $0.2337$& $1.5511$& $0.2263$& $1.5214$& $1.3623$& $0.0050$
  & $1.8255$& $0.0034$& $0.0789$\\\hline
 $0.2411$& $1.5630$& $0.2361$& $1.5312$& $1.4544$& $0.0010$
  & $1.6627$& $0.0008$& $0.0813$\\\hline
\end{tabular}\hss}

\end{table}
In Table~\ref{tabpl3h} we present the parameters of some of these solutions.
Starting from $r=0$ with an RN~type singularity and $W=W_0$, they have two
regular horizons at $r=r_{1,2}$ with $W=W_{1,2}$, two zeros of $W$, a third
horizon at $r_3$ with $W_3$, and extend to $r=\infty$ with dS~asymptotics
and $W=W_\infty$. There is a similar such family with three zeros of $W$,
but we found no such solution with only one zero.

\subsection{Description of the phase space}

\mkfig{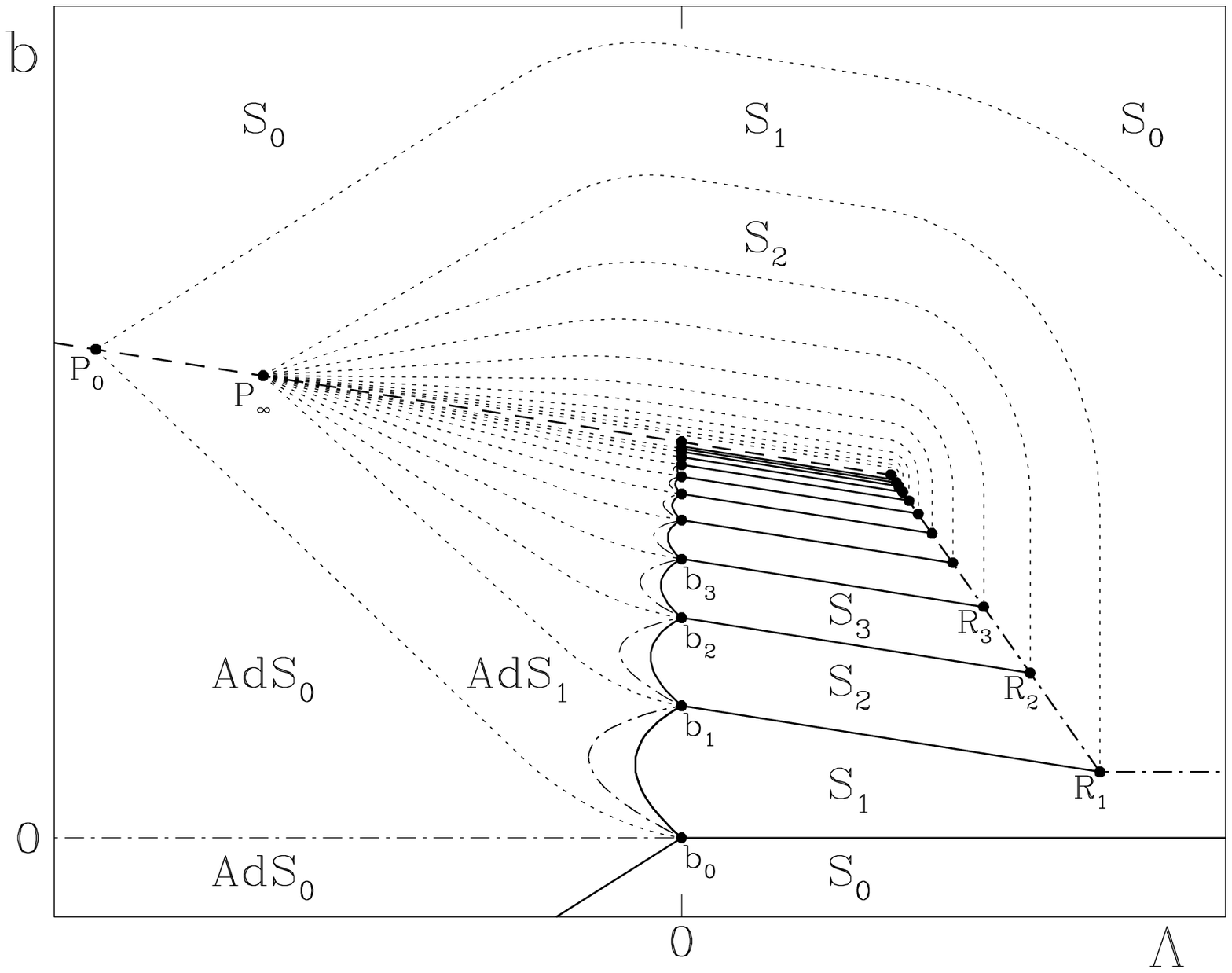}{0.8}{0.9}{63 179 562 565}
{Schematic phase space of solutions for $\sigma=+1$ with a regular origin}
Fig.~\ref{figsketch} schematically shows the phase space for solutions
starting from a regular origin, i.e.\ with $\sigma=+1$, dividing the
$(\Lambda,b)$ plane into various regions according to their type and
including results for $\Lambda=0$ from \cite{BFM} resp.\ for $\Lambda<0$
from \cite{neglam}. All the cases listed in the Classification
Thm.~\ref{class+} are present in the half plane $\Lambda>0$ of
Fig.~\ref{figsketch}. The regions ${\rm AdS}_n$ with $\Lambda<0$ and ${\rm S}_n$
correspond to solutions with $n$ zeros of $W$ and AdS asymptotics or a
singular origin respectively. At the moment this picture is mostly based on
numerical results, but proving the existence of such a picture will
certainly be an essential part in an existence proof for the various
solution types.

\mkfig{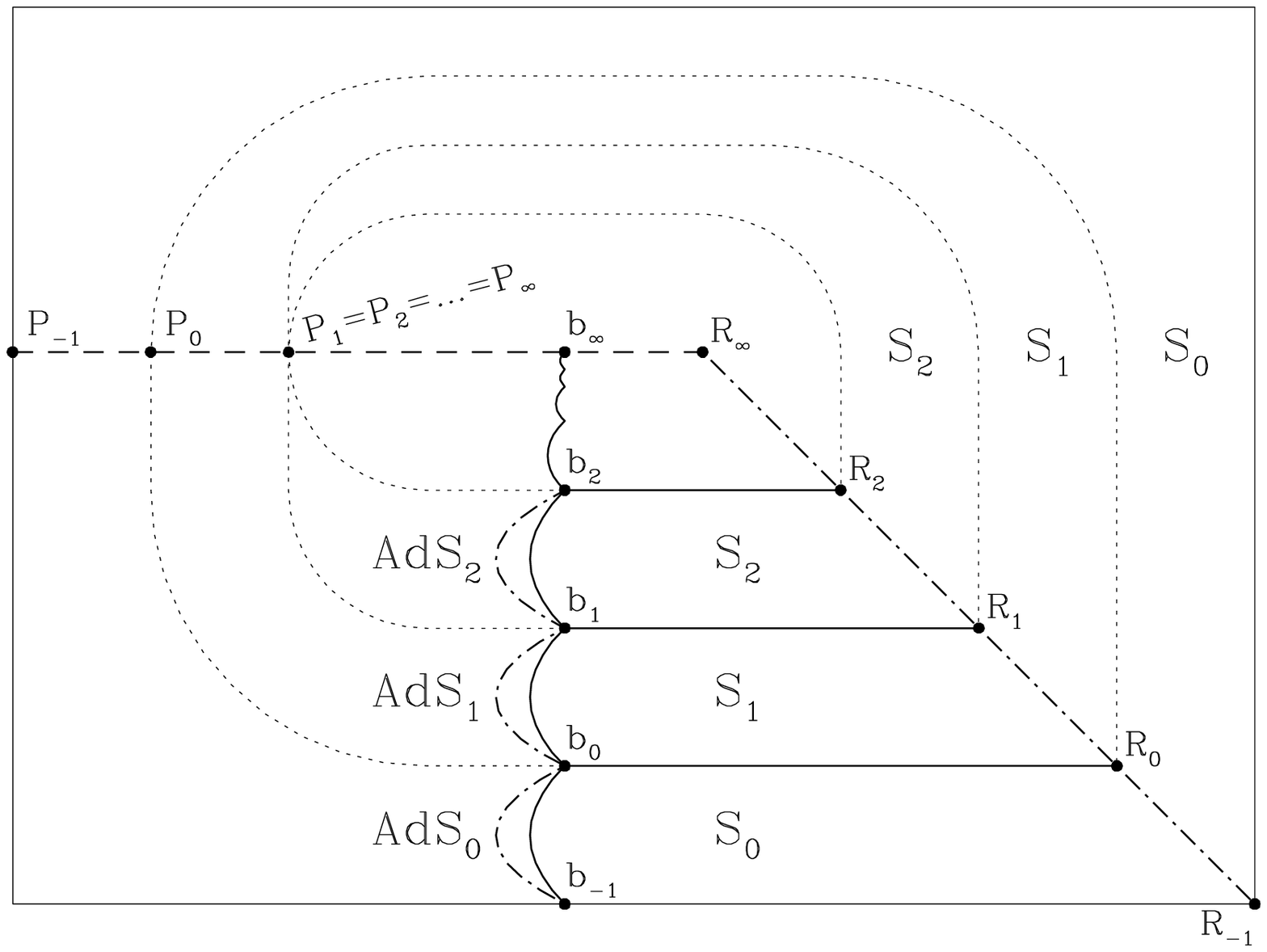}{0.8}{0.9}{89 181 561 539}
{Compactified phase space of solutions for $\sigma=+1$ with a regular
origin}
In order to simplify the description of the phase space we deform and
compactify Fig.~\ref{figsketch} into Fig.~\ref{figcomp} as follows: First we
identify the points $(\Lambda,\pm\infty)$ and add a dashed-dotted line along
$b=\pm\infty$ separating the two previously disconnected parts of the region
${\rm S}_0$. Next we add a point $R_0=(+\infty,0)$ corresponding to the limit
$r_h\to0$ of deSitter space and compactify $(+\infty,b)$ to this point. The
points with constant (finite) $\Lambda$ are then on concentric curves around
$R_0$. We then add the points $b_{-1}$, $P_{-1}$, and $R_{-1}$ as the limits
$\Lambda\to-\infty$ of $(\Lambda,-\sqrt{|\Lambda|/4})$,
$(\Lambda,+\sqrt{|\Lambda|/4})$, and $(\Lambda,\pm\infty)$. And we add points
$P_n$ for $n=1,\ldots$ all identical to $P_\infty$. Finally the boundary of
Fig.~\ref{figcomp} corresponding to $(-\infty,b)$ is compactified to one
point $b_{-1}=P_{-1}=R_{-1}$.

The open region ${\rm AdS}=\cup_{n\ge0}{\rm AdS}_n$ is the complement of the closure of
the open region ${\rm S}=\cup_{n\ge0}{\rm S}_n$. The points $b_n$ at $(0,b_n)$ represent
the asymptotically flat BK solutions (augmented by Minkowski space for $n=0$),
with their limit point $b_\infty$ (compare \cite{BFM}). The points $R_n$
represent the regular 3-sphere solutions, with their limit point $R_\infty$
at $\Lambda=1/4$. ${\rm AdS}_n$ and ${\rm AdS}_{n+1}$ are separated by the dotted curves
from $b_n$ to $P_n$ where $W_\infty=0$; ${\rm S}_n$ and ${\rm S}_{n+1}$ are separated by
the dotted curves from $R_n$ to $P_n$ where $W_0=0$ and the solid curves
from $b_n$ to $R_n$ with a regular horizon. The dashed-dotted curves from
$R_{n-1}$ to $R_n$ with $|W_0|=1$ and an S~type singularity at the origin
divide ${\rm S}_n$ into parts with $|W_0|>1$ and $|W_0|<1$, both with a RN~type
singularity. Similar dashed-dotted curves from $b_{n-1}$ to $b_n$ where
$|W_\infty|=1$ separate ${\rm AdS}_n$ into parts with $|W_\infty|>1$ and
$|W_\infty|<1$. The regions ${\rm AdS}_n$ and ${\rm S}_n$ are separated by the solid
curves from $b_{n-1}$ to $b_n$ with SAdS asymptotics where $W_\infty$ resp.\
$W_0$ diverge (compare \cite{neglam}). Finally the dashed curve from
$R_\infty$ through $b_\infty$, $P_\infty$ with $\Lambda=-15/4$, and $P_0$ to
$P_{-1}$ represents solutions with the RN f.p.\ asymptotics. They oscillate
for $\Lambda>-15/4$ and therefore all regions ${\rm AdS}_n$ and ${\rm S}_n$ for $n\gg1$
are in the vicinity of the curve from $R_\infty$ to $P_\infty$. For
$\Lambda<-15/4$ the RN solutions have a finite number of zeros; numerically
we found $n=1$ from $P_\infty$ to $P_0$ with $\Lambda\approx-5.0646$ and
$n=0$ from $P_0$ to $P_{-1}$. The curves of constant $W_\infty$ in the
region ${\rm AdS}_n$ all connect $b_{n-1}$ and $b_n$ whereas the curves of
constant $W_0$ in the region ${\rm S}_n$ all connect $R_{n-1}$ and $R_n$,
approaching the solid boundary as $W_\infty$ resp.\ $W_0$ diverges or the
dotted and dashed one as $W_\infty$ resp.\ $W_0$ vanishes.

\mkfig{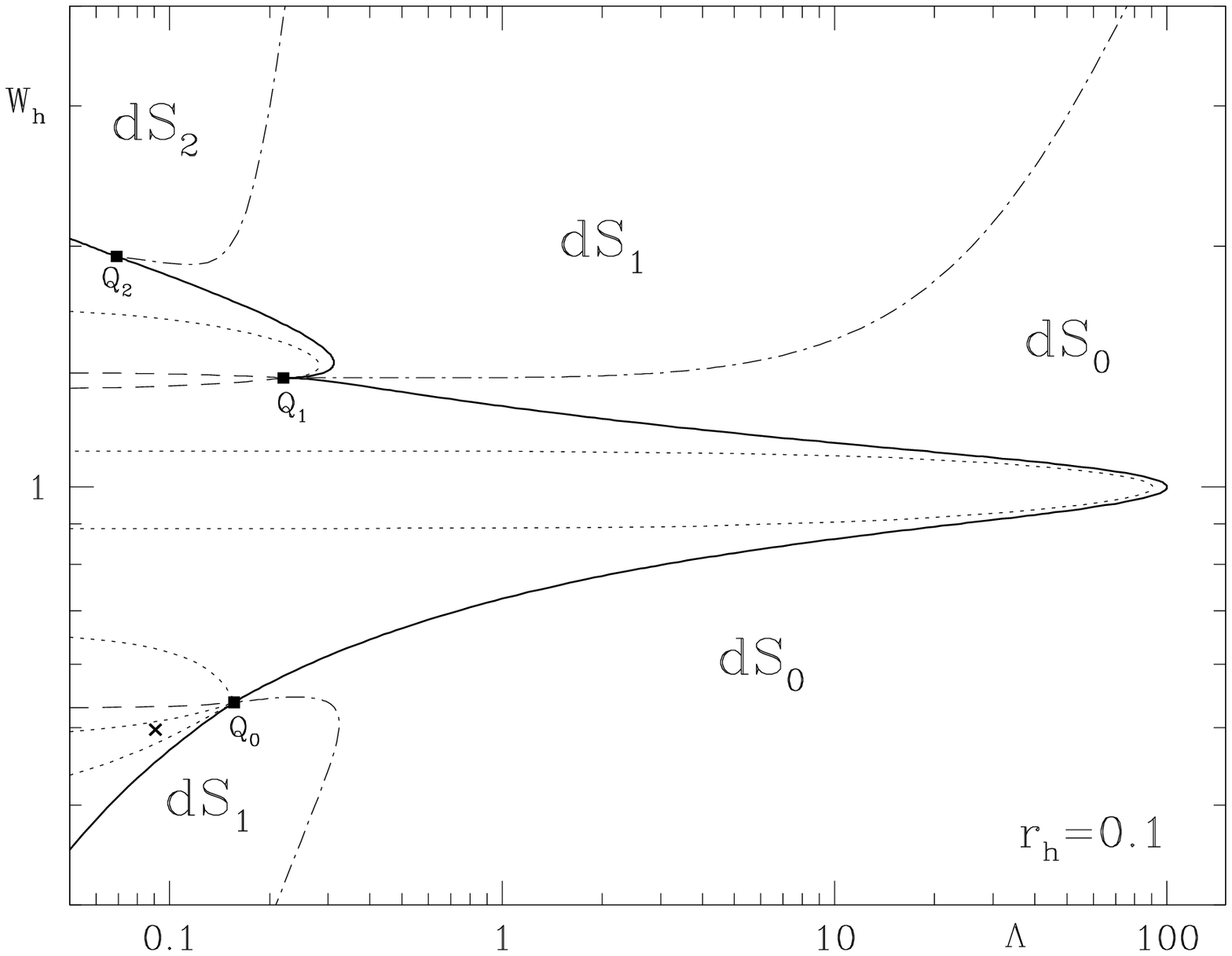}{0.8}{0.9}{63 179 562 565}
{Phase space of solutions for $\sigma=-1$ starting from a regular horizon
with $r_h=0.1$}

Fig.~\ref{figphmr1} shows the phase space of solutions for $\sigma=-1$
starting from a regular horizon with at $r_h=0.1$, again with all the cases
listed in the Classification Thm.~\ref{class-}. The open region
${\rm dS}=\cup_{n\ge0}{\rm dS}_n$ of the $(\Lambda,W_h)$ plane corresponds to solutions
with dS~asymptotics; this region is divided into regions ${\rm dS}_n$ with $n$
zeros of $W$, separated by the dashed-dotted curves where $W_\infty=0$. The
boundary of ${\rm dS}$ represents solutions ending in a f.p.\ with finite $r$: the
SdS~f.p.\ with $W=(-1)^n$ at the solid curves bounding ${\rm dS}_n$ or the
RN~f.p.\ with $W=0$ at the points $Q_i$ connecting these curves.

The complement of the closure of ${\rm dS}$ represents different types of
solutions. They can end at a second regular horizon (dashed curves) or a
singular origin with either an S~type singularity (dotted curves), a
pseudo-RN~type singularity (cross at $(\Lambda,W_h)\approx(0.09,0.5)$), or
unbounded oscillations. In Fig.~\ref{figphmr1} only a selection of such
curves is shown.

\section{Appendix}\label{app}

First we extend Prop.~1 of~\cite{BFM} by a uniqueness statement:
\begin{Prop}{}\label{local}
Suppose a system of differential equations for $m+n$ functions
$u=(u_1,\ldots,u_m)$ and $v=(v_1,\ldots,v_n)$,
\begin{equation}\label{localeq}
t{\dd u_i\over\dd t}=t^{\mu_i}f_i(t,u,v)\;,
\qquad
t{\dd v_i\over\dd t}=-\lambda_i v_i+t^{\nu_i}g_i(t,u,v)\;,
\end{equation}
with constants $\lambda_i$ with ${\rm Re}(\lambda_i)>0$ and integers
$\mu_i,\nu_i\ge1$ and let ${\cal C}$ be an open subset of $R^m$ such that
the functions $f$ and $g$ are analytic in a neighbourhood of $t=0$, $u=c$,
$v=0$ for all $c\in{\cal C}$. Then there exists a unique $m$-parameter
family of solutions of the system~(\ref{local}) with boundary conditions
$(u,v)=(c,0)$ at $t=0$, defined for $c\in{\cal C}$ and $|t|<t_0(c)$ with
some $t_0(c)>0$. These solutions satisfy
\begin{equation}\label{localprop}
u_i(t)=c_i+O(t^{\mu_i})\;,\qquad v_i(t)=O(t^{\nu_i})\;,
\end{equation}
and are analytic in $t$ and $c$.
\end{Prop}
\begin{Prf}{}
The proof of existence and analyticity in \cite{BFM} for real $\lambda_i>0$
applies without any change to complex values with positive real part.

In order to prove uniqueness we assume (without loss of generality) $t\ge0$,
introduce $\tau=-\ln t$ as new independent variable as well as $s=\sqrt{t}$
as additional dependent variable, and replace Eqs.~(\ref{localeq}) by the
autonomous system
\begin{subeqnarray}\label{localhyp}
\dot s&=&-{s\over2}\;,\\
\dot{\bar u}_i&=&{\bar u_i\over2}-s^{2\mu_i-1}f_i(s^2,c+s\bar u,v)\;,\\
\dot v_i&=&\lambda_iv_i-s^{2\nu_i}g_i(s^2,c+s\bar u,v)\;,
\end{subeqnarray}
where $\bar u_i=(u_i-c_i)/s$ vanishes as $\tau\to\infty$, i.e.\ $s\to0$ due
to Eqs.~(\ref{localprop}). This system has the hyperbolic f.p.\ $s=\bar
u=v=0$ with one stable and $m+n$ unstable modes. Thus there exists a unique
1-dimensional stable manifold with $(\bar u,v)$ analytic in $s$. Since
Eqs.~(\ref{localhyp}) are invariant under the substitution $(s,\bar
u,v)\to(-s,-\bar u,v)$ this implies that $(u,v)$ is analytic in $t$.
\end{Prf}

Next we prove some simple Lemmata used in the proof of the classification
theorems.
\begin{Lemma}{}\label{linear}
Consider a solution $y$ of the linear differential equation $\dot y=a+by$ in
some interval $\tau_0\le\tau<\tau_1$ with $|a|$ integrable. If
\begin{equation}
c(\tau',\tau)=\int_{\tau'}^\tau b(\tau'')\dd\tau''\;,
\end{equation}
is bounded from above for $\tau_0\le\tau'\le\tau<\tau_1$ then $y$ is
bounded; if $c(\tau',\tau)$ has a limit as $\tau\to\tau_1$ then $y(\tau)$
has a limit; if $c(\tau',\tau_1)=-\infty$ then $y(\tau_1)=0$.
\end{Lemma}
\begin{Prf}{}
All properties are implied by the explicit form
\begin{equation}
y(\tau)=y(\tau_0)e^{c(\tau_0,\tau)}
  +\int_{\tau_0}^\tau a(\tau')e^{c(\tau',\tau)}\dd\tau'\;.
\end{equation}
\end{Prf}

\begin{Lemma}{}\label{Riccati}
Consider solutions of the Riccati equation $\dot y=a-y^2$.
\begin{itemize}
\item[i)]
Suppose $|a|$ is integrable on some interval $\tau_0\le\tau\le\tau_1$, then
the solution which is unbounded at $\tau_1$ behaves as
$y=1/(\tau-\tau_1)+O(1)$ as $\tau\to\tau_1$. If $a$ is bounded then
$y=1/(\tau-\tau_1)+O(\tau-\tau_1)$.

\item[ii)]
Suppose $a$ is bounded from below and $y<0$ for $\tau_0\le\tau\le\tau_1$ and
$y$ diverges as $\tau\to\tau_1$, then $y\le1/(\tau-\tau_1)+O(\tau-\tau_1)$.

\item[iii)]
Suppose $a$ has a limit $a_1$ for $\tau\to\infty$ and $y$ is bounded, then
$a_1$ must be non-negative and $y(\tau)\to\pm\sqrt{a_1}$ for
$\tau\to\infty$.

\end{itemize}
\end{Lemma}
\begin{PrfItem}{}
\begin{itemize}
\item[i)]
Without restriction we assume $\tau_1=0$, put $y=1/\tau+\eta$,
and get by integration
\begin{equation}
\eta(\tau)=\int_\tau^0\Biggl({\tau'\over\tau}\Biggr)^2
   \Bigl(\eta(\tau')^2-a(\tau')\Bigr)\dd\tau'\;.
\end{equation}
which we solve by iteration starting with $\eta_0=0$. Putting
\\
$A=\int_{\tau_0}^0(\tau'/\tau)^2|a(\tau')|\dd\tau'$ and
$||f||={\rm sup}\{|f(\tau)|,\tau_0\le\tau<0\}$ we can estimate
$||\eta_n||\le A(1-|\tau_0|/3)^{-1}<1$ choosing $|\tau_0|$ small enough. From
this we get $||\eta_{n+1}-\eta_n||\le(2|\tau_0|/3)^n||\eta_1||$ and thus
convergence for $|\tau_0|<3/2$. For $|a|\le B$ and $\tau_0^2B\le1$ we can
estimate $||\eta_n/\tau||\le B/2$.

\item[ii)]
Let $A$ be a constant such that $a\ge A$. We assume again $\tau_1=0$, put
$\eta=1/(\tau y)$, and get by integration
\begin{equation}
\eta(\tau)=1+\int_\tau^0{\tau'^2\over\tau}a\eta^2\,\dd\tau'
  \le1+A\int_\tau^0{\tau'^2\over\tau}\eta^2\,\dd\tau'\;.
\end{equation}
For $A\ge0$ we obtain immediately $\eta\le1$ and thus $y\le1/\tau$. For
$A<0$ we choose $|\tau_0|$ sufficiently small such that $|A|\tau_0^2\le1/2$
and can estimate $\eta\le1+|A|\tau^2$ and consequently $y\le1/\tau-|A|\tau$.

\item[iii)]
Obviously $a_1$ must be non-negative, otherwise $y$ would be unbounded from
below. For any $\epsilon>0$ we can find $\tau_\epsilon$ such that
$|a-a_1|<\epsilon/2$ for $\tau>\tau_\epsilon$. If
$y(\tau_\epsilon)<-\sqrt{a_1+\epsilon}$, then $y$ will be unbounded from
below. Now suppose that $y(\tau_\epsilon)>\sqrt{a_1+\epsilon}$, then we can
find $\tau'_\epsilon>\tau_\epsilon$ such that $y$ decreases for
$\tau_\epsilon<\tau\le\tau_\epsilon'$ and $y<\sqrt{a_1+\epsilon}$
for $\tau>\tau_\epsilon'$. If $a_1=0$ this implies $y\to0$ for
$\tau\to\infty$. For $\epsilon<a_1\ne0$ now suppose that
$y^2(\tau_\epsilon)<a1-\epsilon$, then we can find
$\tau'_\epsilon>\tau_\epsilon$ such that $y$ increases for
$\tau_\epsilon<\tau\le\tau_\epsilon'$ and $y>\sqrt{a_1-\epsilon}$
for $\tau>\tau_\epsilon'$. This implies $y\to\sqrt{a_1}$ for
$\tau\to\infty$. On the other hand, if
$-\sqrt{a_1+\epsilon}<y(\tau_\epsilon)\le-\sqrt{a_1-\epsilon}$ for any
$\epsilon$ then $y\to-\sqrt{a_1}$.
\end{PrfItem}

\begin{Lemma}{}\label{Riccatilin}
Consider a solution of the Riccati equation $\dot y=a+by-y^2$. If $a$ is
bounded from above and $b$ is bounded for $\tau\ge\tau_0$, then $y$ is
either bounded for all $\tau\ge\tau_0$ or diverges to $-\infty$ for some
finite $\tau_1>\tau_0$.
\end{Lemma}
\begin{Prf}{}
Let $A$, $B$ be positive constants such that $a<A$ and $|b|<B$. We can
estimate $\dot y<0$ for $|y|>C=\sqrt{2A}+2B$ and therefore $y$ is bounded
from above. Furthermore $y$ monotonically decreases and $(1/y)\dot{}>1/2$
for $y<-C$, and thus $y\to-\infty$ for some finite $\tau_1$.
\end{Prf}

%\newpage

\section*{Acknowledgements}

We would like to thank H.~Giacomini and M.T.~Grau for helpful discussions on
dynamical systems. We also thank the Theoretical Physics groups of the
Max-Planck-Institute for Physics resp.\ University of Tours for their kind
hospitality during several visits, where part of the work was done.

\newpage

\mkfigs

\end{document}